\documentclass[traditabstract]{aa} % for the abstract without structuration 
\usepackage{graphicx}
\usepackage{txfonts}
\usepackage{natbib}
\usepackage{color}
\usepackage{multirow}
\usepackage{subfigure}
\usepackage{lscape}
\usepackage{latexsym}
\usepackage{changepage}		%for adjustwidth
\bibpunct{(}{)}{;}{a}{}{,} % to follow the A&A style
\usepackage[usenames,dvipsnames,svgnames,table]{xcolor}
\usepackage[breaklinks, colorlinks, citecolor=CornflowerBlue]{hyperref}
\usepackage{url}

\begin{document}
\definecolor{Red}{rgb}{1,0,0}
\authorrunning{Remco F.J. van der Burg et al.,}
%Hakon, Herve, Stefano
%Monique, Gabriel, Jessica, La Palma guys?
%Adam, Sean, other GOGREEN/SpARCS people?
   \title{The stellar mass function of galaxies in \textit{Planck}-selected clusters at $0.5 < z < 0.7$: new constraints on the timescale and location of satellite quenching}
%   \titlerunning{The SMF of galaxies in massive clusters at $0.5<z<0.7$}
   \titlerunning{Environmental quenching of cluster satellite galaxies at $0.5<z<0.7$}
   \author{Remco~F.~J.~van der Burg\inst{1,2,3}\thanks{\email{rvanderb@eso.org}}, Sean~McGee\inst{4},  Herv\'e~Aussel\inst{2,3}, H\aa kon~Dahle\inst{5}, \\Monique~Arnaud\inst{2,3}, Gabriel~W.~Pratt\inst{2,3}, Adam~Muzzin\inst{6}   } %, H.~Aussel\inst{1}, G.~W.~Pratt\inst{1}, M.~Arnaud\inst{1}, J.-B.~Melin\inst{2}, N.~Aghanim\inst{3}, R.~Barrena\inst{4,5}, H.~Dahle\inst{6}, M.~Douspis\inst{3}, A.~Ferragamo\inst{4,5}, S.~Fromenteau\inst{7}, R.~Herbonnet\inst{8}, G.~Hurier\inst{3,9}, E.~Pointecouteau\inst{10,11}, J.~A.~Rubi\~no-Mart\'in\inst{4,5}, A.~Streblyanska\inst{4,5}}
    %R.F.J. van der Burg, H. Aussel, G.W. Pratt, M. Arnaud, J.-B. Melin, N. Aghanim, R. Barrena, H. Dahle, M. Douspis, A. Ferragamo, S. Fromenteau, R. Herbonnet, G. Hurier, E. Pointecouteau, J.A. Rubino-Martin, A. Streblyanska
	\institute{European Southern Observatory, Karl-Schwarzschild-Str. 2, 85748, Garching, Germany
	\and IRFU, CEA, Universit\'e Paris-Saclay, F-91191 Gif-sur-Yvette, France
   \and Universit\'e Paris Diderot, AIM, Sorbonne Paris Cit\'e, CEA, CNRS, F-91191 Gif-sur-Yvette, France	   
\and School of Physics and Astronomy, University of Birmingham, Edgbaston, Birmingham B15 2TT, England
\and Institute of Theoretical Astrophysics, University of Oslo, P.O. Box 1029, Blindern, N-0315 Oslo, Norway
\and Department of Physics and Astronomy, York University, 4700 Keele St., Toronto, Ontario, Canada, MJ3 1P3
                    }

   \date{Submitted 5 June 2018; accepted 29 June 2018}
% \abstract{}{}{}{}{} 
% 5 {} token are mandatory

  \abstract{We study the abundance of star-forming and quiescent galaxies in a sample of 21 clusters at $0.5<z<0.7$, detected with the \textit{Planck} satellite. Thanks to the large volume probed by \textit{Planck}, these systems are extremely massive, and provide an excellent laboratory to study any environmental effects on their galaxies' properties. We measure the cluster galaxy stellar mass function (SMF), which is a fundamental observable to study and constrain the formation and evolution of galaxies. Our measurements are based on homogeneous and deep multi-band photometry spanning $u$- to the $\mathrm{K_s}$-band for each cluster and are supported by spectroscopic data from different programs. The galaxy population is separated between quiescent and star-forming galaxies based on their rest-frame U-V and V-J colours. The SMF is compared to that of field galaxies at the same redshifts, using data from the COSMOS/UltraVISTA survey. We find that the shape of the SMF of star-forming galaxies does not depend on environment, while the SMF of quiescent galaxies has a significantly steeper low-mass slope in the clusters compared to the field. This indicates that a different quenching mechanism is at play in clusters compared to the field, accentuated by a quenched fraction that is much higher in the clusters. We estimate the environmental quenching efficiency ($f_{\mathrm{EQ}}$), i.e.~the probability for a galaxy that would normally be star forming in the field, to be quenched due to its environment. The $f_{\mathrm{EQ}}$ shows no stellar-mass dependence in any environment, but it increases from 40\% in the cluster outskirts to $\sim90\%$ in the cluster centres. The radial signature of $f_{\mathrm{EQ}}$ provides constraints on where the dominant quenching mechanism operates in these clusters and on what timescale. Exploring these using a simple model based on galaxy orbits obtained from an N-body simulation, we find a clear degeneracy between both parameters. For example, the quenching process may either be triggered on a long ($\sim$3 Gyr) time scale at large radii ($r\sim 8R_{500}$), or happen well within 1 Gyr at $r<R_{500}$. The radius where quenching is triggered is at least $r_{\mathrm{quench}}> 0.67R_{500}$ (95\%CL). The ICM density at this location (as probed with \textit{XMM-Newton}), suggests that ram-pressure stripping of the cold gas is a likely cause of quenching. In addition to this cluster-quenching mechanism we find that 20-32\%, depending on the cluster-specific quenching process, of accreted galaxies were already pre-processed (i.e.~quenched by the surrounding overdensities) before they fell into the clusters.} 
   \maketitle
%
%________________________________________________________________

\hyphenation{in-tra-clus-ter}
\hyphenation{rank-or-der}

\section{Introduction}

Over the past decades we have obtained an increasingly clear picture of the formation and evolution of galaxies in the Universe. The galaxy population can be broadly divided in two distinct types. Star-forming galaxies have blue colours, typically disk-like morphologies and a relatively high star formation rate, whereas quiescent galaxies have redder colours, more spheroidal morphologies, and a (near) absence of star formation. Generally speaking, star-forming galaxies are found to dominate in abundance at relatively early times and at low stellar masses \citep{kauffmann04,baldry06,peng10,muzzin13b}. A central question is how galaxies transform from actively star-forming systems to passive quiescent galaxies. An important factor in this quest is to understand how the environment of a galaxy affects this transformation process, since at fixed stellar mass, galaxies in groups and clusters are found to have a higher probability of being quiescent than galaxies in the field \citep{dressler80,blanton05,woo13}. 

In recent years it has been shown that the effect of the environment is largely separable from the quenching processes that act internally, at least in the local universe \citep{baldry06,peng10,kovac2014}. Specifically, \citet{peng10} introduced terms of ``mass quenching'', which would be operating independently of the environment and most effectively quench massive galaxies, and ``environmental quenching'', which would be operating independently of stellar mass and quench galaxies preferentially in overdense regions \citep[also see e.g.][]{davies16,kawinwanichakij17}. The total quenching effect would simply be the product of these two contributions. 

What the physics are behind the two separate quenching routes is a matter of debate. ``Mass quenching'' is often associated with feedback, either from supernovae and galactic winds \citep{oppenheimer10}, or from active galactic nuclei \citep{bower06,croton06,tremonti07}. Many mechanisms have been proposed to be responsible for ``environmental quenching''. In the overdensities of groups and clusters stripping of cold gas \citep[through ram-pressure:][]{Gunn1972}, or hot gas \citep[``strangulation''/``starvation'':][]{Larson1980}, as well as galaxy harassment or (dry) mergers \citep{Moore1996}, could be responsible for the observed trends. To make progress in understanding these environmental processes, it is essential to understand in which environments and on which timescales they operate \citep{balogh04}. 

Most studies that aim to understand the process of environmental quenching are focussed on large cosmological volumes in for instance the COSMOS field \citep{peng10,darvish16}, VIPERS \citep{davidzon16}, 3D-HST \citep{fossati17}, the Subaru Hyper Suprime-Cam survey \citep{jian18}, and ZFOURGE \citep{kawinwanichakij17,papovich18}. Most of these studies separate the galaxy population in four density quartiles, so that the environmental effects can be studied between the different quartiles. At least at high redshift ($z\gtrsim 1$) it is found that the environmental quenching process is not working completely independently of stellar mass \citep{papovich18}. A starvation/strangulation scenario in which the supply of hot gas is cut off from a galaxy would be highly effective at these redshifts, where star formation rates and outflows deplete the cold gas supply \citep[the ``overconsumption model'',][]{mcgee14}. This would likely introduce a mass-dependent effect, since the gas depletion time of higher-mass galaxies ought to be shorter, and this could explain the measured trends. Also for lower mass galaxies ($M_{\star} \lesssim 10^{9.5}\,\mathrm{M_{\odot}}$), measured quenching time scales are comparable to the gas depletion timescale \citep{davies16}.

Whereas such a division in density quartiles allows for a study of environmental quenching in reasonably overdense regions, the most extreme environments are either not probed, or are washed out by more moderate overdensities. And yet it is in these regions where the physics of the quenching may be notably different \citep{balogh16,kawinwanichakij17}. Quenching processes may be more violent, leading to very high quenched fractions of cluster galaxies compared to the field at the same redshift \citep{delucia04,vdB13,annunziatella14,balogh16,nantais16,nantais17,wagner17}. Dynamical processes in clusters may become more important at later time to quench star formation, and these may act in a mass-independent fashion. A notable example of this is stripping by ram pressure, which would directly remove the cold gas supply from a galaxy and quench its star formation on a very short timescale \citep{zinger18,fossati16,bellhouse17,jaffe18}.

To understand where and when the most extreme environmental quenching is taking place in clusters, several studies have focussed on a clustercentric-distance dependent study, some even focussed on projected phase space of different galaxy populations \citep[such as those of galaxies in the ``transition'' phase; ][]{Oman2013,muzzin14,poggianti16,jaffe18}. To study the relative excess of already-quenched galaxies in clusters compared to the field, single values of the environmental quenching efficiency are typically reported for satellite galaxies \citep[cf.~Fig.~7 in][and references therein]{nantais16}, even though there is likely to be a substantial trend with radius. Such a study as a function of radius can probe the ``backsplash'' of already-quenched ejected cluster satellites \citep{wetzel14}, and pre-processing/quenching of future cluster satellites in the surrounding large-scale overdensity \citep{fujita04}.

While new surveys push the frontiers of these studies to higher redshifts \citep[e.g.][]{balogh17}, their samples are thus-far limited in size, and clusters are of moderate mass and over-density. This renders a radial-dependent study of environmental quenching difficult. In this paper we focus instead on highly massive, and thus over-dense, clusters at intermediate redshifts. This allows us to specifically study the environmental quenching excess as a function of radial distance from the cluster centres. The sample we study is composed of 21 massive clusters detected with the \textit{Planck} SZ survey at redshifts $0.5<z<0.7$. Since \textit{Planck} is an all-sky survey (even though we only consider the northernmost 2/3 here), we probe the highest-density environments at these intermediate redshifts, where environmental effects are expected to be substantial. We concentrate on a photometric data set spanning $u$- to the $\mathrm{K_s}$-band for each cluster, using which we estimate stellar masses for individual galaxies, and separate them by type based on their best-fit SED.

Our starting point is a measurement of the galaxy Stellar Mass Function (SMF), which describes the number density of galaxies as a function of their stellar mass, and which is a key observable to study the formation and evolution of galaxies. By comparing the SMF to the underlying halo mass function, the efficiency with which galaxies form can be measured, and this is an essential test and diagnostic tool for large hydrodynamical simulations such as Illustris \citep{genel14} and EAGLE \citep{schaye15}. Measuring the SMF in these massive clusters provides further constraints for the next generation of hydrodynamical simulations, in which large overdensities can be specifically focussed on to study the influence of such environments on the evolution of galaxies \citep[e.g.][]{bahe17}. These SMFs are the main ingredients to estimate environmental quenching locations and timescales, which we describe and discuss in the remainder of this work.

The structure of this paper is as follows. Section \ref{sec:sample} describes the cluster sample and photometric data set we utilise. Section \ref{sec:analysis} lays out the main analysis, ranging from photometric redshift measurements to a statistical accounting of fore- and background galaxies in our cluster galaxy sample. The main results, measurements of the SMF and environmental quenching efficiency, are presented in Sect.~\ref{sec:SMF}~\&~\ref{sec:EQE}, respectively. We discuss our findings in Sect.~\ref{sec:discussion}, and conclude in Sect.~\ref{sec:summary}.

All magnitudes we quote are in the AB magnitudes system, and we adopt $\Lambda$CDM cosmology with $\Omega_{\mathrm{m}}=0.3$, $\Omega_{\Lambda}=0.7$ and $\mathrm{H_0=70\, km\, s^{-1}\,  Mpc^{-1}}$. Uncertainties are given at the 1-$\sigma$ level, unless explicitly stated otherwise.

\section{Cluster Sample \& Data}\label{sec:sample}
The clusters we study are drawn from a sample of 33 clusters that were detected with \textit{Planck}, and confirmed by autumn 2011 to be at $z>0.5$. This sample was the target of an XMM-Newton Large Programme `Unveiling the most massive galaxy clusters at $z > 0.5$ with Planck and XMM-Newton' (PI M. Arnaud), in AO-11. The properties of the sample, in particular regarding their morphological properties, Intra Cluster Medium and the cluster scaling relations, are outlined in Arnaud et al., in prep. 

In this paper, we study the galaxy content in a sub-sample of 21 clusters that make up the northernmost ($\mathrm{Dec} > -25^{\circ}$) part of this parent sample. Several of these clusters were already priorly studied, particularly as part of X-ray selected samples of clusters in this redshift range \citep{bohringer00,ebeling07,piffaretti11}. Several are in optical catalogues constructed using SDSS data \citep{wen12,rykoff14}. 

Table~\ref{tab:dataoverview} presents the main characteristics of the sample. It makes a comparison of the mass estimated from the \textit{Planck} SZ signal, and mass based on the deep X-ray maps (M-$Y_X$ relation). Even though the SZ mass proxy is blindly extracted \citep[i.e. without prior knowledge on the location of the cluster, cf.][]{psz2}, both proxies are consistent at the massive end (within $\sim$10\% in mass). They slightly diverge at the low-mass end due to Eddington bias in the SZ mass proxy \citep[cf. e.g.][]{vdB16}. The differences between $Y_X$ and $Y_{SZ}$ will be discussed in more detail in Arnaud et al., in prep. The current paper refers to cluster masses as $M_{500}$\footnote{All quoted masses in this paper are defined with respect to the critical density at the cluster redshift. $R_{500}$ is thus defined as the radius at which the mean interior density is 500 times the critical density, and $M_{500}$ is the mass contained within this radius. We will occasionally use an overdensity of 200 in an analogous fashion.}, based on the $Y_X$ scaling relation. 

To support our analysis, we combine different sources of spectroscopic information in the 21 fields we study. Fourteen of the clusters are covered in DR14 of SDSS \citep{sdssDR14}. For nine clusters we have obtained redshifts with the Nordic Optical Telescope \citep[][Dahle et al., in prep.]{psz1} or Gemini \citep{planckxmmvalidation13}. \citet{ebeling14} publish a catalogue with hundreds of spectroscopic redshifts in two of the fields we study (\texttt{PSZ2 G180.25+21.03} and \texttt{PSZ2 G228.16+75.20}). For \texttt{PSZ2 G046.13+30.72} and \texttt{PSZ2 G155.27-68.42} we have obtained spectroscopic redshifts from a program undertaken with the Canary Islands observatories \citep{planckcanary16}. %\footnote{had to redo the astrometry from the original masks}. 
We searched the NED database\footnote{https://ned.ipac.caltech.edu/} for any spec-$z$s we may have missed in the literature. The only two clusters that remain without a previously-measured spectroscopic redshift (\texttt{PSZ2 G193.31-46.13} and \texttt{PSZ2 G219.89-34.39}) have been observed using VLT/FORS2 multi-object spectroscopy (PID=090.A-0925, PI=Bohringer). Using our own custom pipeline we reduced these spectra and measured the redshifts of several member galaxies. The number of (unique) spectroscopic targets and cluster members for all clusters are listed in Table~\ref{tab:dataoverview}.

\begin{table*}%[ht]
\caption{Overview of the cluster sample studied here.} 
\label{tab:dataoverview}
\begin{center}
\begin{adjustwidth}{-0.30cm}{}
\begin{tabular}{l l l l r r l l}
\hline
\hline
&&&& $\mathrm{Y_{X}}-M_{500}$ & $\mathrm{Y_{SZ}}-M_{500}$& $R_{500,\mathrm{Y_{X}}}$&\\
Name & Redshift$^{\mathrm{a}}$ & RA$_\mathrm{J2000}^{\mathrm{BCG}}$ & Dec$_\mathrm{J2000}^{\mathrm{BCG}}$ & [$10^{14}\, \mathrm{M_{\odot}}$]& [$10^{14}\, \mathrm{M_{\odot}}$] & [kpc] & Alternative Name \\
\hline
 \texttt{PSZ2 G044.77-51.30} & 0.503(3/1) & 22:14:57.27 & $-$14:00:12.7 & $7.95^{+0.44}_{-0.43}$ & $8.36^{+0.61}_{-0.62}$&${1175}^{+21}_{-22}$&MACSJ2214.9-1359\\
 \texttt{PSZ2 G045.32-38.46} & 0.589(10/1) & 21:29:26.13 & $-$07:41:27.6 & $7.36^{+0.66}_{-0.64}$ & $7.63^{+0.64}_{-0.68}$&${1107}^{+32}_{-33}$&MACSJ2129.4-0741\\
 \texttt{PSZ2 G045.87+57.70} & 0.609(87/22) & 15:18:20.56 & +29:27:40.2 & $5.82^{+0.22}_{-0.22}$ & $7.03^{+0.66}_{-0.71}$&${1016}^{+13}_{-13}$&\\
 \texttt{PSZ2 G046.13+30.72} & 0.569(67/17) & 17:17:05.55 & +24:04:23.7 & $3.17^{+0.22}_{-0.22}$ & $6.39^{+0.80}_{-0.84}$&${\,\,\,843}^{+19}_{-20}$&\\
 \texttt{PSZ2 G070.89+49.26} & 0.602(94/34) & 15:56:25.24 & +44:40:42.6 & $5.02^{+0.20}_{-0.21}$ & $6.46^{+0.65}_{-0.72}$&${\,\,\,970}^{+13}_{-14}$&\\
 \texttt{PSZ2 G073.31+67.52} & 0.609(110/35) & 14:20:40.11 & +39:55:06.9 & $6.15^{+0.26}_{-0.25}$ & $6.74^{+0.55}_{-0.63}$&${1035}^{+14}_{-14}$&WHL J215.168+39.91\\
 \texttt{PSZ2 G094.56+51.03} & 0.539(47/20) & 15:08:21.98 & +57:55:14.9 & $6.15^{+0.25}_{-0.24}$ & $5.87^{+0.44}_{-0.43}$&${1064}^{+14}_{-14}$&WHL J227.050+57.90\\
 \texttt{PSZ2 G099.86+58.45} & 0.615(104/16) & 14:14:47.20 & +54:47:03.5 & $7.09^{+0.42}_{-0.42}$ & $6.85^{+0.48}_{-0.49}$&${1082}^{+21}_{-22}$&WHL J213.697+54.78\\
 \texttt{PSZ2 G111.61-45.71} & 0.546(388/187) & 00:18:33.58 & +16:26:15.9 & $9.21^{+0.24}_{-0.24}$ & $9.79^{+0.53}_{-0.53}$&${1214}^{+11}_{-11}$&RXC J0018.5+1626\\
 \texttt{PSZ2 G144.83+25.11} & 0.591(4/1) & 06:47:50.65 & +70:14:54.0 & $7.78^{+0.21}_{-0.20}$ & $8.25^{+0.71}_{-0.73}$&${1127}^{+10}_{-10}$&MACSJ0647.7+7015\\
 \texttt{PLCK G147.30-16.60$^{\mathrm{b}}$} & 0.645(8/8) & 02:56:23.45 & +40:17:28.9 & $6.51^{+0.29}_{-0.28}$ & $7.41^{+0.80}_{-0.86}$&${1040}^{+15}_{-15}$&RXC J0254.4+4134\\
 \texttt{PSZ2 G155.27-68.42} & 0.567(68/24) & 01:37:24.98 & $-$08:27:22.9 & $8.01^{+0.46}_{-0.38}$ & $8.93^{+0.65}_{-0.70}$&${1149}^{+22}_{-18}$&WHL J24.3324-8.477\\
 \texttt{PSZ2 G180.25+21.03} & 0.546(1151/529) & 07:17:35.63 & +37:45:17.4 & $12.83^{+0.17}_{-0.17}$ & $11.49^{+0.53}_{-0.55}$&${1356}^{+6}_{-6}$&MACSJ0717.5+3745\\
 \texttt{PSZ2 G183.90+42.99} & 0.559(94/25) & 09:10:51.05 & +38:50:22.3 & $8.44^{+0.60}_{-0.53}$ & $6.95^{+0.73}_{-0.75}$&${1173}^{+27}_{-25}$&WHL J137.713+38.83\\
 \texttt{PSZ2 G193.31-46.13} & 0.634(45/2) & 03:35:52.00 & $-$06:59:23.4 & $5.49^{+0.30}_{-0.32}$ & $6.07^{+0.75}_{-0.83}$&${\,\,\,986}^{+18}_{-19}$&\\
 \texttt{PSZ2 G201.50-27.31} & 0.538(1181/278) & 04:54:10.83 & $-$03:00:51.5 & $7.90^{+0.30}_{-0.29}$ & $8.30^{+0.70}_{-0.73}$&${1157}^{+14}_{-14}$&MACSJ0454.1-0300\\
 \texttt{PSZ2 G208.61-74.39} & 0.718(10/5) & 02:00:16.38 & $-$24:54:51.5 & $5.23^{+0.23}_{-0.23}$ & $6.25^{+0.72}_{-0.79}$&${\,\,\,939}^{+14}_{-14}$&\\
 \texttt{PSZ2 G211.21+38.66} & 0.505(46/18) & 09:11:11.52 & +17:46:29.1 & $5.48^{+0.22}_{-0.22}$ & $6.99^{+0.73}_{-0.79}$&${1038}^{+14}_{-14}$&RXC J0911.1+1746\\
 \texttt{PSZ2 G212.44+63.19} & 0.529(56/14) & 10:52:48.75 & +24:16:11.3 & $4.15^{+0.23}_{-0.23}$ & $5.62^{+0.80}_{-0.90}$&${\,\,\,937}^{+17}_{-18}$&RMJ105252.4+241530.0\\
 \texttt{PSZ2 G219.89-34.39} & 0.734(15/5) & 04:54:45.32 & $-$20:16:58.8 & $6.77^{+0.33}_{-0.29}$ & $7.97^{+0.61}_{-0.67}$&${1016}^{+16}_{-15}$&\\
 \texttt{PSZ2 G228.16+75.20} & 0.544(585/285) & 11:49:35.68 & +22:23:54.7 & $9.36^{+0.64}_{-0.62}$ & $10.42^{+0.52}_{-0.55}$&${1221}^{+27}_{-27}$&RXC J1149.5+2224\\
\hline
\end{tabular}
\end{adjustwidth}
\end{center}
\begin{list}{}{}
\item[$^{\mathrm{a}}$] In parentheses the number of spectroscopic redshifts overlapping with the region for which we have photometry, and the number of spectroscopic cluster members (within 3000 km/s from the cluster mean redshift), respectively.
\item[$^{\mathrm{b}}$] Cluster detected at a significance slightly below the cut-off value used for the PSZ2 catalogue. The $\mathrm{Y_{SZ}}$ is measured on the final Planck maps.
\end{list}
\end{table*}

\subsection{Cluster Photometry}
\begin{table*}%[ht]
\caption{Photometric data set used in this work. The reported magnitudes are median 5-$\sigma$ limits measured on the PSF-homogenized stacked images in circular apertures with a diameter of 2$''$. The values listed are after correction for Galactic dust, so are indicative of the galaxy population we study. The instruments and filters used for the different clusters are indicated. Whenever there are two observations with multiple instruments, both are used in the analysis, but only the magnitude corresponding to the deepest stack is reported below.}
\label{tab:photometry}
\begin{center}
\begin{adjustwidth}{-0.90cm}{}
\begin{tabular}{l l l l l l l l l l l l l l l}
\hline
\hline
Name & $\mathrm{K_{s,det}}$ & PSF &$\mathrm{M_{\star,det}/M_{\odot}}$&$u$&$B$&$g$&$V$&$r$&$R_c$&$i$&$I_c$&$z$&J&$\mathrm{K_s}$\\
 &  & FWHM &dex&&&&&&&&&&&\\
\hline
\texttt{PSZ2 G044.77-51.30}&23.14&0.55$''$&9.25&25.4$^\mathrm{a}$&26.0$^\mathrm{e}$&$-$&25.9$^\mathrm{e}$&24.8$^\mathrm{a}$&26.0$^\mathrm{e}$&$-$&25.4$^\mathrm{e}$&24.9$^\mathrm{f}$&23.4$^\mathrm{g}$&23.3$^\mathrm{g}$\\
\texttt{PSZ2 G045.32-38.46}&23.19&0.57$''$&9.36&25.5$^\mathrm{a}$&25.6$^\mathrm{e}$&$-$&25.3$^\mathrm{e}$&$-$&25.2$^\mathrm{e}$&$-$&24.9$^\mathrm{e}$&24.9$^\mathrm{d}$&23.8$^\mathrm{g}$&23.3$^\mathrm{g}$\\
\texttt{PSZ2 G045.87+57.70}&22.66&0.54$''$&9.61&25.5$^\mathrm{b}$&$-$&26.1$^\mathrm{d}$&$-$&25.9$^\mathrm{d}$&$-$&25.4$^\mathrm{d}$&$-$&23.8$^\mathrm{b}$&23.0$^\mathrm{g}$&22.5$^\mathrm{g}$\\
\texttt{PSZ2 G046.13+30.72}&22.38&0.66$''$&9.68&25.2$^\mathrm{b}$&$-$&25.7$^\mathrm{d}$&$-$&25.8$^\mathrm{d}$&$-$&24.7$^\mathrm{d}$&$-$&24.1$^\mathrm{d}$&22.8$^\mathrm{g}$&22.3$^\mathrm{g}$\\
\texttt{PSZ2 G070.89+49.26}&22.27&0.64$''$&9.75&25.5$^\mathrm{b}$&$-$&26.2$^\mathrm{d}$&$-$&25.8$^\mathrm{d}$&$-$&25.1$^\mathrm{d}$&$-$&24.0$^\mathrm{b}$&23.1$^\mathrm{g}$&22.4$^\mathrm{g}$\\
\texttt{PSZ2 G073.31+67.52}&22.44&0.55$''$&9.69&25.7$^\mathrm{b}$&$-$&26.3$^\mathrm{a}$&$-$&25.4$^\mathrm{a}$&$-$&25.0$^\mathrm{b}$&$-$&23.9$^\mathrm{b}$&23.2$^\mathrm{g}$&22.6$^\mathrm{g}$\\
\texttt{PSZ2 G094.56+51.03}&22.38&0.66$''$&9.62&25.3$^\mathrm{b}$&$-$&26.1$^\mathrm{ad}$&$-$&25.8$^\mathrm{d}$&$-$&25.0$^\mathrm{ad}$&$-$&23.5$^\mathrm{b}$&23.0$^\mathrm{g}$&22.5$^\mathrm{g}$\\
\texttt{PSZ2 G099.86+58.45}&22.31&0.57$''$&9.75&25.6$^\mathrm{a}$&$-$&26.0$^\mathrm{ad}$&$-$&25.4$^\mathrm{ad}$&$-$&24.7$^\mathrm{cd}$&$-$&23.6$^\mathrm{a}$&23.1$^\mathrm{g}$&22.4$^\mathrm{g}$\\
\texttt{PSZ2 G111.61-45.71}
&22.25&0.72$''$&9.68&25.5$^\mathrm{a}$&26.5$^\mathrm{e}$&25.5$^\mathrm{a}$&26.3$^\mathrm{e}$&25.5$^\mathrm{a}$&26.2$^\mathrm{e}$&26.0$^\mathrm{cd}$&25.7$^\mathrm{e}$&25.0$^\mathrm{af}$&23.5$^\mathrm{g}$&22.6$^\mathrm{g}$\\
\texttt{PSZ2 G144.83+25.11}&22.95&0.78$''$&9.46&25.0$^\mathrm{a}$&25.7$^\mathrm{e}$&$-$&25.4$^\mathrm{e}$&$-$&25.5$^\mathrm{e}$&$-$&25.2$^\mathrm{e}$&25.0$^\mathrm{d}$&23.3$^\mathrm{g}$&23.2$^\mathrm{g}$\\
\texttt{PLCK G147.30-16.60}&22.57&0.52$''$&9.68&24.5$^\mathrm{b}$&$-$&25.8$^\mathrm{d}$&$-$&25.4$^\mathrm{d}$&$-$&25.1$^\mathrm{d}$&$-$&23.6$^\mathrm{b}$&23.2$^\mathrm{g}$&22.7$^\mathrm{g}$\\
\texttt{PSZ2 G155.27-68.42}&22.36&0.59$''$&9.69&24.6$^\mathrm{b}$&$-$&25.7$^\mathrm{b}$&$-$&25.3$^\mathrm{b}$&$-$&24.3$^\mathrm{b}$&$-$&23.8$^\mathrm{b}$&23.3$^\mathrm{g}$&22.6$^\mathrm{g}$\\
\texttt{PSZ2 G180.25+21.03}
&23.13&0.65$''$&9.33&25.6$^\mathrm{a}$&25.9$^\mathrm{e}$&25.6$^\mathrm{a}$&25.7$^\mathrm{e}$&25.3$^\mathrm{a}$&25.5$^\mathrm{e}$&24.8$^\mathrm{d}$&$-$&25.0$^\mathrm{d}$&23.1$^\mathrm{g}$&23.3$^\mathrm{g}$\\
\texttt{PSZ2 G183.90+42.99}&22.51&0.64$''$&9.58&25.4$^\mathrm{b}$&$-$&25.9$^\mathrm{d}$&$-$&25.9$^\mathrm{d}$&$-$&25.5$^\mathrm{d}$&$-$&23.9$^\mathrm{b}$&23.5$^\mathrm{g}$&22.9$^\mathrm{g}$\\
\texttt{PSZ2 G193.31-46.13}&22.45&0.59$''$&9.73&24.8$^\mathrm{b}$&$-$&25.8$^\mathrm{b}$&$-$&25.1$^\mathrm{b}$&$-$&24.8$^\mathrm{b}$&$-$&23.6$^\mathrm{b}$&23.1$^\mathrm{g}$&22.7$^\mathrm{g}$\\
\texttt{PSZ2 G201.50-27.31}
&23.08&0.56$''$&9.34&25.7$^\mathrm{a}$&26.2$^\mathrm{e}$&25.6$^\mathrm{a}$&26.0$^\mathrm{e}$&25.6$^\mathrm{a}$&26.0$^\mathrm{e}$&24.7$^\mathrm{c}$&25.5$^\mathrm{e}$&25.1$^\mathrm{ad}$&23.5$^\mathrm{g}$&23.4$^\mathrm{g}$\\
\texttt{PSZ2 G208.61-74.39}&21.98&0.69$''$&10.00&25.0$^\mathrm{b}$&$-$&25.8$^\mathrm{b}$&$-$&25.2$^\mathrm{b}$&$-$&24.5$^\mathrm{b}$&$-$&23.8$^\mathrm{b}$&23.3$^\mathrm{g}$&22.4$^\mathrm{g}$\\
\texttt{PSZ2 G211.21+38.66}
&22.94&0.78$''$&9.33&25.9$^\mathrm{a}$&26.1$^\mathrm{e}$&25.2$^\mathrm{a}$&26.1$^\mathrm{e}$&24.9$^\mathrm{a}$&26.1$^\mathrm{e}$&24.0$^\mathrm{c}$&25.4$^\mathrm{e}$&25.1$^\mathrm{ad}$&23.0$^\mathrm{g}$&23.3$^\mathrm{g}$\\
\texttt{PSZ2 G212.44+63.19}&22.54&0.67$''$&9.52&25.4$^\mathrm{b}$&$-$&26.0$^\mathrm{d}$&$-$&26.0$^\mathrm{d}$&$-$&25.5$^\mathrm{d}$&$-$&23.7$^\mathrm{b}$&23.7$^\mathrm{g}$&22.7$^\mathrm{g}$\\
\texttt{PSZ2 G219.89-34.39}&22.08&0.74$''$&9.96&24.9$^\mathrm{b}$&$-$&25.7$^\mathrm{b}$&$-$&25.4$^\mathrm{b}$&$-$&24.4$^\mathrm{b}$&$-$&23.7$^\mathrm{b}$&22.9$^\mathrm{g}$&22.4$^\mathrm{g}$\\
\texttt{PSZ2 G228.16+75.20}&22.82&0.73$''$&9.45&26.0$^\mathrm{a}$&26.6$^\mathrm{e}$&$-$&26.2$^\mathrm{e}$&$-$&26.3$^\mathrm{e}$&25.3$^\mathrm{d}$&$-$&25.4$^\mathrm{d}$&23.1$^\mathrm{g}$&23.2$^\mathrm{g}$\\
\hline
\end{tabular}
\end{adjustwidth}
\end{center}
\begin{list}{}{}
\item[$^{\mathrm{a}}$]  CFHT/MegaCam filters used until Jan 2015, $i$-band after Oct 2007
\item[$^{\mathrm{b}}$]  CFHT/MegaCam filters used after Feb 2015
\item[$^{\mathrm{c}}$]  CFHT/MegaCam $i$-band filter used until June 2007
\item[$^{\mathrm{d}}$]  Subaru/Suprime-Cam fully-depleted back-illuminated CCDs (installed Jan 2009) with SDSS-like filters
\item[$^{\mathrm{e}}$]  Subaru/Suprime-Cam MIT/LL CCDs (used until Dec 2008) with Johnson/Bessell-like filters
\item[$^{\mathrm{f}}$]  Subaru/Suprime-Cam MIT/LL CCDs (used until Dec 2008) with standard $z$-band filter
\item[$^{\mathrm{g}}$]  CFHT/WIRCam
\end{list}
\end{table*}

Deep archival follow-up imaging data are available for many of these clusters \citep[in particular data retrieved from the SMOKA science archive;][]{SMOKA}, but these tend to be drawn from X-ray selected samples. Our analysis benefits from a homogeneous wavelength coverage of the full sample of 21 clusters. Studying the full sample not only statistically enhances the results presented in this study, but also ensures that we sample the full range of cluster dynamical properties, approximating a mass-selected sample. Since the spectroscopic data is of varying quality and completeness, we base our analysis almost entirely on the photometry. To be able to measure accurate and precise photometric redshifts of galaxies in the cluster fields, we require photometric coverage with at least 7 filters per cluster, ranging from the $u$-band ($\sim3000\AA$) to the $\mathrm{K_s}$-band ($\sim22,000\AA$). Accounting for the deep archival data for a sub-set of the sample, we obtained the remaining photometry through different time allocations on the wide-field imagers at CFHT (PI vdBurg, PIDs: 15AF006, 15BF005, 16BF013) and Subaru (PI Dahle, PIDs: S12B-164S, S13A-120, S15A-118).

An overview of all imaging data is given in Table~\ref{tab:photometry}. The photometric data we use have been taken over more than 10 years, and some of the instruments and filter sets have been upgraded over this time span. These differences are indicated in the footnotes of Table~\ref{tab:photometry}, and the different wavelength-responses of each filter and instrument are taken into account in our analysis.

We perform basic steps to reduce the optical imaging data, such as bias, flat-field correction and cosmic-ray removal. As an additional step, we remove background patterns, particularly fringe residuals, by using the dithered pattern of observations to differentiate signals that are fixed in position on the ccd array from sky-bound signals. This is explained and illustrated in Fig. 1 of \citet{vdB16}. For the near infrared data from our own WIRCam program (PI vdBurg) we have followed an extended dither pattern, where the cluster centre is dithered from chip to chip. This ensures a very clean background subtraction, even on scales of the intra-cluster light (ICL).

Astrometric registering has been performed with \texttt{SCAMP} \citep{scamp} using the USNO-B1 catalogue as reference, or with external catalogues from Pan-STARRS \citep{chambers16} for clusters that have been observed after the public release of their data. The astrometric precision between filters is better than 0.10$''$, ensuring that colour measurements can be done accurately. 

We automatically place masks on bright stars based on their locations in the guide star catalogue 2.3 \citep{lasker08}. After this the images are inspected manually, and masks are placed on additional diffraction spikes, reflective haloes, and image artefacts. During our analysis we take account of the reduced effective areas after masking.

\subsubsection{Object detection and colours}\label{sec:objectdetection}
We perform object detection in the original $\mathrm{K_s}$-band. Since the range in M/L between the different galaxy types is smallest in this band, this ensures a catalogue that is close to being stellar-mass selected. We use \texttt{SExtractor} to detect objects, following the criterion that at least 5 adjacent pixels have a flux density that is $>$ 1.5$\sigma$ relative to the local background RMS. 

To be able to measure colours of the same intrinsic part of the galaxies we study, differences in the PSF between clusters and filters need to be accounted for. We use \texttt{PSFEx} \citep{psfex} to determine convolution kernels that homogenise the PSF for each cluster between different filters. We then measure colours for the $\mathrm{K_s}$-band-selected sources by performing photometry in circular apertures with a 2$''$ diameter on these PSF-matched images. 

A benefit of these wide-field images is that they contain a large population of Galactic stars that can be used to calibrate the flux scale \citep[e.g.][]{SLR,gilbank11,vdB13,kelly14}. We use the effective wavelength-response curves of each detector, filter\footnote{Most of these are compiled at http://svo2.cab.inta-csic.es/theory/fps/index.php} and atmosphere model in combination with the stellar spectral library of \citet{pickles98} to create a reference stellar locus. The stellar spectral library we use\footnote{see http://www.eso.org/sci/facilities/paranal/decommissioned/isaac/ tools/lib.html} is updated with additional information in the near-infrared from \citet{ivanov04}, in addition to the original library from \citet{pickles98}. After applying magnitude offsets to our instrumental magnitudes to match this reference locus, we reach a photometric calibration with an uncertainty that depends on the filter, but is generally $\lesssim0.05$ magnitude. While this calibrates the colours of sources in our catalogues, we perform absolute flux calibration in the $\mathrm{K_s}$-band with respect to the 2MASS all-sky reference catalogue \citep{2MASS}. 

Uncertainties on aperture flux measurements of faint galaxies are dominated by fluctuations in the sky background. We estimate this noise component by randomly placing apertures on sky positions that do not overlap with sources detected in the $\mathrm{K_s}$-band. This procedure takes into account the correlated noise properties between adjacent pixels that originated from the convolution and the re-binning of the data to a common grid and PSF.

\subsection{UltraVISTA Reference Field}
There are two reasons to study a reference field, i.e. a field without a massive cluster, in our work. First, to study the impact on the evolution of galaxies by their massive host haloes, properties of galaxies at the same redshift between cluster and field are compared. Second, to study the properties of galaxies that are part of the massive clusters in our sample, we have to consider projection effects; i.e. fore- and background galaxies that enter our sample of cluster galaxies. By performing the same selection criteria on galaxies in a parallel field, we can take these projection effects into account statistically. We refer to this as ``background subtraction''. The more information (photometric redshift, distance from cluster centre) one can use, the more cleanly the background can be accounted for.

We make use of the COSMOS/UltraVISTA field, which has been extensively studied and for which a multi-band photometric catalogue is publicly available \citep{muzzin13a}. How we use this catalogue for the background correction, taking cosmic variance uncertainties due to this single field into account, is described in the Sect.~\ref{sec:bgcorrection}.

\section{Analysis}\label{sec:analysis}
\subsection{Photometric Redshifts}
\begin{figure*}
\resizebox{\hsize}{!}{\includegraphics{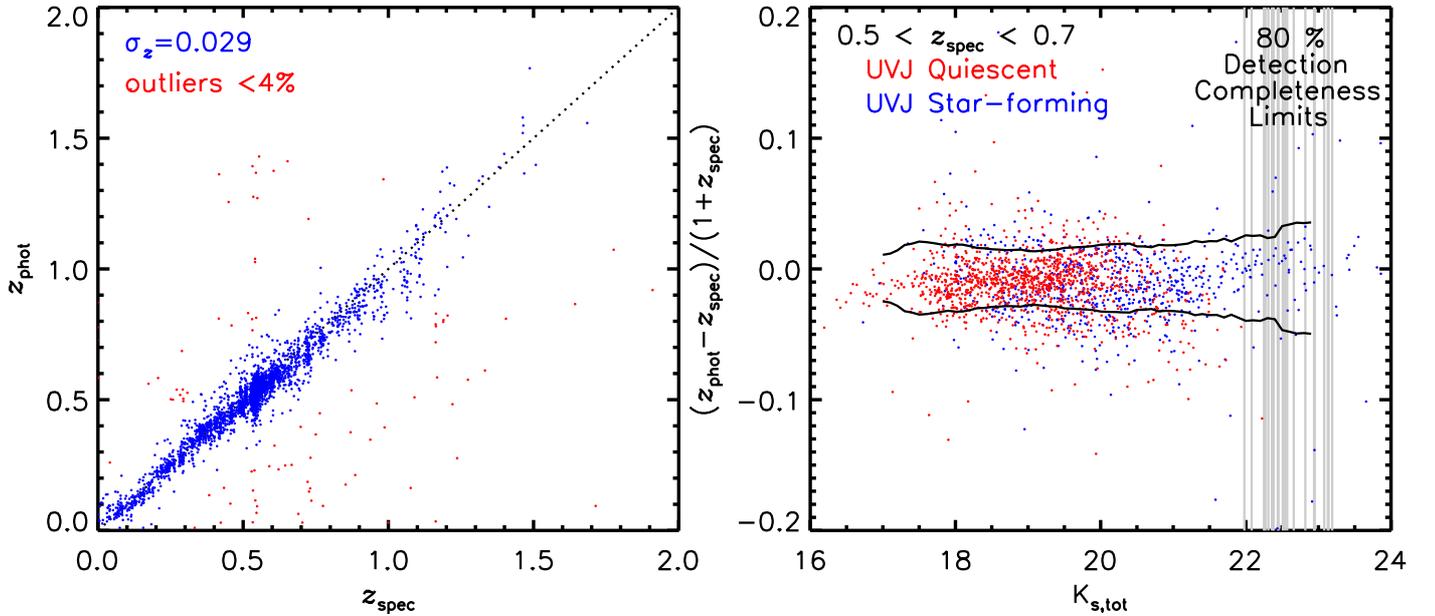}}
\caption{\textit{Left:} Spectroscopic versus photometric redshifts for the 21 cluster fields. Outliers, objects for which $|\Delta z| > 0.15$, are marked in red. The outlier fraction is $3.5\%$, the scatter of the remaining objects is $\sigma_z = 0.029$. \textit{Right:} Scatter in $\Delta z$ as a function of $\mathrm{K_s}$-band magnitude, for sources with $0.5<z_{\mathrm{spec}}<0.7$. Quiescent and star-forming galaxies are marked in red and blue, respectively. The source detection limits of the 21 $\mathrm{K_s}$-band stacks are indicated.}
\label{fig:speczphotz}
\end{figure*}

We use the template-fitting code \texttt{EAZY} \citep{brammer08} to estimate photometric redshifts (photo-$z$s) for each object. The photo-$z$s correspond to the peak ($z_\mathrm{peak}$) of the posterior probability distribution P($z$) given by \texttt{EAZY}. Figure \ref{fig:speczphotz}a shows a comparison between the spectroscopic redshift and photometric redshift. We define a relative scatter  $\Delta z = \frac{z_{\mathrm{phot}}-z_{\mathrm{spec}}}{1+z_{\mathrm{spec}}}$ for each object with a reliable spectroscopic redshift $z_{\mathrm{spec}}$. There are $\sim3.5 \%$ outliers, defined as objects for which $|\Delta z |> 0.15$. For the remaining galaxies we measure the mean of $\Delta z$ and the scatter around this mean, $\sigma_z$, finding a bias of $|\Delta z |= 0.008$ and scatter of $\sigma_z = 0.029$. 

Since the galaxies for which spectroscopic redshifts have been measured are generally bright, and preferentially have emission lines, it is not immediately clear if the reported performance is representative for the entire galaxy sample down to the detection limit. In Fig.~\ref{fig:speczphotz}b we separate the galaxies by class (star-forming versus quiescent, cf. Sect.~\ref{sec:rfcolours}), and plot the differences as a function of $\mathrm{K_s}$-band magnitude, restricted to the redshift range where our clusters are located, $0.5<z_\mathrm{spec}<0.7$. Within this spectroscopic redshift range, the photo-$z$ scatter is $\sigma_z = 0.028$ for star-forming galaxies, and $\sigma_z = 0.023$ for quiescent galaxies.

This Figure shows that the success rate of measuring spectroscopic redshifts of faint galaxies is higher for star-forming than for quiescent galaxies, since the former have typically strong emission lines. The scatter, shown in the solid curve, based on a running bin width of 1 magnitude, does not significantly depend on magnitude. This suggest that the flux measurements that define the SED are precise, and the photo-$z$ determination is limited by the representativity of the templates, the accuracy of the filter curves, and the accuracy of the flux calibration. Since photo-$z$s for quiescent galaxies are slightly more precise than for star-forming galaxies (due to a stronger spectral feature around the characteristic 4000$\AA$ break), we expect the photo-$z$ performance of quiescent galaxies to be at least similar to that of star-forming galaxies at faint magnitudes. 

We use the broad-band colours to identify and flag stars in our catalogue \citep[e.g.][]{whitaker11,vdB13}, without having to make a selection based on size or morphology. That is because galaxies have very different Spectral Energy Distributions (SEDs) from stars, particularly towards near-IR wavelengths. 

We use the following colour criterion, which is similar to the ones used in aforementioned studies, to select the sample of galaxies: 
\begin{equation}
 \mathrm{J-K_s} >0.18\cdot(u- \mathrm{J} )-0.60 \cup \mathrm{J-K_s} >0.08\cdot(u- \mathrm{J} )-0.30
\end{equation}

\subsection{Identifying BCGs}
We select the brightest cluster galaxy (BCG) in the $\mathrm{K_s}$-band that is located within 1$'$ from the X-ray peak, and that has a photometric redshift within 0.10 from the cluster redshift. In all but one case this automatic identification corresponds to what we would have selected by hand based on the colour images in Appendix~\ref{sec:colourimages}. For \texttt{PSZ2 G219.89-34.39} there is a mis-identification since the apparent BCG has a blue core, and is likely contaminated by blue light from a nearby source (cf.~\ref{fig:gallery4}, as is also apparent from the VLT/FORS2 spectrum of this galaxy). The photo-$z$ is 0.56, which is 0.17 lower than the cluster redshift. We select only this BCG by hand, and all the others following the criteria above. The BCGs are marked in the Figures of Appendix~\ref{sec:colourimages}, and their positions are reported in Table~\ref{tab:dataoverview}.

\subsection{Stellar masses and background correction}\label{sec:bgcorrection}
We measure stellar masses for each galaxy using the SED-fitting code FAST \citep{kriek09}. We use stellar population synthesis models from \citet{bc03}, and assume a \citet{chabrier03} IMF, solar metallicity, and the \citet{calzetti00} dust law. The star formation history is parametrised as $SFR \propto e^{-t/\tau}$, where the time-scale $\tau$ is allowed to range between 10 Myr and 10 Gyr. These settings are identical to those used to measure stellar masses of the UltraVISTA sample, which we use to provide a field comparison. For an appropriate analysis that relies on a statistical subtraction of galaxies in the clusters' fore- and background, as described next, we fix the redshift of each individual galaxy to the clusters' mean redshift (cf. Table~\ref{tab:dataoverview}). 

We also construct tailored catalogues from the UltraVISTA main catalogue for each cluster, where we select only 8 filters: $uBVriz\mathrm{JK_s}$, and add artificial noise to the aperture flux measurements to match the depth of the data in the cluster fields. We verify that the performance of EAZY (scatter in photo-$z$s versus spec-$z$s) is then similar to that of the cluster fields for similarly bright galaxies. To perform the statistical field subtraction of fore- and background galaxies, we run FAST on the tailored catalogues to estimate stellar masses, while also fixing the redshift of each galaxy to the cluster's mean redshift. Identical settings are used when running FAST on the cluster fields, as on the reference field.

Cluster galaxies are initially selected to have a photometric redshift within 0.07 from the cluster mean spectroscopic redshift, which is several times larger than the photo-$z$ scatter. In Appendix \ref{sec:appphotozsel} we test the effect of this choice, and show that the reported results are robust. 

The uncertainty due to cosmic variance in the reference field is taken into account following the prescription of \citet{moster11}, based on the volume subtended by the UltraVISTA area in the redshift range $0.5<z<0.7$. We find that this has no significant effect on the cluster SMF, except in the outskirts, where the overdensity of the cluster field with respect to the background decreases. The estimated uncertainties are shown in the main figures presented in this work.

\subsection{Completeness correction}\label{sec:completeness}
We identify and correct for two observational effects that affect galaxies around the detection limit in the $\mathrm{K_s}$-band stack. First, the detection rate of objects drops towards fainter magnitudes due to noise fluctuations. Second, the objects that \textit{are} detected may have a flux measurement that is biased compared to their intrinsic brightness. 

To measure the influence of these effects, we study the recovered fraction and fluxes of simulated sources in the detection band. We inject sources with an exponential (i.e. S\'ersic-$n$=1) profile and half-light radii between 1-3 kpc (uniform distribution), ellipticities uniformly drawn between 0.0 and 0.2, and a uniform magnitude distribution. These values are appropriate for sources around our detection limit. Since the depth of the detection image is not uniform, we consider the region within 6$'$ radial distance from the cluster centres. This corresponds to 2.2 Mpc (2.6 Mpc) at $z=0.50$ ($z=0.70$) and covers the parts of the clusters that are relevant for this study. In one occasion we will probe the properties of galaxies up to larger cluster-centric distances; in Appendix~\ref{sec:appraddependence} we study the effect of the reduced depth in the cluster outskirts on this result. 

We run exactly the same source detection algorithm as for the main analysis (cf.~Sect.~\ref{sec:objectdetection}) on the $\mathrm{K_s}$-band stacks that include the simulated sources. The $\mathrm{K_s}$-band magnitude limits at which 80\% of the simulated sources are recovered, are listed in Table~\ref{tab:photometry} and also indicated in the right hand panel of Fig.~\ref{fig:speczphotz}. We find that around this limit, sources are measured to be 0.12 magnitudes fainter than they are intrinsically, while for the brightest sources this difference is negligible ($\sim 0.01$). We correct the measured fluxes by these magnitude-dependent corrections. 

Stellar mass limits that correspond to the 80\% completeness limit in the $\mathrm{K_s}$-band are also listed in Table~\ref{tab:photometry}. We base these on a \citet{bc03} template with a formation redshift of $z$=3.0. Since such an old stellar population is relatively faint for their stellar mass, this corresponds to a conservative limit when galaxies with more recent star formation are also considered.

\subsection{Star-forming versus Quiescent galaxies}\label{sec:rfcolours}
We measure rest-frame U-V and V-J colours of the best-fit SEDs from \texttt{EAZY}, while fixing the redshift to the cluster mean redshift. These colours have been shown to be effective to separate star-forming from quiescent galaxies, even in the presence of dust reddening \citep[e.g.][]{wuyts07,williams09,patel12}. 

There are small offsets in the UVJ colour distribution between the different clusters, and with respect to the UltraVISTA field sample. Such differences are not uncommon, and several studies have applied corrections to the selection criteria to rectify this \citep[][also see the discussion in Appendix A of \citet{leebrown17}]{whitaker11,muzzin13b,skelton14}. Rather than adapting the selection criteria of quiescent galaxies for each cluster, we applied offsets to the rest-frame colours to shift them back to the UltraVISTA reference in redshift range $0.5<z<0.7$. The mean absolute shift applied is 0.06 magnitude to the U-V colour, and 0.04 to the V-J colour. This brings the colour distribution of cluster galaxies in good agreement with those of field galaxies in UltraVISTA, see Fig.~\ref{fig:UVJ}. In Appendix~\ref{sec:appuvjdiv} we test the effect of a possible residual shift in colour between the cluster fields and the reference field to the main results presented in this paper.  

Inspecting the bimodal galaxy distribution by-eye, we select a sample of quiescent galaxies following the following criteria:
\begin{equation}
 \mathrm{U-V} > 1.3 \,\,\,\cap\,\,\, \mathrm{V-J} < 1.6 \,\,\,\cap\,\,\, \mathrm{U-V} > 0.55+(\mathrm{V-J})
\end{equation}  

Note that Fig.~\ref{fig:UVJ} shows the UVJ colour distribution of the UltraVISTA galaxies based on all the photometric information. For the purpose of a statistical background subtraction we run a similar analysis on a subset of eight UltraVISTA filters, with noise added to resemble the photometric quality of the cluster fields (cf.~Sect.~\ref{sec:bgcorrection}). 

\begin{figure}
\resizebox{\hsize}{!}{\includegraphics{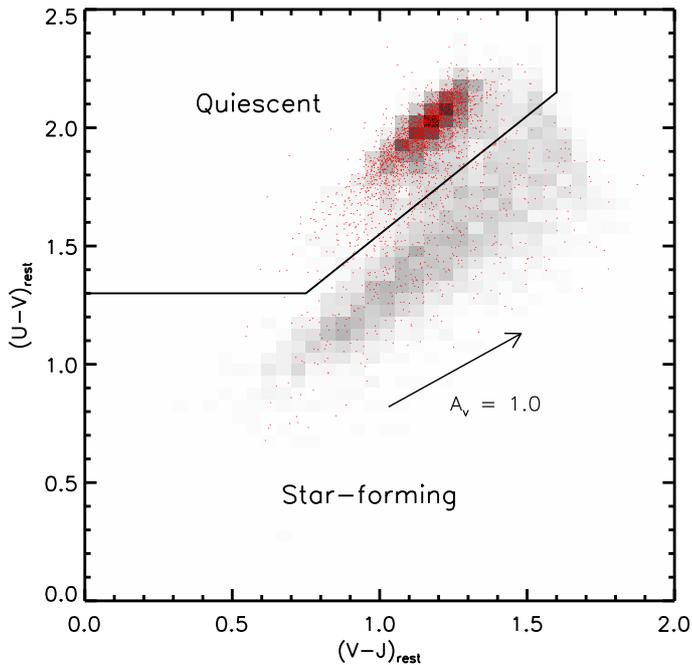}}
\caption{Rest-frame U-V versus V-J diagram for galaxies with stellar masses $M_{\star}\geq 10^{10}\,\mathrm{M_{\odot}}$. \textit{Grey distribution:} UltraVISTA field galaxies with redshifts $0.5<z<0.7$. \textit{Red points:} Galaxies from the cluster sample studied here, within redshift $|\Delta z| \leq 0.05$ from the cluster mean redshift, and within projected $R \leq R_{500}$ from the cluster centres.}
\label{fig:UVJ}
\end{figure}

\section{The Stellar Mass Function}\label{sec:SMF}

\begin{figure*}
\resizebox{\hsize}{!}{\includegraphics{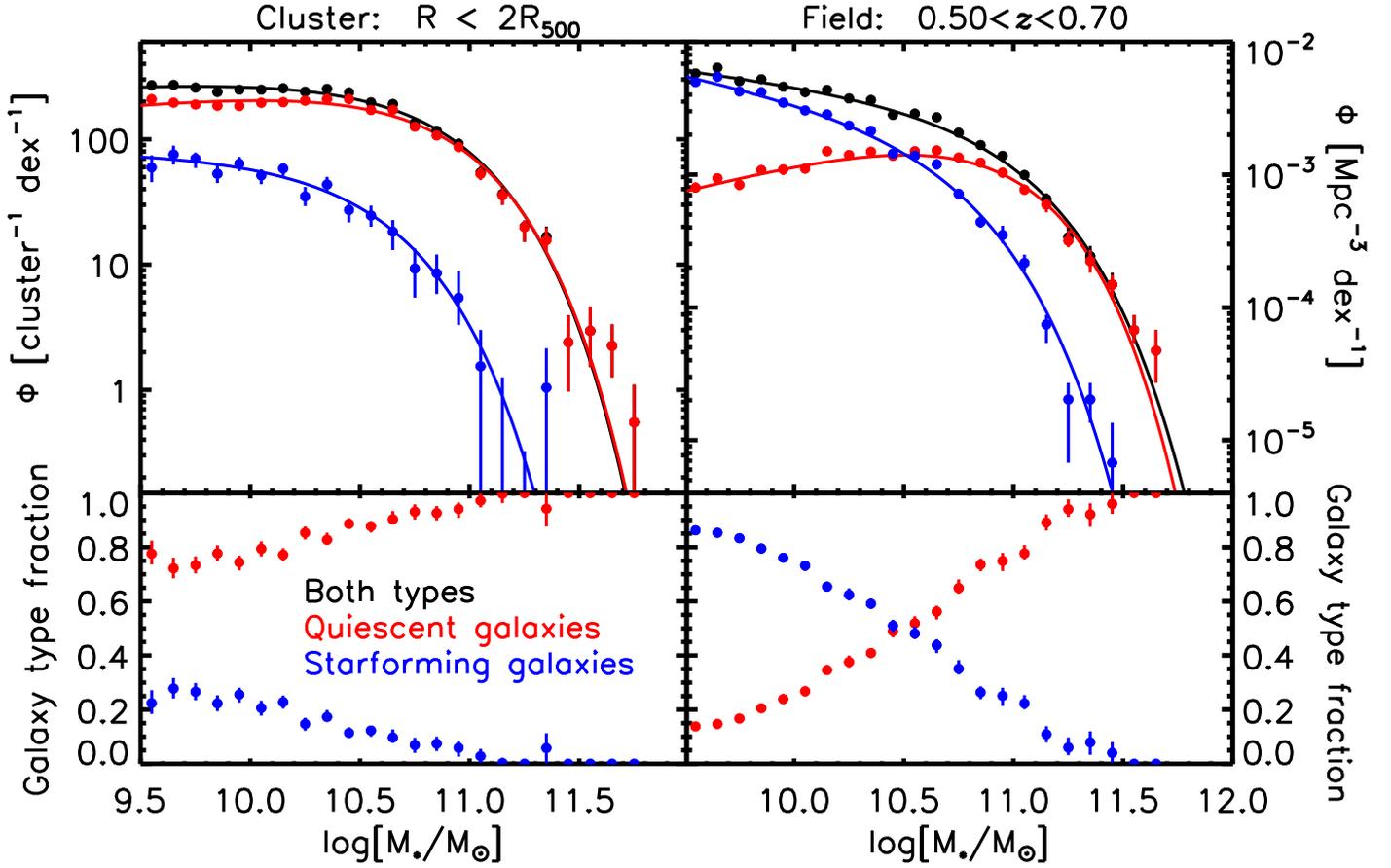}}
\caption{\textit{Top row, left panel:} SMF of cluster galaxies within $R \leq 2\times R_{500}$ from the cluster centres. \textit{Blue and red data points:} The population of star-forming and quiescent galaxies, respectively. \textit{Black points:} Full galaxy population. \textit{Top row, right panel:} SMF of field galaxies from the COSMOS/UltraVISTA survey at the same redshifts ($0.5<z<0.7$). \textit{Lower panels:} Relative fraction of quiescent galaxies and star-forming galaxies as a function of stellar mass.}
\label{fig:SMF_clustervsfield}
\end{figure*}

\begin{table*}%[ht]
\caption{Data points of the SMFs plotted in Figs.~\ref{fig:SMF_clustervsfield}~\&~\ref{fig:SMF_clusterradbins}.}
\label{tab:SMFpoints}
\begin{center}
%\begin{adjustwidth}{-0.90cm}{}
\begin{tabular}{c l | l l l l l | l}
\hline
\hline
&&\multicolumn{5}{c|}{Clusters}&\multicolumn{1}{c}{Field}\\
&&\multicolumn{5}{c|}{$\Phi\, [$cluster$^{-1}$ dex$^{-1}]$}&\multicolumn{1}{c}{$\Phi\, [10^{-5}$ Mpc$^{-3}$ dex$^{-1}]$}\\
&log[M$_\star$/$M_{\odot}$]&R$<$2R$_{500}$&R$<$0.5R$_{500}$&0.5R$_{500}<$R$<$R$_{500}$&R$_{500}<$R$<$1.5R$_{500}$&1.5R$_{500}<$R$<$2R$_{500}$&\multicolumn{1}{c}{0.5$<z<$0.7}\\
\hline
\parbox[t]{2mm}{\multirow{23}{*}{\rotatebox[origin=c]{90}{All galaxies}}}&9.55&$270.4^{+23.3+10.2}_{-23.2-10.2}$&$67.73^{+9.76+0.63}_{-7.06-0.63}$&$86.75^{+15.26+1.90}_{-8.97-1.90}$&$45.34^{+9.68+3.17}_{-8.13-3.17}$&$64.16^{+12.98+4.44}_{-8.35-4.44}$&$577.8^{+16.9}_{-14.2}$\\
&9.65&$272.0^{+17.4+10.3}_{-18.6-10.3}$&$65.95^{+8.54+0.64}_{-7.99-0.65}$&$98.68^{+11.22+1.94}_{-10.19-1.94}$&$55.37^{+10.82+3.23}_{-8.44-3.23}$&$51.56^{+15.92+4.53}_{-9.64-4.53}$&$639.3^{+18.9}_{-20.3}$\\
&9.75&$258.8^{+15.9+8.7}_{-14.3-8.7}$&$68.02^{+6.58+0.54}_{-5.32-0.54}$&$84.51^{+8.48+1.63}_{-8.44-1.63}$&$63.29^{+6.33+2.72}_{-9.62-2.72}$&$43.49^{+8.61+3.81}_{-7.83-3.81}$&$506.2^{+21.6}_{-18.2}$\\
&9.85&$238.5^{+14.8+9.5}_{-14.5-9.5}$&$68.28^{+8.60+0.60}_{-6.34-0.60}$&$70.43^{+5.75+1.79}_{-9.00-1.79}$&$56.29^{+5.65+2.98}_{-6.85-2.98}$&$41.40^{+8.14+4.17}_{-5.79-4.17}$&$522.4^{+25.0}_{-20.3}$\\
&9.95&$248.8^{+13.7+8.1}_{-15.2-8.1}$&$69.72^{+8.28+0.51}_{-5.98-0.51}$&$79.01^{+7.48+1.52}_{-8.22-1.52}$&$50.37^{+6.21+2.53}_{-6.99-2.53}$&$42.79^{+6.85+3.55}_{-5.77-3.55}$&$458.9^{+14.2}_{-16.9}$\\
&10.05&$249.0^{+11.1+7.8}_{-13.3-7.8}$&$67.56^{+4.22+0.49}_{-6.63-0.49}$&$79.23^{+8.03+1.47}_{-6.96-1.47}$&$52.95^{+6.29+2.45}_{-6.02-2.45}$&$45.16^{+6.99+3.43}_{-5.49-3.43}$&$417.6^{+10.1}_{-14.9}$\\
&10.15&$255.4^{+14.1+8.6}_{-14.5-8.6}$&$77.62^{+6.52+0.54}_{-6.26-0.54}$&$74.86^{+6.95+1.61}_{-5.67-1.61}$&$56.22^{+5.47+2.68}_{-6.60-2.68}$&$41.55^{+7.43+3.75}_{-5.25-3.75}$&$435.2^{+14.9}_{-16.9}$\\
&10.25&$240.0^{+10.7+7.6}_{-11.5-7.6}$&$82.06^{+4.90+0.47}_{-7.44-0.47}$&$76.57^{+8.50+1.42}_{-6.20-1.42}$&$40.23^{+6.51+2.37}_{-5.71-2.37}$&$37.70^{+6.63+3.02}_{-5.40-2.10}$&$374.4^{+16.9}_{-16.2}$\\
&10.35&$252.9^{+17.0+7.7}_{-15.2-7.7}$&$85.57^{+6.23+0.48}_{-5.77-0.48}$&$67.31^{+6.20+1.44}_{-6.65-1.44}$&$52.52^{+5.59+2.39}_{-6.36-2.22}$&$44.74^{+6.50+3.35}_{-5.47-3.35}$&$364.3^{+18.9}_{-16.2}$\\
&10.45&$236.7^{+12.6+6.4}_{-13.7-6.4}$&$73.37^{+7.55+0.40}_{-6.77-0.40}$&$62.11^{+6.20+1.19}_{-4.32-1.19}$&$61.17^{+5.71+1.99}_{-5.99-1.99}$&$34.88^{+5.68+2.78}_{-5.48-2.78}$&$281.8^{+14.9}_{-11.5}$\\
&10.55&$197.1^{+11.0+6.2}_{-12.6-6.2}$&$61.12^{+5.39+0.17}_{-6.06-0.20}$&$63.43^{+4.28+1.15}_{-5.78-1.15}$&$39.87^{+7.27+1.92}_{-5.73-1.92}$&$30.35^{+4.67+2.69}_{-4.67-2.69}$&$287.9^{+11.5}_{-14.2}$\\
&10.65&$191.1^{+6.9+6.1}_{-10.3-6.1}$&$55.54^{+7.38+0.38}_{-6.72-0.34}$&$60.70^{+5.20+1.15}_{-4.99-1.15}$&$47.44^{+6.30+1.48}_{-4.68-1.28}$&$25.73^{+5.49+2.68}_{-4.37-2.68}$&$270.3^{+16.2}_{-14.2}$\\
&10.75&$135.3^{+8.4+4.5}_{-8.1-4.5}$&$44.03^{+5.78+0.28}_{-4.35-0.28}$&$35.68^{+3.86+0.85}_{-4.40-0.85}$&$31.59^{+4.16+1.42}_{-3.41-1.42}$&$21.96^{+3.73+1.93}_{-2.82-1.76}$&$206.8^{+8.8}_{-8.8}$\\
&10.85&$116.8^{+7.4+4.3}_{-9.6-4.3}$&$41.62^{+5.69+0.27}_{-2.60-0.27}$&$30.98^{+4.58+0.66}_{-3.42-0.46}$&$24.82^{+3.62+1.35}_{-4.57-1.35}$&$15.00^{+2.80+1.81}_{-3.46-1.74}$&$166.9^{+12.8}_{-9.5}$\\
&10.95&$92.26^{+8.30+2.93}_{-7.31-2.93}$&$27.66^{+3.55+0.18}_{-3.93-0.18}$&$32.27^{+4.50+0.55}_{-3.46-0.55}$&$14.46^{+3.64+0.91}_{-3.01-0.90}$&$16.20^{+4.26+1.03}_{-4.04-0.87}$&$138.5^{+10.8}_{-7.4}$\\
&11.05&$54.26^{+6.71+2.02}_{-5.49-1.73}$&$18.72^{+3.42+0.13}_{-3.11-0.12}$&$11.42^{+2.56+0.38}_{-2.95-0.38}$&$12.21^{+2.50+0.44}_{-2.25-0.44}$&$13.09^{+3.34+0.85}_{-2.91-0.62}$&$99.34^{+6.76}_{-8.11}$\\
&11.15&$36.58^{+5.66+1.20}_{-6.54-1.27}$&$11.57^{+3.04+0.09}_{-2.78-0.09}$&$10.10^{+2.81+0.21}_{-2.18-0.21}$&$\,\,\,7.01^{+2.47+0.35}_{-2.15-0.35}$&$\,\,\,6.80^{+2.92+0.57}_{-2.63-0.48}$&$65.55^{+6.76}_{-5.41}$\\
&11.25&$20.02^{+2.51+0.64}_{-4.76-0.64}$&$\,\,\,7.63^{+1.93+0.04}_{-2.04-0.04}$&$\,\,\,6.98^{+2.16+0.14}_{-1.83-0.14}$&$\,\,\,3.18^{+1.19+0.20}_{-1.44-0.20}$&$\,\,\,1.29^{+1.47+0.28}_{-1.07-0.28}$&$33.79^{+6.08}_{-4.05}$\\
&11.35&$16.69^{+3.46+0.54}_{-3.62-0.54}$&$\,\,\,5.84^{+1.71+0.03}_{-2.09-0.03}$&$\,\,\,5.25^{+1.81+0.10}_{-1.45-0.09}$&$\,\,\,2.53^{+1.48+0.14}_{-1.48-0.14}$&$\,\,\,2.76^{+1.07+0.20}_{-1.48-0.20}$&$24.33^{+4.73}_{-4.05}$\\
&11.45&$\,\,\,2.40^{+1.55+0.18}_{-1.43-0.18}$&$\,\,\,0.44^{+0.51+0.01}_{-0.44-0.01}$&$\,\,\,1.77^{+1.01+0.03}_{-1.03-0.03}$&$\,\,\,0.74^{+0.55+0.06}_{-0.60-0.06}$&-&$14.87^{+3.38}_{-3.38}$\\
&11.55&$\,\,\,2.96^{+1.65+0.10}_{-1.44-0.10}$&$\,\,\,3.41^{+1.33+0.01}_{-1.15-0.01}$&-&-&-&$\,\,\,6.76^{+2.03}_{-1.35}$\\
&11.65&$\,\,\,2.25^{+1.09+0.05}_{-0.99-0.05}$&$\,\,\,0.47^{+0.49+0.00}_{-0.47-0.00}$&$\,\,\,0.93^{+0.98+0.01}_{-0.50-0.01}$&$\,\,\,0.91^{+0.99+0.01}_{-0.91-0.02}$&-&$\,\,\,4.73^{+2.03}_{-2.03}$\\
&11.75&$\,\,\,0.55^{+0.55+0.00}_{-0.55-0.00}$&$\,\,\,0.67^{+0.67+0.00}_{-0.67-0.00}$&-&-&-&-\\
\hline
\parbox[t]{2mm}{\multirow{23}{*}{\rotatebox[origin=c]{90}{Quiescent only}}}&9.55&$208.9^{+15.5+1.0}_{-14.3-1.0}$&$61.36^{+7.11+0.06}_{-6.14-0.06}$&$76.56^{+10.23+0.18}_{-8.45-0.18}$&$31.98^{+8.31+0.30}_{-6.93-0.30}$&$36.66^{+7.53+0.42}_{-5.74-0.42}$&$79.74^{+7.43}_{-5.41}$\\
&9.65&$196.0^{+15.3+1.2}_{-13.2-1.2}$&$59.39^{+6.04+0.07}_{-8.69-0.07}$&$68.76^{+7.72+0.22}_{-8.66-0.22}$&$38.39^{+8.51+0.36}_{-6.05-0.36}$&$28.49^{+9.56+0.51}_{-5.77-0.51}$&$93.94^{+6.76}_{-8.11}$\\
&9.75&$189.4^{+13.2+1.1}_{-12.5-1.1}$&$56.59^{+4.75+0.07}_{-5.68-0.07}$&$66.27^{+7.49+0.22}_{-7.63-0.22}$&$36.91^{+4.40+0.36}_{-4.17-0.36}$&$27.87^{+4.69+0.50}_{-4.82-0.50}$&$83.80^{+7.43}_{-7.43}$\\
&9.85&$185.4^{+13.0+1.5}_{-11.6-1.5}$&$58.36^{+6.96+0.09}_{-6.11-0.09}$&$54.21^{+6.41+0.27}_{-6.09-0.27}$&$45.99^{+6.66+0.46}_{-4.08-0.46}$&$24.60^{+5.13+0.64}_{-4.35-0.64}$&$108.1^{+8.1}_{-8.8}$\\
&9.95&$184.5^{+11.4+1.5}_{-10.1-1.5}$&$61.24^{+6.61+0.09}_{-5.01-0.09}$&$64.33^{+4.62+0.28}_{-7.37-0.28}$&$33.14^{+5.62+0.47}_{-4.14-0.47}$&$22.77^{+5.86+0.65}_{-4.73-0.65}$&$109.5^{+8.8}_{-10.1}$\\
&10.05&$195.5^{+13.9+1.6}_{-11.4-1.7}$&$54.46^{+4.36+0.10}_{-5.43-0.10}$&$64.31^{+5.31+0.31}_{-6.57-0.31}$&$40.65^{+6.06+0.52}_{-5.30-0.52}$&$33.50^{+5.39+0.72}_{-4.21-0.72}$&$110.8^{+8.1}_{-7.4}$\\
&10.15&$197.6^{+10.3+2.3}_{-11.7-2.3}$&$67.70^{+6.61+0.14}_{-5.56-0.14}$&$57.65^{+7.12+0.43}_{-5.08-0.43}$&$39.00^{+5.20+0.72}_{-5.50-0.72}$&$27.86^{+4.97+1.00}_{-4.82-1.00}$&$150.0^{+8.1}_{-8.8}$\\
&10.25&$203.3^{+12.6+2.4}_{-9.0-2.4}$&$72.08^{+4.81+0.15}_{-5.76-0.15}$&$59.82^{+6.77+0.45}_{-6.26-0.45}$&$31.61^{+4.76+0.76}_{-3.54-0.76}$&$35.54^{+4.95+1.06}_{-4.57-1.06}$&$140.6^{+9.5}_{-10.8}$\\
&10.35&$209.9^{+11.1+2.8}_{-12.4-2.8}$&$76.81^{+5.61+0.17}_{-5.41-0.17}$&$55.44^{+5.46+0.52}_{-5.53-0.52}$&$48.01^{+5.22+0.86}_{-6.06-0.86}$&$27.57^{+3.98+1.20}_{-3.51-1.20}$&$148.7^{+10.1}_{-9.5}$\\
&10.45&$209.2^{+11.1+2.5}_{-11.9-2.5}$&$71.39^{+6.97+0.16}_{-7.24-0.16}$&$55.60^{+5.05+0.47}_{-5.27-0.47}$&$50.44^{+6.26+0.78}_{-3.96-0.78}$&$26.87^{+3.99+1.09}_{-3.85-1.09}$&$138.5^{+8.8}_{-9.5}$\\
&10.55&$171.3^{+12.1+2.7}_{-9.8-2.7}$&$60.52^{+5.64+0.17}_{-6.16-0.17}$&$56.03^{+4.08+0.50}_{-5.23-0.50}$&$33.58^{+4.78+0.84}_{-4.82-0.84}$&$19.98^{+3.95+1.17}_{-3.95-1.17}$&$150.0^{+8.8}_{-9.5}$\\
&10.65&$171.8^{+9.3+2.8}_{-7.6-2.8}$&$52.72^{+7.25+0.18}_{-5.05-0.18}$&$50.89^{+5.19+0.53}_{-3.07-0.53}$&$46.88^{+5.27+0.88}_{-5.24-0.88}$&$19.83^{+4.11+1.23}_{-4.21-1.23}$&$152.1^{+12.2}_{-12.2}$\\
&10.75&$126.0^{+6.0+2.4}_{-6.6-2.4}$&$42.84^{+4.95+0.15}_{-5.22-0.15}$&$34.77^{+3.35+0.45}_{-4.92-0.45}$&$27.09^{+2.92+0.74}_{-3.12-0.74}$&$19.31^{+3.92+1.04}_{-3.34-1.04}$&$134.5^{+7.4}_{-6.8}$\\
&10.85&$107.3^{+7.0+2.4}_{-8.9-2.4}$&$38.84^{+4.71+0.15}_{-3.35-0.15}$&$30.28^{+4.23+0.46}_{-3.29-0.46}$&$21.01^{+3.54+0.76}_{-3.28-0.76}$&$13.05^{+3.05+1.07}_{-2.13-1.07}$&$123.0^{+10.1}_{-10.1}$\\
&10.95&$86.82^{+7.28+1.98}_{-8.12-1.98}$&$25.63^{+3.61+0.12}_{-2.85-0.12}$&$29.82^{+4.57+0.37}_{-3.20-0.37}$&$12.61^{+3.73+0.62}_{-3.40-0.62}$&$15.77^{+4.59+0.87}_{-3.62-0.87}$&$103.4^{+9.5}_{-8.1}$\\
&11.05&$52.98^{+6.69+1.42}_{-5.66-1.42}$&$18.30^{+3.22+0.09}_{-3.08-0.09}$&$\,\,\,9.85^{+2.78+0.27}_{-2.26-0.27}$&$11.96^{+2.52+0.44}_{-2.11-0.44}$&$12.32^{+3.11+0.62}_{-2.20-0.62}$&$77.04^{+6.76}_{-6.76}$\\
&11.15&$35.66^{+5.50+1.11}_{-5.70-1.11}$&$10.77^{+2.95+0.07}_{-2.43-0.07}$&$\,\,\,9.97^{+2.64+0.21}_{-2.27-0.21}$&$\,\,\,7.01^{+2.47+0.35}_{-2.15-0.35}$&$\,\,\,6.67^{+2.83+0.48}_{-2.52-0.48}$&$59.47^{+6.08}_{-7.43}$\\
&11.25&$19.90^{+2.50+0.64}_{-4.78-0.64}$&$\,\,\,7.63^{+1.93+0.04}_{-2.04-0.04}$&$\,\,\,6.19^{+2.41+0.12}_{-1.42-0.12}$&$\,\,\,3.18^{+1.19+0.20}_{-1.44-0.20}$&$\,\,\,1.29^{+1.47+0.28}_{-1.07-0.28}$&$31.76^{+6.08}_{-3.38}$\\
&11.35&$15.64^{+3.93+0.46}_{-3.26-0.46}$&$\,\,\,5.19^{+1.46+0.03}_{-1.92-0.03}$&$\,\,\,5.01^{+1.41+0.09}_{-1.61-0.09}$&$\,\,\,2.05^{+1.46+0.14}_{-1.46-0.14}$&$\,\,\,2.76^{+1.07+0.20}_{-1.48-0.20}$&$22.30^{+4.73}_{-4.05}$\\
&11.45&$\,\,\,2.40^{+1.55+0.18}_{-1.43-0.18}$&$\,\,\,0.44^{+0.51+0.01}_{-0.44-0.01}$&$\,\,\,1.77^{+1.01+0.03}_{-1.03-0.03}$&$\,\,\,0.74^{+0.55+0.06}_{-0.60-0.06}$&-&$14.87^{+2.70}_{-3.38}$\\
&11.55&$\,\,\,2.96^{+1.65+0.10}_{-1.44-0.10}$&$\,\,\,3.41^{+1.33+0.01}_{-1.15-0.01}$&-&-&-&$\,\,\,6.76^{+2.03}_{-1.35}$\\
&11.65&$\,\,\,2.25^{+1.09+0.05}_{-0.99-0.05}$&$\,\,\,0.47^{+0.49+0.00}_{-0.47-0.00}$&$\,\,\,0.93^{+0.98+0.01}_{-0.50-0.01}$&$\,\,\,0.91^{+0.99+0.01}_{-0.91-0.02}$&-&$\,\,\,4.73^{+2.03}_{-2.03}$\\
&11.75&$\,\,\,0.55^{+0.55+0.00}_{-0.55-0.00}$&$\,\,\,0.67^{+0.67+0.00}_{-0.67-0.00}$&-&-&-&-\\
\hline
\parbox[t]{2mm}{\multirow{19}{*}{\rotatebox[origin=c]{90}{Star-forming only}}}&9.55&$59.59^{+14.93+9.19}_{-14.11-9.19}$&$\,\,\,7.59^{+2.97+0.57}_{-4.42-0.57}$&$10.71^{+5.75+1.72}_{-4.77-1.72}$&$13.26^{+6.21+2.87}_{-5.77-2.87}$&$28.35^{+7.58+4.02}_{-9.32-4.02}$&$497.4^{+17.6}_{-13.5}$\\
&9.65&$75.72^{+13.39+9.18}_{-12.83-9.18}$&$\,\,\,7.92^{+4.18+0.57}_{-3.73-0.57}$&$30.11^{+8.43+1.72}_{-6.42-1.72}$&$17.61^{+5.90+2.87}_{-6.77-2.87}$&$23.35^{+7.92+4.02}_{-6.99-4.02}$&$545.4^{+16.9}_{-18.9}$\\
&9.75&$70.13^{+8.25+7.56}_{-11.11-7.56}$&$12.19^{+3.50+0.47}_{-4.03-0.47}$&$19.02^{+4.43+1.42}_{-4.95-1.42}$&$25.68^{+5.21+2.36}_{-6.45-2.36}$&$15.47^{+5.73+3.31}_{-4.47-3.31}$&$423.0^{+19.6}_{-17.6}$\\
&9.85&$53.06^{+9.29+8.08}_{-8.34-8.09}$&$10.30^{+2.51+0.51}_{-2.59-0.51}$&$15.02^{+4.76+1.52}_{-3.84-1.52}$&$\,\,\,8.79^{+4.00+2.53}_{-4.40-2.53}$&$17.56^{+4.30+3.54}_{-5.73-3.54}$&$417.6^{+18.2}_{-19.6}$\\
&9.95&$63.89^{+8.45+6.61}_{-8.22-6.61}$&$\,\,\,8.60^{+2.65+0.41}_{-2.08-0.41}$&$15.27^{+3.66+1.24}_{-3.76-1.24}$&$16.63^{+4.13+2.07}_{-3.72-2.07}$&$19.82^{+5.03+2.89}_{-5.55-2.89}$&$348.0^{+14.2}_{-14.9}$\\
&10.05&$51.46^{+6.05+6.19}_{-7.61-6.19}$&$11.91^{+3.03+0.39}_{-2.34-0.39}$&$15.32^{+3.13+1.16}_{-2.72-1.16}$&$11.96^{+2.92+1.93}_{-3.55-1.94}$&$11.57^{+4.26+2.71}_{-3.29-2.71}$&$304.1^{+14.9}_{-10.8}$\\
&10.15&$58.46^{+5.74+6.29}_{-7.40-6.29}$&$\,\,\,9.26^{+2.45+0.39}_{-2.44-0.39}$&$16.94^{+3.67+1.18}_{-2.85-1.18}$&$17.33^{+3.88+1.97}_{-4.12-1.97}$&$14.74^{+3.58+2.75}_{-3.98-2.75}$&$283.8^{+12.8}_{-14.2}$\\
&10.25&$35.01^{+6.63+5.15}_{-5.68-5.15}$&$\,\,\,9.59^{+1.91+0.32}_{-2.21-0.32}$&$17.04^{+3.50+0.97}_{-3.89-0.97}$&$\,\,\,8.64^{+2.71+1.61}_{-3.62-1.61}$&$\,\,\,1.42^{+2.98+2.25}_{-1.42-1.42}$&$233.8^{+12.2}_{-12.8}$\\
&10.35&$43.59^{+6.49+4.91}_{-6.04-4.91}$&$\,\,\,9.09^{+2.42+0.31}_{-2.11-0.31}$&$12.25^{+3.50+0.92}_{-3.60-0.92}$&$\,\,\,5.10^{+2.60+1.53}_{-3.22-1.54}$&$16.68^{+4.51+2.15}_{-3.92-2.15}$&$213.6^{+14.2}_{-10.1}$\\
&10.45&$27.28^{+4.98+3.86}_{-5.58-3.86}$&$\,\,\,1.80^{+2.28+0.24}_{-1.16-0.24}$&$\,\,\,7.08^{+2.58+0.72}_{-1.92-0.72}$&$\,\,\,9.79^{+2.78+1.21}_{-3.01-1.21}$&$\,\,\,7.68^{+3.65+1.69}_{-2.69-1.69}$&$143.9^{+10.1}_{-7.4}$\\
&10.55&$24.65^{+4.92+3.47}_{-4.61-3.47}$&$\,\,\,0.00^{+0.86+0.10}_{-0.00-0.00}$&$\,\,\,6.95^{+3.83+0.65}_{-1.99-0.65}$&$\,\,\,6.67^{+2.68+1.09}_{-2.98-1.09}$&$10.12^{+3.30+1.52}_{-3.25-1.52}$&$138.5^{+8.8}_{-10.1}$\\
&10.65&$18.31^{+4.38+3.33}_{-5.26-3.33}$&$\,\,\,2.18^{+1.38+0.21}_{-1.70-0.21}$&$\,\,\,9.67^{+2.52+0.62}_{-2.30-0.62}$&$\,\,\,0.00^{+2.76+0.80}_{-0.00-0.00}$&$\,\,\,6.41^{+2.41+1.46}_{-2.92-1.46}$&$119.6^{+8.1}_{-10.8}$\\
&10.75&$\,\,\,9.30^{+4.17+2.17}_{-3.86-2.17}$&$\,\,\,1.52^{+1.01+0.14}_{-0.96-0.14}$&$\,\,\,1.08^{+1.03+0.41}_{-1.08-0.41}$&$\,\,\,4.49^{+2.37+0.68}_{-2.20-0.68}$&$\,\,\,2.48^{+2.67+0.95}_{-2.01-0.95}$&$71.63^{+6.76}_{-6.76}$\\
&10.85&$\,\,\,8.52^{+3.52+1.88}_{-2.70-1.88}$&$\,\,\,3.10^{+1.48+0.12}_{-1.47-0.12}$&$\,\,\,1.04^{+1.03+0.35}_{-1.04-0.35}$&$\,\,\,3.30^{+2.51+0.59}_{-2.07-0.59}$&$\,\,\,1.05^{+1.55+0.82}_{-1.05-0.82}$&$43.93^{+5.41}_{-4.05}$\\
&10.95&$\,\,\,5.43^{+3.48+0.95}_{-2.14-0.95}$&$\,\,\,1.44^{+1.17+0.06}_{-1.17-0.06}$&$\,\,\,2.24^{+1.46+0.18}_{-1.45-0.18}$&$\,\,\,2.13^{+1.47+0.30}_{-1.58-0.30}$&$\,\,\,0.00^{+0.82+0.21}_{-0.00-0.00}$&$35.14^{+6.08}_{-4.73}$\\
&11.05&$\,\,\,1.54^{+1.47+0.60}_{-1.51-0.60}$&$\,\,\,0.26^{+0.50+0.04}_{-0.26-0.04}$&$\,\,\,1.21^{+1.00+0.11}_{-0.97-0.11}$&$\,\,\,0.00^{+0.70+0.00}_{-0.00-0.00}$&$\,\,\,0.41^{+1.08+0.26}_{-0.41-0.26}$&$21.63^{+3.38}_{-2.70}$\\
&11.15&$\,\,\,0.12^{+1.14+0.37}_{-0.12-0.12}$&$\,\,\,0.81^{+0.49+0.02}_{-0.49-0.02}$&$\,\,\,0.04^{+0.50+0.07}_{-0.04-0.04}$&-&$\,\,\,0.00^{+1.00+0.14}_{-0.00-0.00}$&$\,\,\,7.43^{+1.35}_{-2.03}$\\
&11.25&$\,\,\,0.00^{+0.32+0.00}_{-0.00-0.00}$&-&$\,\,\,0.36^{+0.48+0.02}_{-0.36-0.02}$&-&-&$\,\,\,2.03^{+0.68}_{-1.35}$\\
&11.35&$\,\,\,1.04^{+1.10+0.08}_{-1.00-0.08}$&$\,\,\,0.64^{+0.67+0.00}_{-0.64-0.00}$&$\,\,\,0.41^{+0.50+0.01}_{-0.41-0.01}$&$\,\,\,0.33^{+0.48+0.02}_{-0.33-0.02}$&-&$\,\,\,2.03^{+0.68}_{-0.68}$\\
&11.45&-&-&-&-&-&$\,\,\,0.68^{+0.68}_{-0.68}$\\                                                                        
\hline
\end{tabular}
%\end{adjustwidth}
\end{center}
\end{table*}

The galaxy stellar mass function (SMF) is a fundamental measure of any population of galaxies, and a critical measurement against which galaxy-formation models are tested. The SMF is measured for the cluster galaxies up to a cluster-centric radius of 2$\times R_{500}$ and shown in the left panel of Fig.~\ref{fig:SMF_clustervsfield}. For this measurement all 21 clusters are stacked, excluding the BCGs. The background is subtracted and an incompleteness correction is applied. For each bin we only use clusters down to their 80\% stellar mass completeness limit (cf.~Table~\ref{tab:photometry}). To compensate for clusters missing in the lowest-mass bins, we increase the weight of galaxies in the remaining clusters. To weigh each cluster properly, we base these on the richnesses measured for each cluster. Richnesses are measured following the definition given in \citet{rykoff14}, and we discuss this mass proxy in van der Burg et al., in prep.. 

We measure the ``average'' cluster all the way down to the stellar mass limit of $10^{9.5}\,\mathrm{M_{\odot}}$. Error bars denote uncertainties estimated from 100 bootstrap re-samplings of all galaxies in which we draw galaxies with replacement. The number of galaxies we draw in each realisation follow a Poisson distribution with mean equal to the number of galaxies in the stack. To make sure that the uncertainties are not dominated by the (perhaps too low) number of clusters in our sample, we perform 25 additional re-samplings of the clusters themselves. We find that both bootstrap procedures result in comparable uncertainties; the sample of clusters is thus large enough that we would have obtained the same results as presented here, if we would have observed 21 different clusters taken from a similar parent sample. 

From the separation between quiescent and star-forming galaxies in the left panel of Fig.~\ref{fig:SMF_clustervsfield}, we find that the galaxy population in these massive clusters is completely dominated by quiescent galaxies, all the way down to the stellar mass limit ($10^{9.5}\,\mathrm{M_{\odot}}$). To interpret our findings of the cluster SMF further, we make a comparison with the SMF of field galaxies at the same redshift as the clusters, in the right hand panel of Fig.~\ref{fig:SMF_clustervsfield}. From a comparison of cluster and field, it is clear that the fraction of quenched galaxies in the clusters is substantially higher than in the field, at each stellar mass. This is quantified further in the lower panels, where the relative fractions of star-forming and quiescent galaxies are presented as a function of stellar mass.

We model the SMF by fitting a Schechter \citep{schechter76} function to the data. This function is parameterized as 
\begin{equation}
\Phi(M)= \ln (10) \Phi^{*} \left[ 10^{(M-M^{*})(1+\alpha)}\right] \exp \left[ -10^{(M-M^{*})}\right],
\end{equation}
with $M^{*}$ being the characteristic mass, $\alpha$ the low-mass slope, and $\Phi^{*}$ the overall normalisation. We follow a maximum likelihood approach to estimate the parameters that define the shape of the Schechter functions, $M^{*}$  and $\alpha$, along with their uncertainties. For this we use the un-binned data points, and include the completeness correction to the individual galaxies, by setting their weights in the likelihood maximisation. For low stellar masses, these weights also compensate for clusters that are not complete down to these limits. For this purpose, each cluster is scaled by its richness. The background galaxies, from the reference field, are included in the same likelihood and have a negative weight. 

The normalisation of the Schechter function, $\Phi^{*}$,  is evaluated by requiring that the integral over the considered stellar mass range (i.e.~stellar masses larger than $10^{9.5}\,\mathrm{M_\odot}$) equals the number of all cluster galaxies (or more specifically, the sum of all weights). The best-fitting Schechter parameters are listed in Table~\ref{tab:Schechter}, and the corresponding functions are shown in the top panels of Fig.~\ref{fig:SMF_clustervsfield}. The reported Goodness-of-Fit values indicate that the Schechter functions provide reasonable fits to the data points. However, there seem to be some systematic residuals, especially towards the low-mass end of the SMF. Indeed, some literature studies fit double Schechter functions, but since we primarily work with the data points from now on, this paper does not discuss whether a fit can be improved with more degrees of freedom. The data points themselves are listed in Table~\ref{tab:SMFpoints}.

\begin{table*}%[ht]
\caption{Best-fitting Schechter parameters and their 68\% confidence limits for different environments and galaxy types. For the cluster data we quote, in addition to the statistical uncertainty (first error given), the systematic uncertainty due to cosmic variance in the reference field (second error given).}
\label{tab:Schechter}
\begin{center}
%\begin{adjustwidth}{-0.90cm}{}
\begin{tabular}{l l l l l l}
\hline
\hline
& Environment & $\mathrm{log_{10}[}M^{*}/\mathrm{M_{\odot}}]$ & $\alpha$ & $\Phi^{*a}$ &GoF$^b$\\
\hline
\parbox[t]{2mm}{\multirow{6}{*}{\rotatebox[origin=c]{90}{All galaxies}}}&$R<2R_{500}$&$10.81^{+0.02+0.00}_{-0.02-0.00}$&$-0.91^{+0.02+0.00}_{-0.02-0.00}$&$355.97\pm 4.22^{+9.26}_{-8.28}$&1.19\\
&$R<0.5R_{500}$&$10.81^{+0.02+0.02}_{-0.02--0.02}$&$-0.81^{+0.03+-0.02}_{-0.02-0.02}$&$119.89\pm 2.63^{+-4.29}_{-5.75}$&1.12\\
&$0.5R_{500}<R<R_{500}$&$10.85^{+0.03+0.00}_{-0.03-0.01}$&$-1.00^{+0.04+0.00}_{-0.03-0.00}$&$93.39\pm 2.00^{+2.36}_{-1.85}$&1.01\\
&$R_{500}<R<1.5R_{500}$&$10.76^{+0.03+-0.00}_{-0.04-0.01}$&$-0.85^{+0.05+0.02}_{-0.03--0.01}$&$86.44\pm 2.22^{+6.20}_{-2.33}$&0.88\\
&$1.5R_{500}<R<2R_{500}$&$10.79^{+0.06+0.01}_{-0.03-0.00}$&$-1.00^{+0.04+0.00}_{-0.06-0.00}$&$55.48\pm 1.58^{+3.98}_{-4.41}$&0.73\\
&Average field&$10.98^{+0.02}_{-0.02}$&$-1.20^{+0.02}_{-0.02}$&$320.59\pm 3.49$&0.61\\
\hline
\parbox[t]{2mm}{\multirow{6}{*}{\rotatebox[origin=c]{90}{Quiescent}}}&$R<2R_{500}$&$10.81^{+0.01+0.00}_{-0.02-0.00}$&$-0.83^{+0.03+0.00}_{-0.02-0.00}$&$325.41\pm 4.27^{+3.99}_{-4.87}$&1.22\\
&$R<0.5R_{500}$&$10.82^{+0.03+-0.01}_{-0.03-0.01}$&$-0.78^{+0.04+0.02}_{-0.03--0.02}$&$111.98\pm 2.58^{+3.29}_{--2.61}$&1.22\\
&$0.5R_{500}<R<R_{500}$&$10.85^{+0.04+0.00}_{-0.03-0.00}$&$-0.95^{+0.04+0.00}_{-0.04--0.00}$&$82.55\pm 1.95^{+1.18}_{-0.26}$&0.88\\
&$R_{500}<R<1.5R_{500}$&$10.75^{+0.03+0.00}_{-0.03--0.00}$&$-0.71^{+0.05+0.00}_{-0.05-0.01}$&$80.44\pm 2.33^{+1.32}_{-2.82}$&1.05\\
&$1.5R_{500}<R<2R_{500}$&$10.80^{+0.05+0.01}_{-0.05-0.01}$&$-0.85^{+0.06+0.01}_{-0.06-0.01}$&$47.16\pm 1.62^{+1.79}_{-2.30}$&0.86\\
&Average field&$10.86^{+0.02}_{-0.02}$&$-0.55^{+0.03}_{-0.03}$&$313.85\pm 5.83$&0.72\\
\hline
\parbox[t]{2mm}{\multirow{6}{*}{\rotatebox[origin=c]{90}{Star-forming}}}&$R<2R_{500}$&$10.50^{+0.04+0.02}_{-0.04-0.05}$&$-1.02^{+0.06+0.04}_{-0.06-0.02}$&$76.58\pm 2.12^{+5.14}_{-2.79}$&0.80\\
&$R<0.5R_{500}$&$10.69^{+0.13+0.00}_{-0.11-0.01}$&$-1.11^{+0.15+0.00}_{-0.14-0.01}$&$8.65\pm 0.61^{+0.35}_{-0.31}$&1.32\\
&$0.5R_{500}<R<R_{500}$&$10.54^{+0.09+0.01}_{-0.07-0.01}$&$-1.00^{+0.12+0.02}_{-0.11-0.00}$&$22.24\pm 1.13^{+1.46}_{-1.01}$&1.03\\
&$R_{500}<R<1.5R_{500}$&$10.42^{+0.09+0.06}_{-0.07-0.08}$&$-0.92^{+0.14+0.11}_{-0.14-0.07}$&$23.10\pm 1.29^{+0.31}_{--0.19}$&1.09\\
&$1.5R_{500}<R<2R_{500}$&$10.35^{+0.09+0.09}_{-0.07-0.17}$&$-0.95^{+0.14+0.23}_{-0.15-0.10}$&$28.89\pm 1.48^{+4.80}_{-1.30}$&0.91\\
&Average field&$10.69^{+0.03}_{-0.03}$&$-1.33^{+0.03}_{-0.03}$&$238.72\pm 3.20$&0.75\\
\hline
\end{tabular}
%\end{adjustwidth}
\end{center}
\begin{list}{}{}
\item[$^{\mathrm{a}}$]  Normalisation is reported in the same units as the data points were presented in Table~\ref{tab:SMFpoints}, i.e. [cluster$^{-1}$] for the cluster data, and [$10^{-5}$ Mpc$^{-3}$] for the average field.
\item[$^{\mathrm{b}}$]  Even though we perform a maximum likelihood fit to the unbinned data, we report Goodness of Fits as $\chi^2/\mathrm{d.o.f.}$, where the best-fit models are compared to the binned data.
\end{list}
\end{table*}

\subsection{Radial-dependence of SMF}
\begin{figure*}
\resizebox{\hsize}{!}{\includegraphics{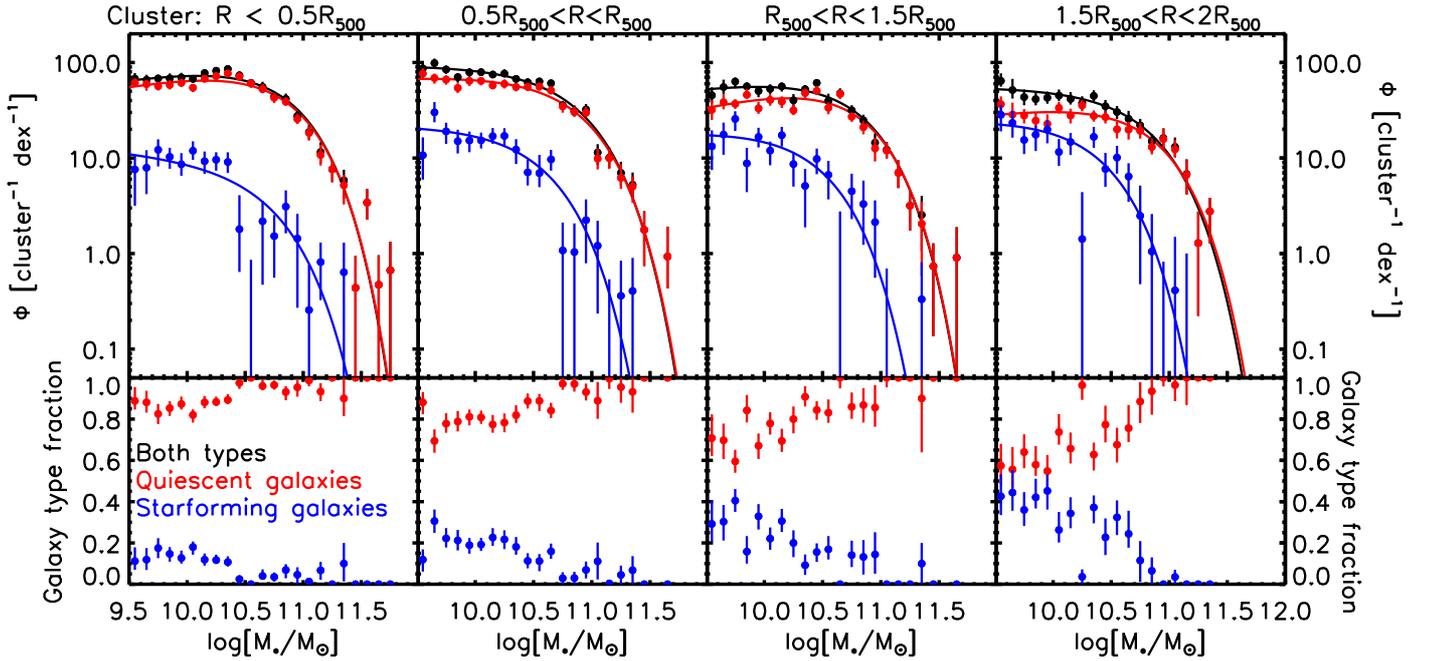}}
\caption{\textit{Top panels:} SMF of cluster galaxies at different radial distances from the cluster centres. \textit{Blue and red data points:} The population of star-forming and quiescent galaxies, respectively. \textit{Black points:} Full galaxy population. \textit{Lower panels:} Relative fraction of quiescent galaxies and star-forming galaxies as a function of stellar mass.}
\label{fig:SMF_clusterradbins}
\end{figure*}

A study of the SMF of galaxies at different radial distances from the cluster centres would allow a more detailed understanding of what happens to the galaxies as they are accreted by the clusters. With the current sample we have the statistics to do this, and Fig.~\ref{fig:SMF_clusterradbins} shows the former cluster SMF, split in four radial bins. The best-fitting Schechter functions are overplotted. Qualitatively, a strong trend is immediately visible; the quenched fraction of galaxies drops with radial distance from the cluster centre, at each stellar mass. 

\begin{figure*}
\resizebox{\hsize}{!}{\includegraphics{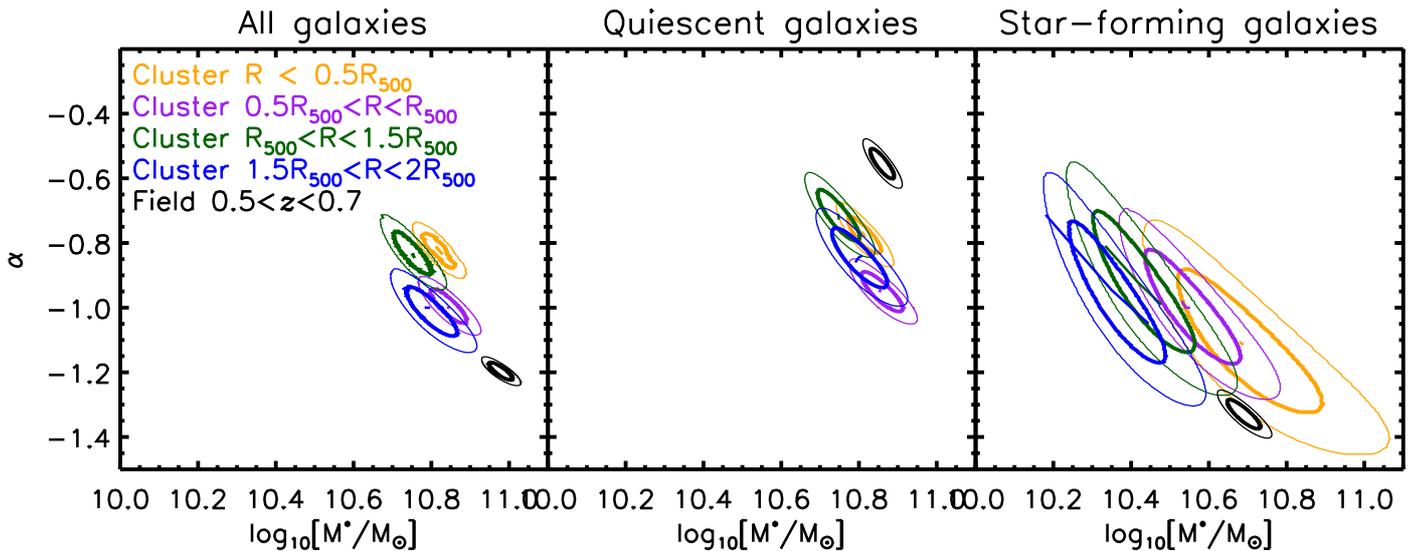}}
\caption{68\% and 95\% likelihood contours for the Schechter parameters $\alpha$ and $M^*$, for the different galaxy types in different panels. \textit{Black:} Schechter parameters of field galaxy population. \textit{Other colours:} Schechter parameters of the cluster galaxies, in different radial bins, as indicated. The lines within the contours indicate how cosmic variance (+/-1$\sigma$) affects the best-fit Schechter parameters. That effect is strongest for the star-forming galaxies at large clustercentric distances, where the overdensity compared to the background is lowest.}
\label{fig:ellipses_master}
\end{figure*}

Realising that the uncertainties on the best-fitting Schechter parameter are not independent of each other, we plot the 2-dimensional 68\% and 95\% confidence regions on $\alpha$ and $M^*$ in Fig.~\ref{fig:ellipses_master}. We note that for star-forming galaxies the over-density compared to the reference field is low, especially in the outskirts. This results in large uncertainties on their SMF, and on the fitted Schechter parameters. The cosmic variance uncertainty is visualised by the lines that are superimposed on the ellipses; these connect the $\pm 1\sigma$ systematic uncertainties due to cosmic variance with the nominal measurement. These systematics are so large that the shape of the SMF of star-forming galaxies is consistent with being independent of environment. 

The quiescent galaxies have a much higher over-density compared to the reference field, and their SMF can thus be measured more accurately since cosmic variance plays a negligible role. The shape of the SMF of quiescent galaxies does not vary significantly with the radial distance from the cluster centre. Also the shape of the total SMF, which is always completely dominated by quiescent galaxies, does not vary significantly between radial bins. There is a significant difference, however, between the SMF of quiescent galaxies in the clusters and in the average field; there are relatively more low-mass quiescent galaxies in the clusters. Some quenching models \citep[e.g.][]{peng10} interpret this as indicative of another quenching mechanism of galaxies in the field as in clusters. If galaxies are ``environmentally'' quenched, one would expect a steeper SMF of quenched galaxies at low masses. That is because the SMF of star-forming galaxies is steep, and because quenching due to environment is supposed to be largely mass-independent. Our findings are broadly consistent with that picture.

\subsection{Normalisation of the SMF}
\begin{figure}
\resizebox{\hsize}{!}{\includegraphics{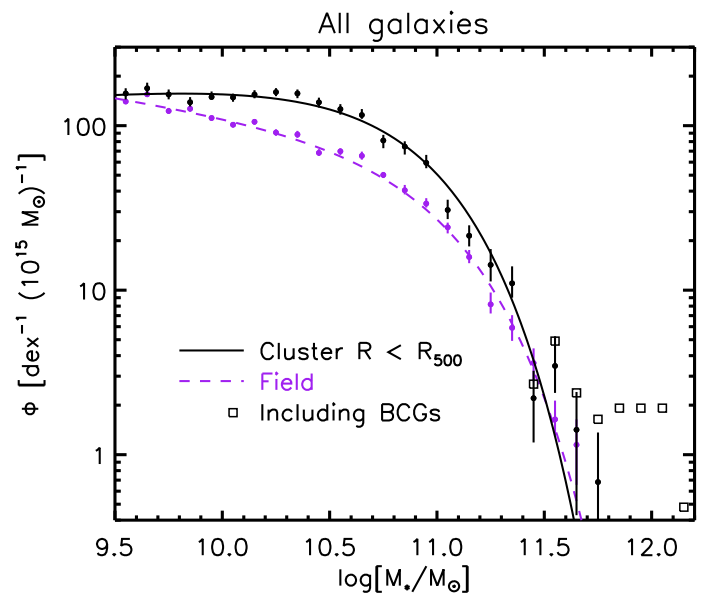}}
\caption{SMF of galaxies in clusters and field, normalised by the total mass associated with the cluster population (black) and the total mass in the volume from which the field SMF is measured (purple). \textit{Squares:} SMF including BCGs, to provide a full accounting of stellar mass in the cluster galaxies.}
\label{fig:SMF_massnorm}
\end{figure}

The cluster SMFs presented in Figs.~\ref{fig:SMF_clustervsfield}~\&~\ref{fig:SMF_clusterradbins} are normalised in number of galaxies per cluster. To understand and compare the efficiency of galaxy formation in clusters to the field, we make a more direct comparison in the normalisations between the different environments. Since clusters have, by definition, a very high volume density of galaxies compared to the field, normalising the SMF over the total volume may not be insightful. Instead, following \citet{vdB13}, we normalise the SMF of field and clusters to the total amount of matter associated with each galaxy population in Fig.~\ref{fig:SMF_massnorm}.  

For the field sample we take the total comoving volume in the redshift range $0.5<z<0.7$ covered by the UltraVISTA unmasked survey area of 1.62 square degrees \citep{muzzin13a}, amounting to $1.5\cdot 10^6\,\mathrm{Mpc^{3}}$. Multiplying this with the average matter density of the Universe in our cosmology, we find a total amount of matter (i.e.~dark matter + baryonic) of $6.1\cdot 10^{16}\,\mathrm{M_{\odot}}$. 

For the clusters we take the total mass within a projected radius of $R_{500}$, but integrated along the line-of-sight. For this we assume that the galaxies follow the total mass, approximated by an NFW profile with concentration $c_{500}\approx 2-3$ \citep[e.g.][]{dutton14,klypin16,vdB18b}. To go from the mass within a sphere of radius $R_{500}$ (which equals $M_{500}$, by definition) to the mass within this cylinder, one multiplies $M_{500}$ with a constant factor of $\sim 1.43$. The total mass associated with the 21 clusters, within a projected radius of $R_{500}$ is then $2.0\cdot 10^{16}\,\mathrm{M_{\odot}}$. 

Fig.~\ref{fig:SMF_massnorm} shows the resulting SMF, normalised by the total mass. Per unit total mass there is a clear overdensity of galaxies in the clusters compared to the field \citep[qualitatively similar to what was found by][]{vdB13}. This shows that it is impossible to create clusters from simply accreting an average field population of galaxies, since the latter includes low-density regions such as voids, where the star formation efficiency is very low. Since groups are typically found in the vicinity of clusters, it is likely that the accretion of these systems caused a galaxy excess in clusters compared to the field \citep[cf. discussions in][]{hennig17,chiu18}.

\subsection{Literature comparison}
The general trends we observe are in line with previous measurements. \citet{vulcani13} study the SMF of cluster galaxies in the same redshift range as we do. They do, for each galaxy type, find no significant difference in the shape of their SMF between cluster and field. Contrary to this work, however, we do find a significant difference between the SMF of quiescent cluster and field galaxies. It is possible that this is owing to our data being substantially deeper ($\sim$ 1 dex), allowing us to probe the low-mass regions where the differences are most pronounced. 

\citet{papovich18} also study the SMF over a redshift range that overlaps with ours, in different environments probed with ZFOURGE and in the NMBS. They have data with similar depth, or even slighly deeper than ours, and base their density estimates on a nearest-neighbour approach. Similar to this work, they find a steepening in the low-mass slope of the SMF of quiescent galaxies in overdensities compared to the field. However, \citet{papovich18} also find an increase in $M^*$ towards higher densities \citep[also see][who base their study on the VIPERS data set]{davidzon16}, which we do not find in the cluster environments. A possible explanation for the apparent discrepancy with these studies is that they probe more moderate overdensities, and include galaxies that are central to their own haloes, while we have deliberately not taken the BCGs into account. Central galaxies are expected to grow from the merging of in-falling satellites, and this may explain the increase in $M^*$. Also in the more moderate environments probed by ZFOURGE and VIPERS, mergers between satellites may be more frequent than in the cluster environment, where relative velocities are expected to be too high for mergers to occur.

\citet{annunziatella14} study the SMF of a massive CLASH cluster at $z=0.44$. They find that the galaxy population is dominated by quiescent galaxies, but only for stellar masses $M_{\star} \gtrsim 10^{10}\,\mathrm{M_\odot}$. This apparent lack of cluster quiescent galaxies may be the result of a different way of separating star-forming from quiescent galaxies, compared to our method. Also cluster-to-cluster variations may play a role here, as suggested by a similar study of the more local cluster Abell 209 \citep{annunziatella16}.

At higher redshift, $z\sim 1$, \citet{vdB13} measure the SMF of 10 clusters, and find that these systems are already dominated by quiescent galaxies down to stellar masses of $\sim10^{10}\,\mathrm{M_\odot}$. Contrarily to more local studies, they find that the shape of the SMF of quiescent is similar between clusters and the field. A likely explanation, apart from a possible evolution with redshift, is that a stellar mass limit of $10^{10}\,\mathrm{M_\odot}$ may not be low enough to probe any differences in the SMF were they are expected \citep[cf.][for a study at even higher redshift]{nantais16}.

\section{Environmental Quenching Efficiency}\label{sec:EQE}
A significant result of this work is that the fraction of quiescent galaxies is much higher in the cluster environment than in the field, at the same redshift, and at a given stellar mass. Here we quantify this using the environmental quenching efficiency ($f_{\mathrm{EQ}}$), which can be thought of as the fraction of galaxies that would normally be star-forming in the field, but are quenched by their environment. Specifically, 

\begin{equation}\label{eq:EQE}
f_{\mathrm{EQ}}=\frac{f_\mathrm{q,cluster}-f_\mathrm{q,field}}{1-f_\mathrm{q,field}},
\end{equation}
where $f_\mathrm{q,cluster}$ and $f_\mathrm{q,field}$ are the quiescent fraction of galaxies in the cluster and field environment, respectively. The quenched fractions are a function of both stellar mass and environment, and therefore also $f_{\mathrm{EQ}}$ may depend on stellar mass and environment. 
Such a term was already introduced by \citet{vandenbosch2008}, and sometimes the term ``conversion fraction'' is used for a similarly defined quantity \citep{balogh16,fossati17}. 

A compilation of $f_{\mathrm{EQ}}$ in groups and clusters is shown in Figure 7 of \citet{nantais16}. The general trend is that $f_{\mathrm{EQ}}$ increases with increasing halo mass (broadly speaking, from group to cluster environments). There is also a hint that $f_{\mathrm{EQ}}$ increases towards the local universe, at a given halo mass. This is in line with results from e.g. \citet{darvish16}, who find no evidence for environmental quenching in more moderate overdensities in the COSMOS field at redshift $z\gtrsim 1$.

\subsection{Environmental quenching efficiency versus stellar mass}\label{sec:eqemass}
\begin{figure}
\resizebox{\hsize}{!}{\includegraphics{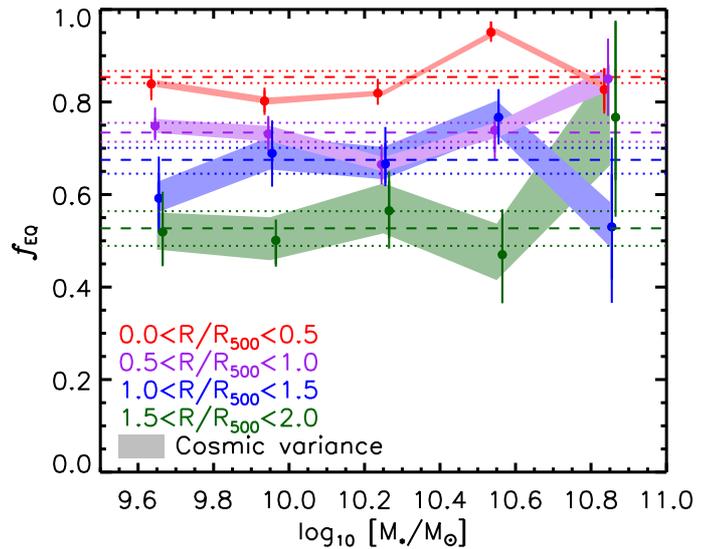}}
\caption{Environmental efficiency as a function of stellar mass, for different radial bins. While there are wiggles around a flat relation, there is no significant monotonic mass-dependent trend.}
\label{fig:eqe_vsstelmass}
\end{figure}

Figure~\ref{fig:eqe_vsstelmass} shows the $f_{\mathrm{EQ}}$ as a function of stellar mass, in four radial bins from the cluster centres. Error bars are the 68\% confidence regions from 100 bootstrap resamplings, where galaxies within the clusters are drawn with replacement. In addition, we perform 25 bootstrap resamplings in which the clusters themselves are drawn with replacement. The two bootstrap runs lead to similar uncertainties. We also show a cosmic variance uncertainty (indicated by the shaded regions) increases towards larger clustercentric radii, where the cluster over-density is lower. 

In none of the radial bins does the $f_{\mathrm{EQ}}$ show a systematic trend with stellar mass (either increasing or decreasing). However, there are significant wiggles around the mean values (indicated by the dotted line); in Sect.~\ref{sec:discussion} we quantify these as signatures of merging of cluster galaxies compared to the field, which likely happened in their pre-processing environment. 

Since there is no clear stellar-mass dependence of the $f_{\mathrm{EQ}}$ in any of the radial bins, it suggests that the main quenching process happening in clusters is mass-independent \citep[also see e.g.][]{kawinwanichakij17}. A candidate would be a stripping process of the star-forming gas that is so rapid that it is essentially mass-independent, namely ram pressure stripping. 

\subsection{Environmental quenching efficiency versus radius}
\begin{figure}
\resizebox{\hsize}{!}{\includegraphics{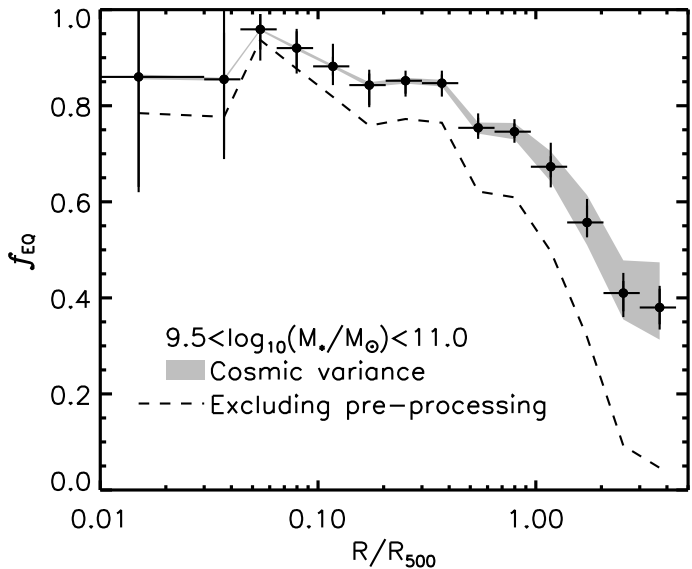}}
\caption{Environmental quenching efficiency as a function of radial distance from the cluster centres, measured in a single broad mass bin. We can do this because environmental quenching, at least in this regime, seems to be a process that is largely mass-independent (cf.~Sect.~\ref{sec:eqemass}). \textit{Dashed curve:} Same but increasing the quenched fraction in the field (cf.~Eq.~\ref{eq:EQE}) so that $f_{\mathrm{EQ}}$ is consistent with zero in the outermost bin.}
\label{fig:eqe_vsradius}
\end{figure}

The $f_{\mathrm{EQ}}$ being independent of stellar mass, we study its dependence on clustercentric radius. We combine the $f_{\mathrm{EQ}}$ of all galaxies with stellar masses in the range $10^{9.5} \leq M_{\star}/\mathrm{M_{\odot}}\leq 10^{11}$ in Figure~\ref{fig:eqe_vsradius}. Using logarithmic bins, we study the $f_{\mathrm{EQ}}$ from $0.01\times R_{500}$ to $4\times R_{500}$. There is a clear signature of environmental quenching that depends on radial distance (which scales with local density). The $f_{\mathrm{EQ}}$ in the cluster centres (where $R/R_{500}<0.1$) are extremely high, $\sim90\%$, even in projection where galaxies on the cluster periphery are mixed along the line-of-sight. There is a steep drop in the $f_{\mathrm{EQ}}$ with radial distance, especially in the range $0.2<R/R_{500}<2.0$.

Interestingly, the $f_{\mathrm{EQ}}$ does not drop all the way to zero towards the cluster periphery, but rather converges to a value of $\sim 0.35$ at the largest clustercentric distances we probe. In Appendix~\ref{sec:appraddependence} we demonstrate that this measurement is robust, even though we study galaxies detected in a slightly shallower region of the $\mathrm{K_s}$-band stack. Using a cosmological N-body simulation, \citet{wetzel14} show that this excess quenching at large radii may be the result of ejected cluster satellites, which orbit even beyond the clusters' virial radii. Another possible contributor to this observed excess quenching on larger scales comes from galaxies that have been pre-processed in the rich group environment that surrounds galaxy clusters \citep[e.g.][]{haines15,bianconi18}. If we re-define the environmental quenching efficiency in Eq.~\ref{eq:EQE} with respect to the quenched fraction in the cluster periphery, i.e.~substitute $f_\mathrm{q,field}$ with $f_\mathrm{q,periphery}$, we obtain the dashed curve shown in Fig.~\ref{fig:eqe_vsradius}. This curve is based on a pre-processed value of 0.35, and illustrates the effect of the main quenching mechanism in the cluster.

At first glance, the measured strong dependence of $f_{\mathrm{EQ}}$ on radius suggest that, whatever physical process is responsible, quenching must happen on a reasonably rapid timescale, at least when galaxies approach the cluster centres. If quenching were a slow process, freshly accreted star-forming galaxies would have time to migrate to the cluster centres while still forming stars, and this would lower the observed $f_{\mathrm{EQ}}$ in the cluster centres. We quantify these statements in the following subsection, in which we employ a simple quenching model to put the observations into context.

\subsection{A simple quenching model}
We consider a model to identify the approximate timescale over which a galaxy is environmentally quenched in the cluster, and the location where this quenching process is triggered. Our basis is a set of N-body simulations of four galaxy clusters, introduced in \citet{taranu14}. The four most massive clusters were identified from a large cosmological N-body simulation with 256$^3$ particles in a cosmological box of side length 512$h^{-1}$ Mpc. Particles in the re-simulation have masses of $6.16\times 10^8\, \mathrm{M_{\odot}}$, meaning that subhaloes down to relatively low halo mass can be resolved and traced in time from $z=3$ to $z=0$. 

Using this simulation, we investigate at which distances from the cluster centres a quenching transformation process is likely to start, and how long it would take for a galaxy to show a signature of quenched star formation. Following a similar approach as in \citet{muzzin14}, in which phase-space distribution of specific subhaloes in these simulations were tracked, we now only consider the projected clustercentric distances of a population of subhaloes in the simulation. Subhaloes are marked that have passed, for the first time, a clustercentric distance $r_{\mathrm{3D,quench}}/R_{500}$ at least a time of $T_{\mathrm{quench}}$ Gyr ago. Projecting each cluster in three directions (x,y,z), we mark the fraction of subhaloes that satisfy these criteria, as a function of projected clustercentric radius. 

\begin{figure}
\resizebox{\hsize}{!}{\includegraphics{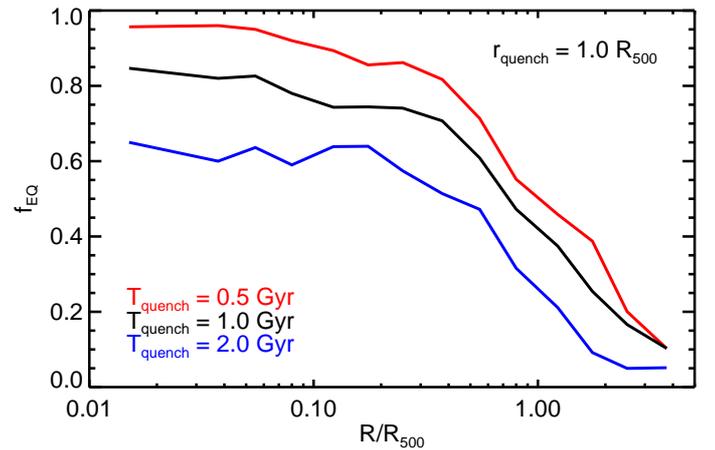}}
\caption{Radial dependence of $f_{\mathrm{EQ}}$ from the model, where we vary the quenching time (as indicated) while fixing the quenching location to $R_{500}$.}
\label{fig:eqe_sim_varT}
\end{figure}

\begin{figure}
\resizebox{\hsize}{!}{\includegraphics{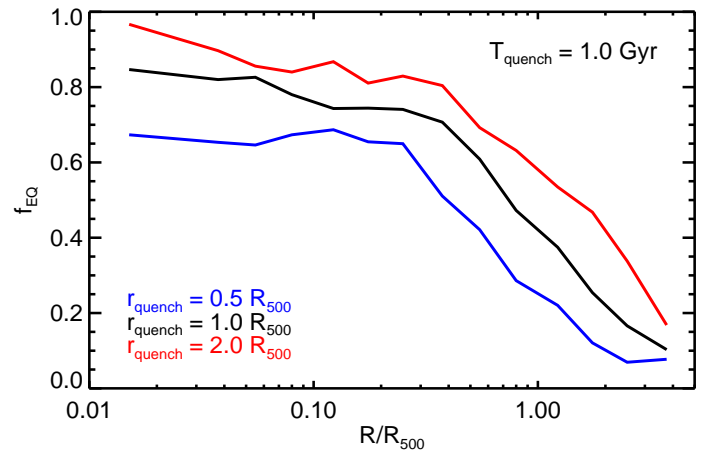}}
\caption{Radial dependence of $f_{\mathrm{EQ}}$ from the model, where we vary the quenching location (as indicated) while fixing the quenching time to 1 Gyr.}
\label{fig:eqe_sim_varR}
\end{figure}

The results are in Figs.~\ref{fig:eqe_sim_varT}~\&~\ref{fig:eqe_sim_varR}, where one parameter in the model is kept constant, while the other is varied. We note that $T_{\mathrm{quench}}$ has to be interpreted as a delay time + quenching time, and that the quenching time itself is supposed to be a rapid process due to the absence of a significant fraction of green valley (transition) galaxies \citep{peng10,wetzel13}. The similarity between Figs.~\ref{fig:eqe_sim_varT}~\&~\ref{fig:eqe_sim_varR} indicate that there is a degeneracy between the quenching radius and time scale. 

\begin{figure}
\resizebox{\hsize}{!}{\includegraphics{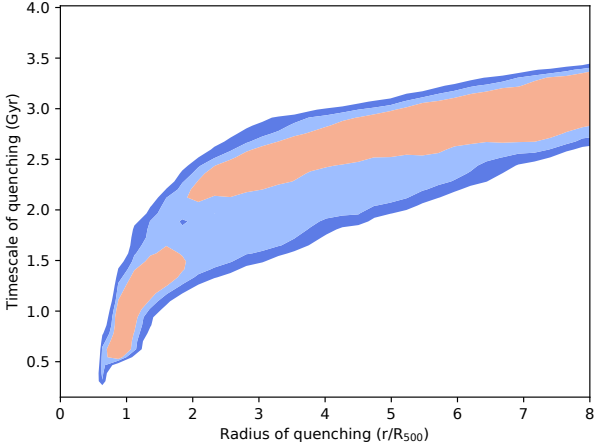}}
\caption{The 1-, 2-, and 3-$\sigma$ contours surrounding the best-fitting location and timescale of quenching, so that our quenching model reproduces the observed radial-dependence of the environmental quenching efficiency. Pre-processing is left as a free parameter in the likelihood maximisation.}
\label{fig:maxlikewithmodel}
\end{figure}

While any ejected satellites that have been quenched by the same (cluster-specific) mechanism would show up in the projected distributions, we note that any possible pre-processing of galaxies in the large-scale overdensity surrounding the clusters is by definition not shown in this simplistic model. We perform a maximum likelihood comparison between the results from the simulation and the actual data, as a function of quenching timescale and -radius, while we include the pre-processed fraction of satellites separately as an extra free parameter. For each value of $r_{\mathrm{3D,quench}}/R_{500}$ and $T_{\mathrm{quench}}$ in the model, we thus marginalise over this pre-processed fraction of galaxies. As an example, the data points for a pre-processed fraction of 0.35 are shown in Fig.~\ref{fig:eqe_vsradius}. The resulting confidence regions on the parameters that represent the best model, are shown in Fig.~\ref{fig:maxlikewithmodel} and these confirm the strong degeneracy between $r_{\mathrm{3D,quench}}/R_{500}$ and $T_{\mathrm{quench}}$.

The contours shown in Fig.~\ref{fig:maxlikewithmodel} follow an intriguing degeneracy; quenching appears to happen either on a short timescale at small clustercentric radii, or over a longer time scale at larger clustercentric radii. Both scenarios reproduce the data similarly well. We note that the bimodality in the 1-$\sigma$ contour is likely due to the limited number of four clusters that we used in the simulation (albeit as studied from 3 orthogonal sight lines). 
The pre-processed fractions of galaxies are not shown in Fig.~\ref{fig:maxlikewithmodel}, and this fraction gradually decreases when going from small radii (and corresponding short times), to larger radii (and longer times). For example, the pre-processing level at $r_{\mathrm{3D,quench}}=1\times R_{500}$ and $T_{\mathrm{quench}}$=1 Gyr is $31.7^{+2.1}_{-2.4}$, while it decreases to $19.8^{+3.2}_{-2.9}$ at $r_{\mathrm{3D,quench}}=7\times R_{500}$ and $T_{\mathrm{quench}}$=3 Gyr.

Even though there is a strong degeneracy between the parameters, we can put a firm lower limit to the radial distance from the cluster centre where quenching is triggered; $r_{\mathrm{quench}}> 0.67R_{500}$ (95\%CL). We can link this to an ICM density where quenching occurs, using our deep XMM-Newton observations. Leaving a more detailed study of the radial gas density distribution of this cluster sample to \citet{vdB18b}, the gas density at $R=0.7R_{500}$ is around $\rho_{\mathrm{ICM}}\approx 2\cdot10^4\,\mathrm{M_{\odot}\,kpc^{-3}}$.

Following their Equation 62, \citet{Gunn1972} estimate at which ICM density ram pressure stripping is expected to become effective in the stripping of the interstellar material of a typical infalling galaxy. They find this to happen at an ICM density of $\sim 5\times 10^{-4}\,\mathrm{atoms\,cm^{-3}}$.  The density at $R=0.7R_{500}$ is already about twice this value, making ram pressure stripping a likely contributor given our constraints on location and time scale of the main stripping process. Similarly, based on a cosmological simulation \citet{zinger18} find that a significant removal of star-forming gas happens at $r \lesssim 0.5R_{\mathrm{vir}}$, which is a similar fraction of $R_{500}$.

Another likely contributor to the quenching process is strangulation/starvation, which is a cut-off from the cosmological accretion of hot gas after a galaxy is accreted by the cluster main halo. Star formation is then expected to quench after the reservoir of molecular gas is depleted. This time scale is on the order of 1 Gyr for local low-mass galaxies \citep{tacconi18}, and likely shorter at higher redshift. This is a slower process than ram pressure stripping, but its time scale is also consistent with our quenching constraints in Fig.~\ref{fig:maxlikewithmodel}. Since this process will be triggered at larger distance (or equivalently, at earlier times) than ram pressure stripping, both processes may be contributing to the observed elevated fraction of quenched galaxies.

\subsection{Caveats}\label{sec:caveats}
Our modelling has shown that the radial distribution of $f_{\mathrm{EQ}}$ combined with a realistic simulation of the orbits of cluster galaxies can provide meaningful constraints on the $r_{\mathrm{3D,quench}}/R_{500}$ and $T_{\mathrm{quench}}$ of cluster galaxies. There are several caveats in our analysis that could play a role in the interpretation, and these are highlighted below.

In this paper we measure the $f_{\mathrm{EQ}}$ as the excess quenching with respect to the field, i.e.~a representative section of the universe at the same redshift. The field sample therefore already includes regions of higher densities, such as groups. Several studies have quantified the $f_{\mathrm{EQ}}$ (or conversion fraction) as the excess quenching with respect to the lowest mass density found in the field \citep[typically the lowest-density quartile, cf.~][]{papovich18}. The quenched fractions we report for the field include some ``pre-quenching'' in moderate overdensities, and the reported $f_{\mathrm{EQ}}$ values would have been even higher if defined with respect to the lowest-density regions. However, since our modelling marginalises over this pre-processing component, this assumption does not affect our estimated parameters that describe the main quenching process in the clusters.

We have defined $f_{\mathrm{EQ}}$ with respect to the field \textit{at the same epoch as the clusters are observed at}. An alternative approach is to consider the field quenched fraction at the time of galaxy accretion \citep[see the discussion in][]{balogh16}, as e.g.~\citet{foltz18} have done in their modelling of the quenching time scales.  Both approaches have their uses; we have chosen the former so that we can isolate what happens in clusters separately, and in addition to, what would have happened to the star-forming properties of the galaxies if they had remained centrals in their own haloes. 

Furthermore, our definition of $f_{\mathrm{EQ}}$ is interpreted in the absence of mergers between galaxies. We know mergers are happening, especially for galaxies in group scales, and mergers may even contribute to the quenching of satellites \citep[e.g.][]{peng10,darvish16}. Mergers may also have affected the observed $f_{\mathrm{EQ}}$ dependence of clustercentric radius. Moreover, mergers lead eventually to growth of the central BCGs. In the next section we discuss these limitations and present a ``transformation function'' that describes how additional processes, such as mergers, are affecting the galaxy population to lead to the SMF we observe for cluster galaxies.

\section{Discussion}\label{sec:discussion}
\begin{figure}
\resizebox{\hsize}{!}{\includegraphics{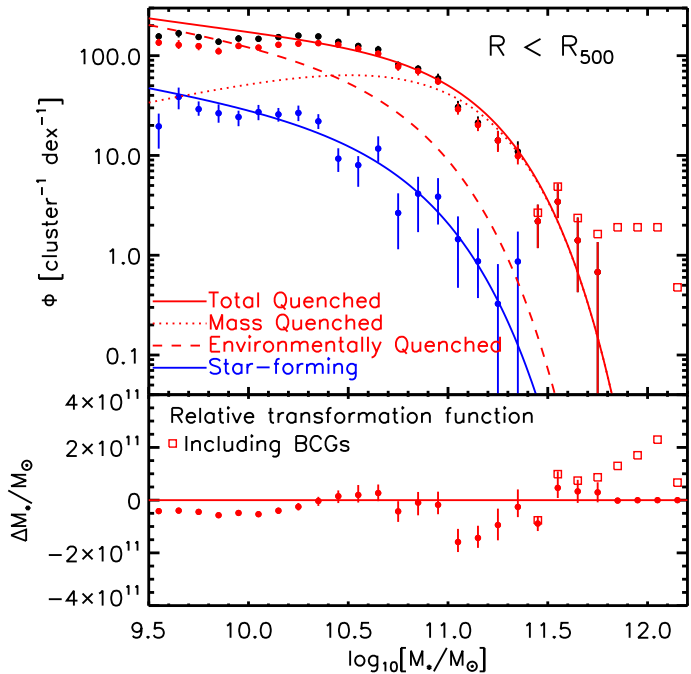}}
\caption{\textit{Top:} Cluster data points for the different galaxy types within $R_{500}$. The curves show a basic quenching model inspired by the formalism presented in \citet{peng10}, with as starting point the UltraVISTA SMF in the redshift range $0.5<z<0.7$. \textit{Bottom:} Difference between the data (total distribution of quenched galaxies) in the cluster compared to the prediction from the model. This suggests a relative removal/destruction of low-mass galaxies and the expected build-up of BCGs.}
\label{fig:simplemodel}
\end{figure}

The previous Section explored the stellar-mass dependence, and the radial dependence of $f_{\mathrm{EQ}}$ to infer a simple quenching scenario for which we have constrained the approximate location and time scale. In this Section we study the impact of a basic environmental-quenching scenario on the SMF, which we measured and studied in Sect.~\ref{sec:SMF}. 

The simplest environmental quenching model that works in a mass-independent fashion (as suggested by Fig.~\ref{fig:eqe_vsstelmass}) leaves the shape of the SMF of star-forming galaxies independent of environment \citep{peng10}. This is consistent with our findings, considering the measurement uncertainties associated with measuring the SMF of star-forming galaxies in the cluster periphery (Fig.~\ref{fig:ellipses_master}). The population of quenched cluster galaxies is then a combination of galaxies that would also have quenched outside of the cluster (mass-quenched galaxies), and the environmentally-quenched galaxies, which in principle follow a similar mass distribution as the star-forming galaxies. We start by employing this quenching model in its basic form, following \citet{vdB13}. 

Figure~\ref{fig:simplemodel} shows the SMF of cluster galaxies within $R \leq R_{500}$, i.e.~a combination of the first two panels of Fig.~\ref{fig:SMF_clusterradbins}, or a differently-normalised version of Fig.~\ref{fig:SMF_massnorm}. The plotted curves are not fits to the plotted data, but rather adapted versions (only in normalisation) of the best-fitting Schechter functions to the field data from the UltraVISTA survey. The total normalisation of the (red+blue) curve is set by the total stellar mass in cluster galaxies (all data points, including the BCGs). Since we assume that the environmental quenching process affects the star-forming population in a mass-independent manner, we set this by parameter $f_{\mathrm{EQ}}$, so that the relative normalisation of the star-forming SMF is (1-$f_{\mathrm{EQ}}$) compared to the one in the field. Again requiring that the normalisation (total stellar mass) of the blue Schechter function is the same as that of the blue cluster galaxies (data points), we find that $f_{\mathrm{EQ}}=0.80$. The quenched part of the star-forming population gives the ``environmentally quenched galaxies'', and when these are combined with the ``mass quenched galaxies'', we arrive at the total population of quenched cluster members, as shown by the solid red curve in Fig.~\ref{fig:simplemodel}.

In its basic form, this quenching model over-predicts the abundance of quiescent low-mass galaxies with $M_{\star} \lesssim 10^{10.2}\,\mathrm{M_{\odot}}$. Interestingly, a similar trend is revealed in Fig.~10 of \citet{vdB13}, in which clusters at slightly higher redshift were studied. However, contrary to this earlier work, we now have the statistics to explore this regime in more detail. That low-mass galaxies show a deficit in clusters compared to this simple quenching model may have to do with their destruction, potentially leading to a build-up of the intra-cluster light (ICL). Given the negative colour gradients that are observed in the ICL of massive clusters, dwarf galaxies are likely contributors to the ICL at large clustercentric distances \citep{demaio15}.

Furthermore, milder interactions and mergers between galaxies \citep[likely also in their pre-processing environment, cf.][]{tomczak17} may also affect the SMF of galaxies, and make it diverge from the field. For instance, \citet{rudnick12} invoke a model that includes mergers between galaxies to reproduce the luminosity function of clusters in the local universe (starting from a distant cluster at $z=1.62$). We attempt to encompass all these processes in a ``transformation function'', shown in the lower panel of Fig.~\ref{fig:simplemodel}. Plotted is the difference between the data points for the quiescent galaxies, and the simple quenching model (solid red line). The plot is normalised in stellar mass per bin and per cluster, and highlights again the relative destruction of low-mass galaxies, in favour of the growth of more massive galaxies such as the central BCGs. Since this plot sums up to 0, by construction, it does not include the build-up of ICL.

\subsection{How to build a massive cluster of galaxies?}
Given our study of 21 massive clusters at $0.5<z<0.7$, we summarise some of the steps required to assemble the galaxy population observed within these systems, as opposed to the general field:
\begin{itemize}
\item As shown in Fig.~\ref{fig:SMF_massnorm}, where we have normalised the SMFs of clusters and field with respect to the total amount of mass associated with the respective galaxy populations, galaxies form relatively efficiently in (future) cluster environments compared to the average Universe. The galaxy abundance in the clusters we study, per unit total mass, is about twice average.

\item The quenched fraction is much higher in these clusters than in the field, at the same redshift, for each stellar mass we probe. It is also elevated compared to the pre-processing we find happening in the cluster surroundings. This quenching process happens in a largely mass-independent fashion (cf.~Fig.~\ref{fig:eqe_vsstelmass}). 

\item There is a significant and strong radial trend in the quenched fraction of cluster galaxies, which we describe as the environmental quenching efficiency $f_{\mathrm{EQ}}$ (cf.~Fig.~\ref{fig:eqe_vsradius}). A comparison with a model that is based on orbits taken from an N-body simulation suggests that the quenching process likely involves strangulation/starvation after cut-off from cosmological accretion, and ram pressure stripping at smaller clustercentric distances to ``finish the job''. Each of these processes ought to happen on time scales that are roughly consistent with what we find in the model.

\item Additional transformations are required to reproduce the observed cluster SMF. These are likely largely caused by merger events, the possible destruction of low-mass galaxies in the clusters, and an effective build-up of the BCGs.  Mergers are happening likely \textit{before} galaxies are being accreted into the clusters, since relative velocities in the cluster environments are too high for galaxies to merge there. We have quantified the combination of these effects in the lower panel of Fig.~\ref{fig:simplemodel}.
\end{itemize}

\section{Summary and conclusions}\label{sec:summary}
We have studied the galaxy population in a sample of 21 high-mass clusters at $0.5<z<0.7$, found in the \textit{Planck} SZ survey. Using multi-band photometry spanning $u$- to the $\mathrm{K_s}$-band for each cluster, we have defined a sample of cluster galaxies, which are highly overdense compared to the back- and foreground. 

The data allow for a precise measurement of the galaxy SMF in clusters of intermediate redshift. We have identified differences in the SMF between the cluster population, and galaxies in the field at the same redshift. Normalising the SMF to the total amount of matter associated with each galaxy population, we find that clusters have a higher galaxy content, per unit total mass, than the average field. 

The most significant differences between the galaxy population in cluster and field arise when we separate the galaxy population between star-forming and quiescent galaxies by means of their rest-frame U-V and V-J colour distributions. The shape of the SMF of star-forming galaxies does not depend on environment. On the contrary, the SMF of quiescent galaxies is significantly different between the cluster and field; there is a relatively higher fraction of low-mass quiescent galaxies in the clusters. Moreover, the fraction of passive galaxies is much higher in the cluster than in the field, and we quantify how this fraction rises steeply with decreasing cluster-centric radius.

We measured the environmental quenching efficiency ($f_{\mathrm{EQ}}$), which describes the fraction of galaxies that would be forming stars in the field, but are quenched solely due to their environment. At fixed radial distance from the cluster centre, the $f_{\mathrm{EQ}}$ does not depend on stellar mass. Contrarily, the $f_{\mathrm{EQ}}$ shows a strong radial dependence within the cluster environment. 

We interpret the observed radial-dependence of the $f_{\mathrm{EQ}}$ with a simple quenching model based on an N-body simulation using which we constrain the characteristic location and time scale of the main environmental quenching process. We find a strong degeneracy between those two parameters. According to the model, quenching may already be triggered at $r_{\mathrm{3D,quench}}\approx 7\times R_{500}$, and would then happen on a long time scale $T_{\mathrm{quench}}\approx$3 Gyr. If quenching is triggered at shorter radial distances $r_{\mathrm{3D,quench}}\approx 1\times R_{500}$, it happens on roughly the molecular gas depletion time scale, $T_{\mathrm{quench}}\approx$ 1 Gyr. Part of the observed quenching excess is thus likely due to ``starvation''/``strangulation'' of galaxies using up their cold gas supply. Interestingly, the model rules out a quenching location $r_{\mathrm{quench}}< 0.67R_{500}$ at 95\% confidence. Our \textit{XMM-Newton} data show that the gas density at this clustercentric distance is so large that ram pressure stripping should be effective, and is likely responsible for the satellite quenching there. This process may thus ``finish the job'' whenever the starvation mechanism does not operate rapidly enough.

\begin{acknowledgements}
We thank Andrea Biviano and Gabriella De Lucia for insightful discussions, and Michael Balogh for constructive feedback on the manuscript. The research leading to these results has received funding from the European Research Council under the European Union's Seventh Framework Programme (FP7/2007-2013) / ERC grant agreement n$^{\circ}$ 340519. HD acknowledges financial support from the Research Council of Norway.

Based on observations obtained with MegaPrime/MegaCam, a joint project of CFHT and CEA/DAPNIA, at the Canada-France-Hawaii Telescope (CFHT) which is operated by the National Research Council (NRC) of Canada, the Institute National des Sciences de l'Univers of the Centre National de la Recherche Scientifique of France, and the University of Hawaii. This work is also based on data collected at Subaru Telescope, which is operated by the National Astronomical Observatory of Japan. Based in part on data collected at Subaru Telescope and obtained from the SMOKA, which is operated by the Astronomy Data Center, National Astronomical Observatory of Japan. Based on observations obtained with \textit{XMM-Newton}, an ESA science mission with instruments and contributions directly funded by ESA Member States and NASA. The development of Planck has been supported by: ESA; CNES and CNRS/INSU-IN2P3-INP (France); ASI, CNR, and INAF (Italy); NASA and DoE (USA); STFC and UKSA (UK); CSIC, MICINN and JA (Spain); Tekes, AoF and CSC (Finland); DLR and MPG (Germany); CSA (Canada); DTU Space (Denmark); SER/SSO (Switzerland); RCN (Norway); SFI (Ireland); FCT/MCTES (Portugal); and PRACE (EU).

This research made use of the following databases: the NED and IRSA databases, operated by the Jet Propulsion Laboratory, California Institute of Technology, under contract with the NASA; SIMBAD, operated at CDS, Strasbourg, France; SZ-cluster database operated by IDOC at IAS under contract with CNES and CNRS.
\end{acknowledgements}

\bibliographystyle{aa} 
\bibliography{MasterRefs} 

\begin{appendix}

\section{Robustness tests}
In this Section we study the effect of several assumptions we had to make in our analysis on the final results. 

\subsection{Photo-$z$ selection}\label{sec:appphotozsel}
One of the choices in our analysis is the initial photo-$z$ selection window of cluster members, before the background subtraction is performed. If such a window is chosen too small, uncertainties in photometric redshifts of cluster members may cause them to scatter out of the selection window. On the other hand, if the photo-$z$ selection window is chosen larger than necessary, we introduce additional noise while performing the statistical background subtraction. Since we use a single reference field to do the statistical background subtraction, especially the contribution from cosmic variance in that field would increase with a larger photo-$z$ cut (since the overdensity will go down).

From all 1527 spectroscopically-confirmed cluster members in our sample, we find that 89.6\% satisfy the photo-$z$ cut of 0.07 (i.e.~only 10.4\% have scattered out). So to first order, the normalisation of the cluster SMF is higher by 1/0.896 compared to our measurement. The scattering may also bias the shape of the SMF, since the photo-$z$ scatter depends (in principle, though only slightly in practise) on the galaxy stellar mass (cf.~Fig.~\ref{fig:speczphotz}). To test the effect of this, we have performed the analysis with photo-$z$ selections of $|\Delta z|<$0.10 and 0.13. This leads to percentages of galaxies that scatter out, of 5.7\% and 4.5\% respectively, approaching the percentage of catastrophic outliers. We checked that the main results in this paper, i.e.~the SMF and the behaviour of the $f_{\mathrm{EQ}}$ do not change significantly (i.e.~by more than the reported uncertainties) when a broader photo-$z$ selection is chosen. The only exception is the SMF of star-forming galaxies in the outskirts ($R>R_{500}$) of the clusters, where the overdensity is very low. We note, however, that since the overdensity of cluster galaxies with respect to the background drops for a broader photo-$z$ selection, uncertainties of all measurements grow substantially. We have therefore chosen a cut of 0.07 in $|\Delta z|$.

\subsection{UVJ division}\label{sec:appuvjdiv}
\begin{figure}
\resizebox{\hsize}{!}{\includegraphics{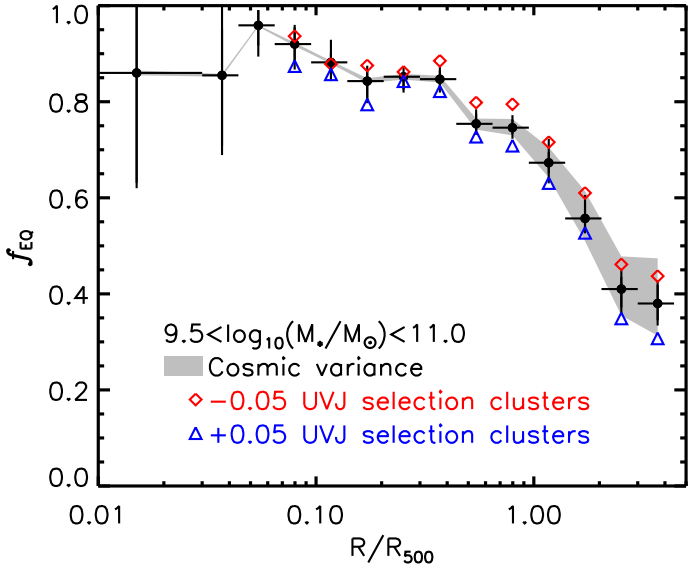}}
\caption{Environmental quenching efficiency as a function of radial distance from the cluster centres, measured in a single broad mass bin. This is a similar plot as Fig.~\ref{fig:eqe_vsradius}, but now we explore the effect of a rather extreme residual shift in rest-frame colour between the field and cluster data.}
\label{fig:eqe_vsradius_testUVJdiff}
\end{figure}

The division between star-forming and quiescent galaxies is a critical part of the analysis. Our analysis makes use of the U-V and V-J rest-frame colours, which ensures that we separate the effect of dust reddening from the reddening due to lack of star formation. In Sect.~\ref{sec:rfcolours} we described the small corrections (on average 0.04-0.06) we made to the rest-frame U-V and V-J colours of the cluster galaxies, to match them to the colour distribution of the reference field. Here we perform a test to check how a residual systematic colour offset between field and cluster data would impact our results. We increase all U-V rest-frame colours of cluster galaxies, and reference background galaxies, by $\pm$0.05, and re-measure the quenched fractions of galaxies in the cluster. Even such a large offset, compared to the residuals we expect, changes that quenched fractions of cluster galaxies by at most 10\%. When we measure $f_{\mathrm{EQ}}$ this has a larger effect, since we have not changed the rest-frame colours of the field galaxies. The result on the radial dependence of $f_{\mathrm{EQ}}$ is shown in Fig.~\ref{fig:eqe_vsradius_testUVJdiff}. Even though a significant change is notable, we note that this is based on a rather extreme systematic between rest-frame colours of cluster and field galaxies. 

\subsection{Radial dependence of $f_{\mathrm{EQ}}$}\label{sec:appraddependence}
The main analysis of this paper studies the properties of galaxies within $2\times R_{500}$ from the cluster centres. These regions fall on the deep part of the $\mathrm{K_s}$-band stacks, and we have characterised the completeness in Sect.~\ref{sec:completeness}. Figure~\ref{fig:eqe_vsradius}, however, shows the environmental quenching efficiency up to $4\times R_{500}$, which covers part of the shallower regions. Due to the dither strategy we chose for the $\mathrm{K_s}$-band imaging, the depth in the $\mathrm{K_s}$-band drops by a maximum amount of 0.7 magnitudes towards a distance of $4\times R_{500}$ from the cluster centres, corresponding to 0.3\texttt{dex} in stellar mass. We note that the optical data, which are essential for precise estimates of photometric redshifts, extend to a larger region around the clusters at uniform depth. We studied the impact of a slight decrease in depth of the detection band on the results plotted in Fig.~\ref{fig:eqe_vsradius}. The measurements move within the plotted uncertainties, when galaxies in the range $10^{9.8} \leq M_{\star}/\mathrm{M_{\odot}}\leq 10^{11}$ are considered (instead of $10^{9.5} \leq M_{\star}/\mathrm{M_{\odot}}\leq 10^{11}$). In particular, we verified that the ``plateau'' in $f_{\mathrm{EQ}}$ at radii $R \gtrsim 2\times R_{500}$ is robust, and not an effect of this decrease in depth.

\section{Cluster gallery}\label{sec:colourimages}
\begin{figure*}
\centering
\begin{minipage}{.495\textwidth}
  \centering
  \includegraphics[width=.90\linewidth]{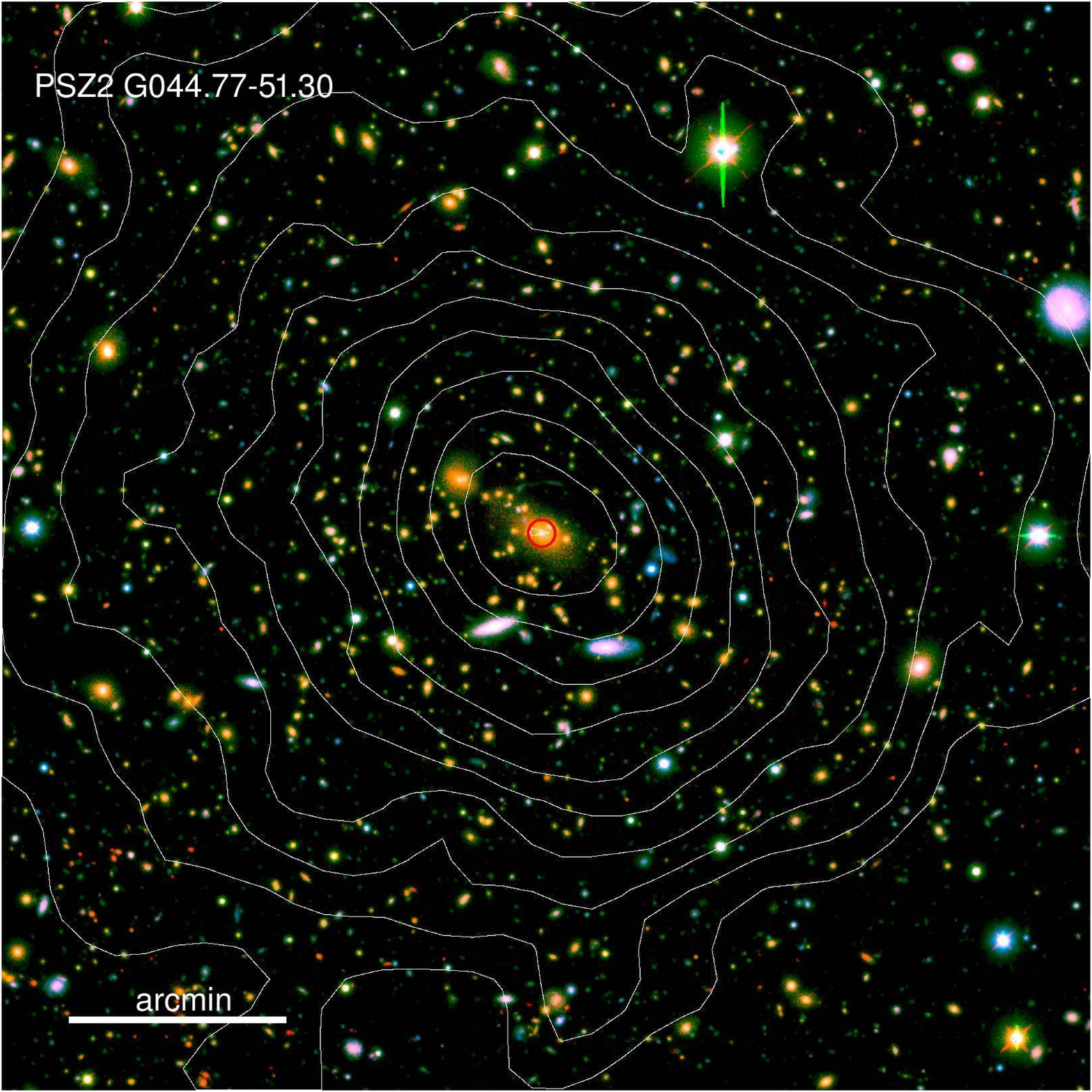}
\end{minipage}
\begin{minipage}{.495\textwidth}
  \centering
  \includegraphics[width=.90\linewidth]{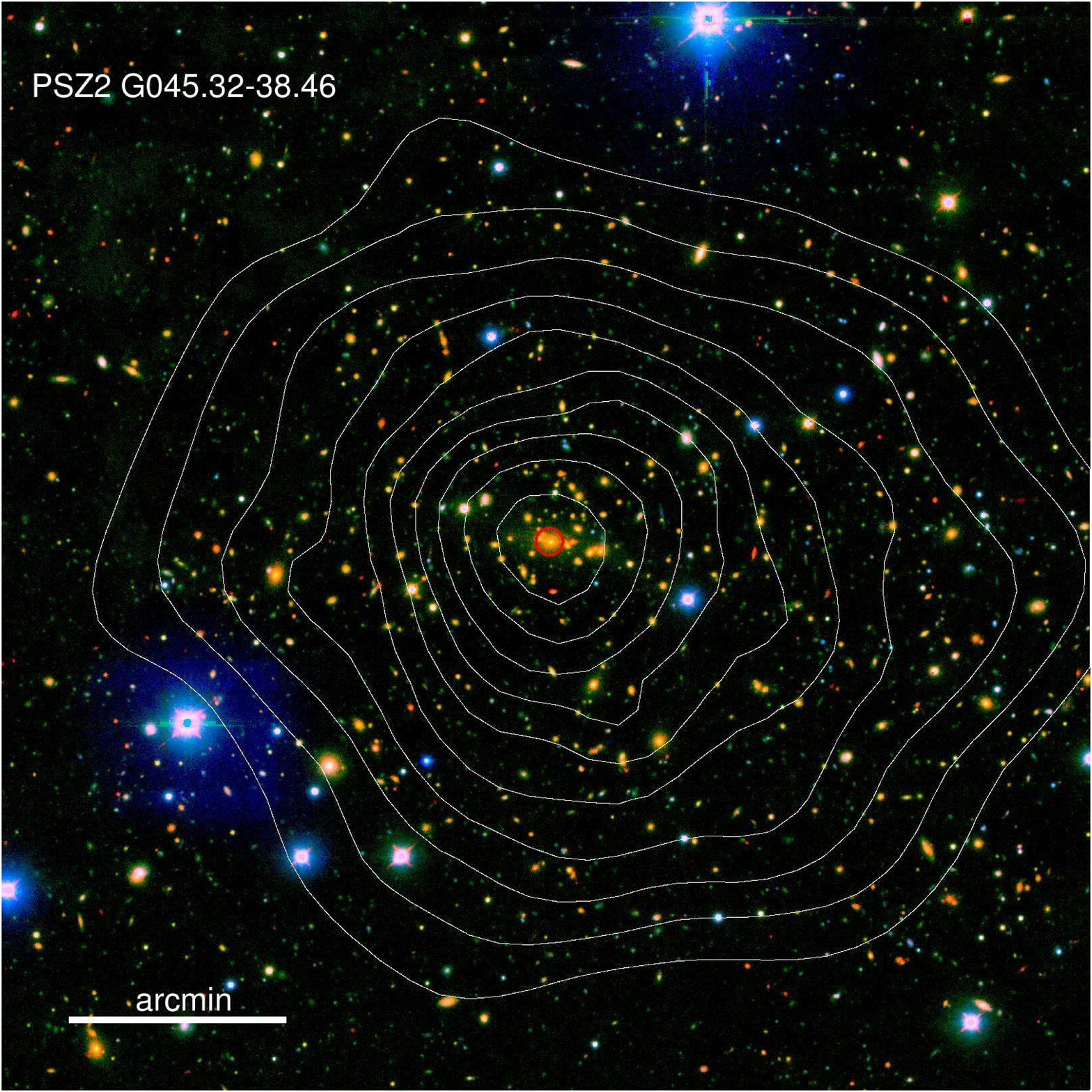}
\end{minipage}
\begin{minipage}{.495\textwidth}
  \centering
  \includegraphics[width=.90\linewidth]{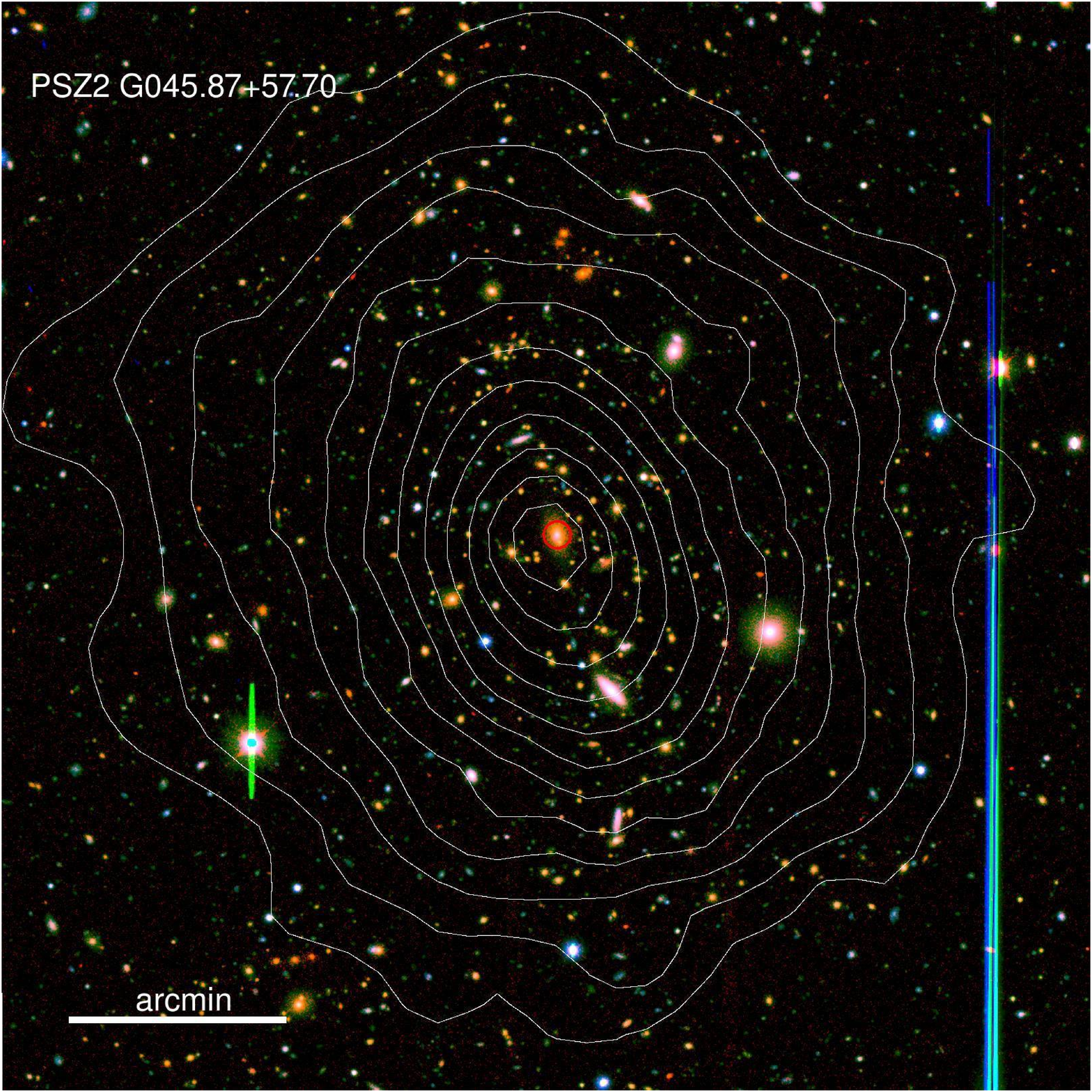}
\end{minipage}
\begin{minipage}{.495\textwidth}
  \centering
  \includegraphics[width=.90\linewidth]{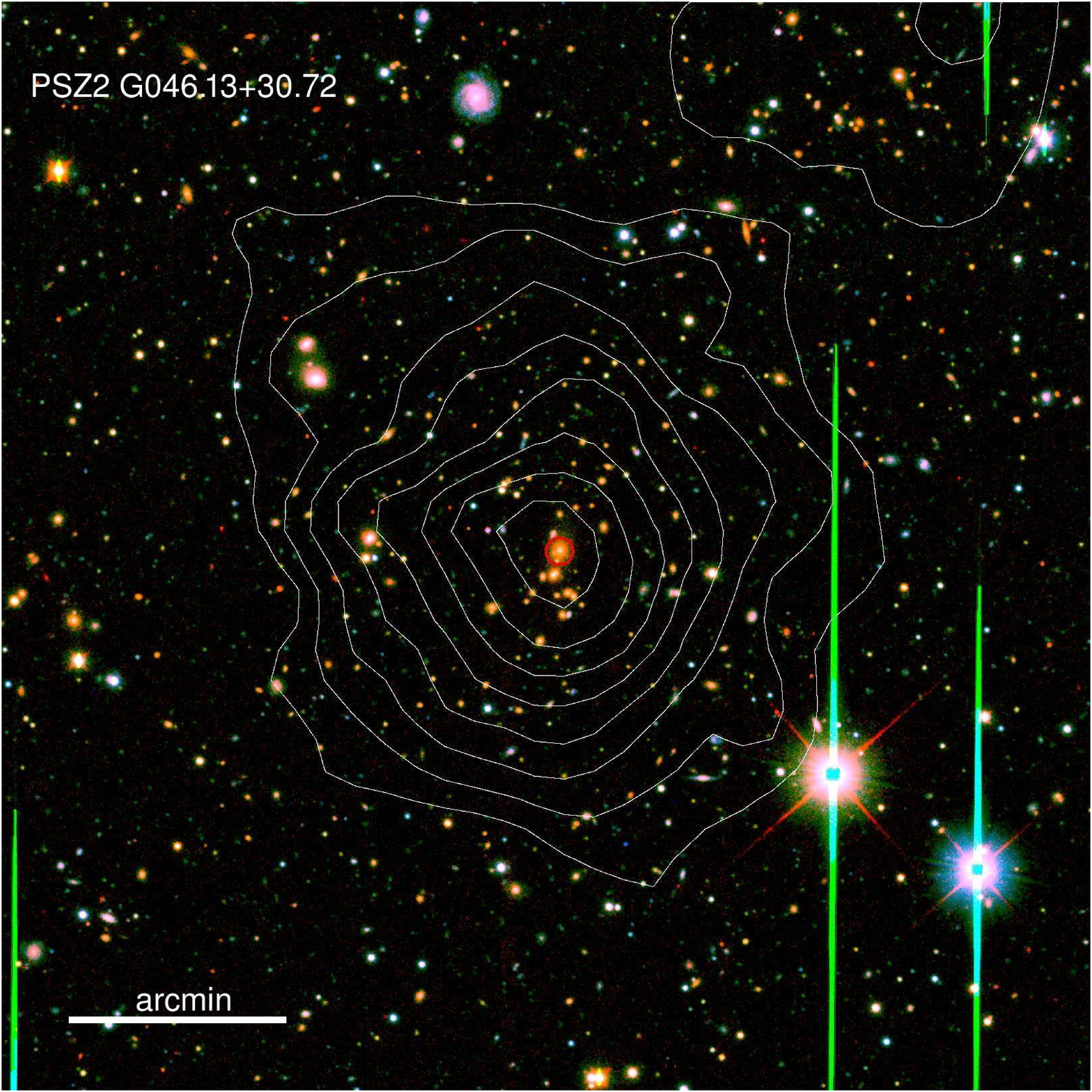}
\end{minipage}
\begin{minipage}{.495\textwidth}
  \centering
  \includegraphics[width=.90\linewidth]{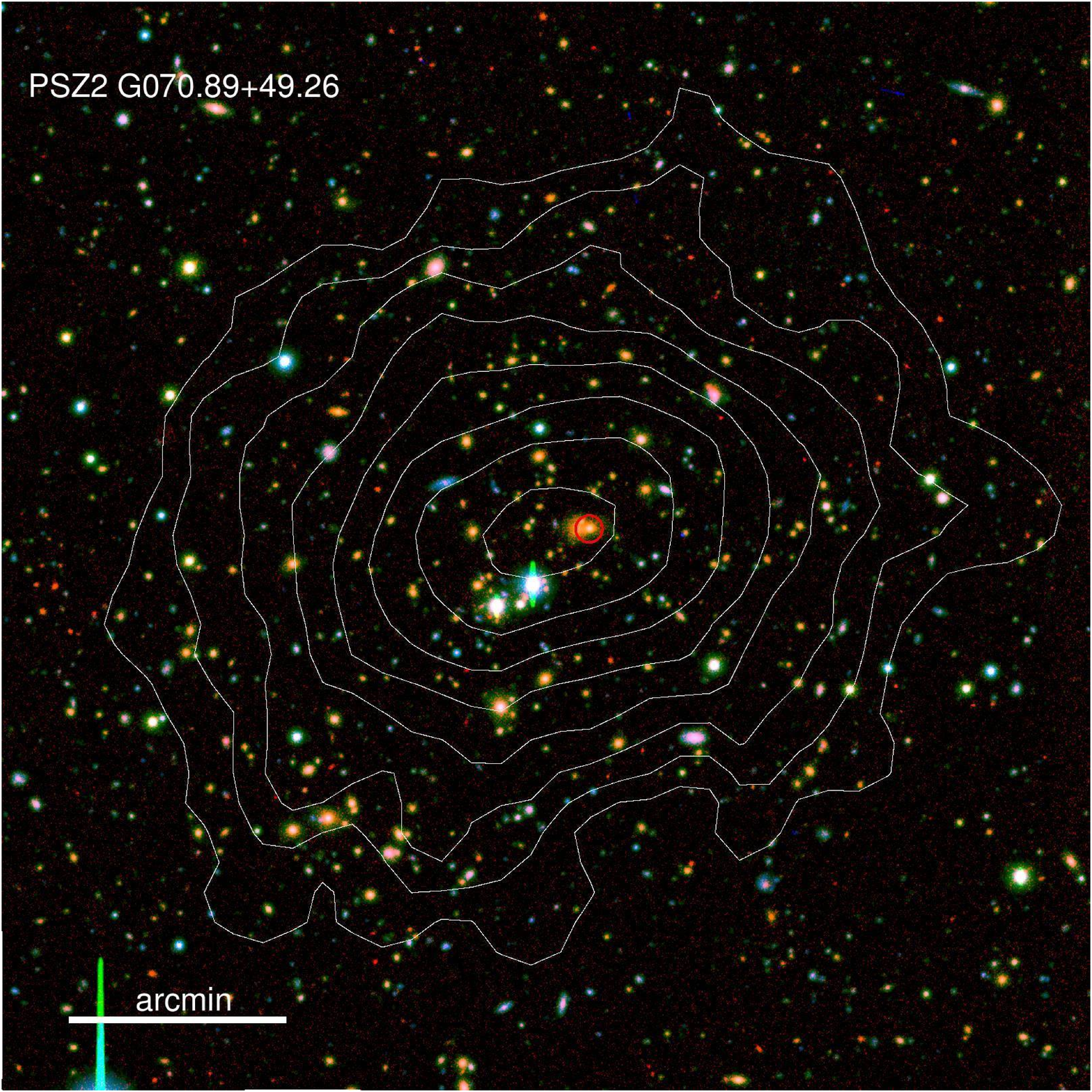}
\end{minipage}
\begin{minipage}{.495\textwidth}
  \centering
  \includegraphics[width=.90\linewidth]{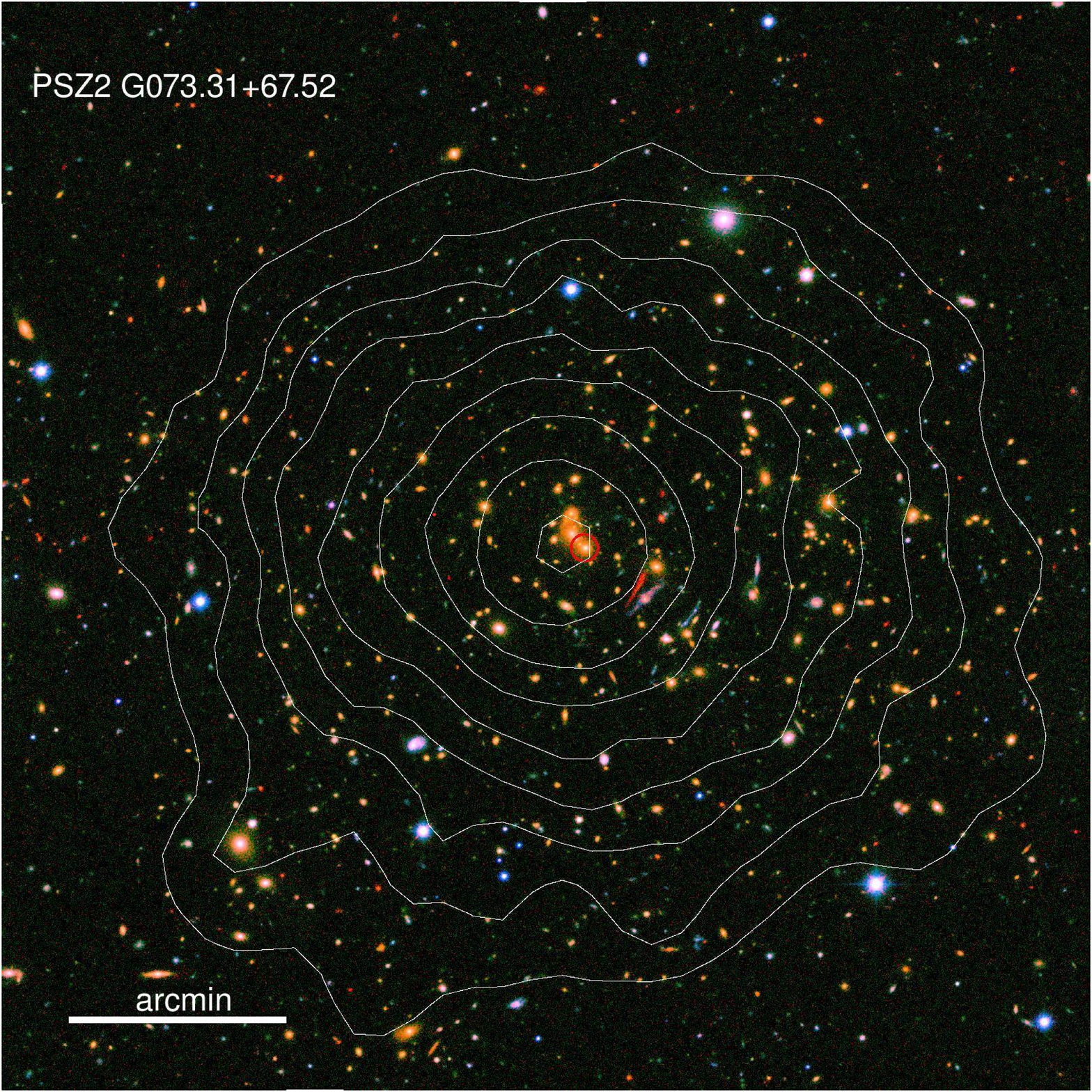}
\end{minipage}
\caption{Colour composite images of the clusters in our sample, based on $g$- or $B$-, $i$- or $I_\mathrm{c}$-, and $\mathrm{K_s}$-band imaging.}
\label{fig:gallery1}
\end{figure*}

\begin{figure*}
\centering
\begin{minipage}{.495\textwidth}
  \centering
  \includegraphics[width=.90\linewidth]{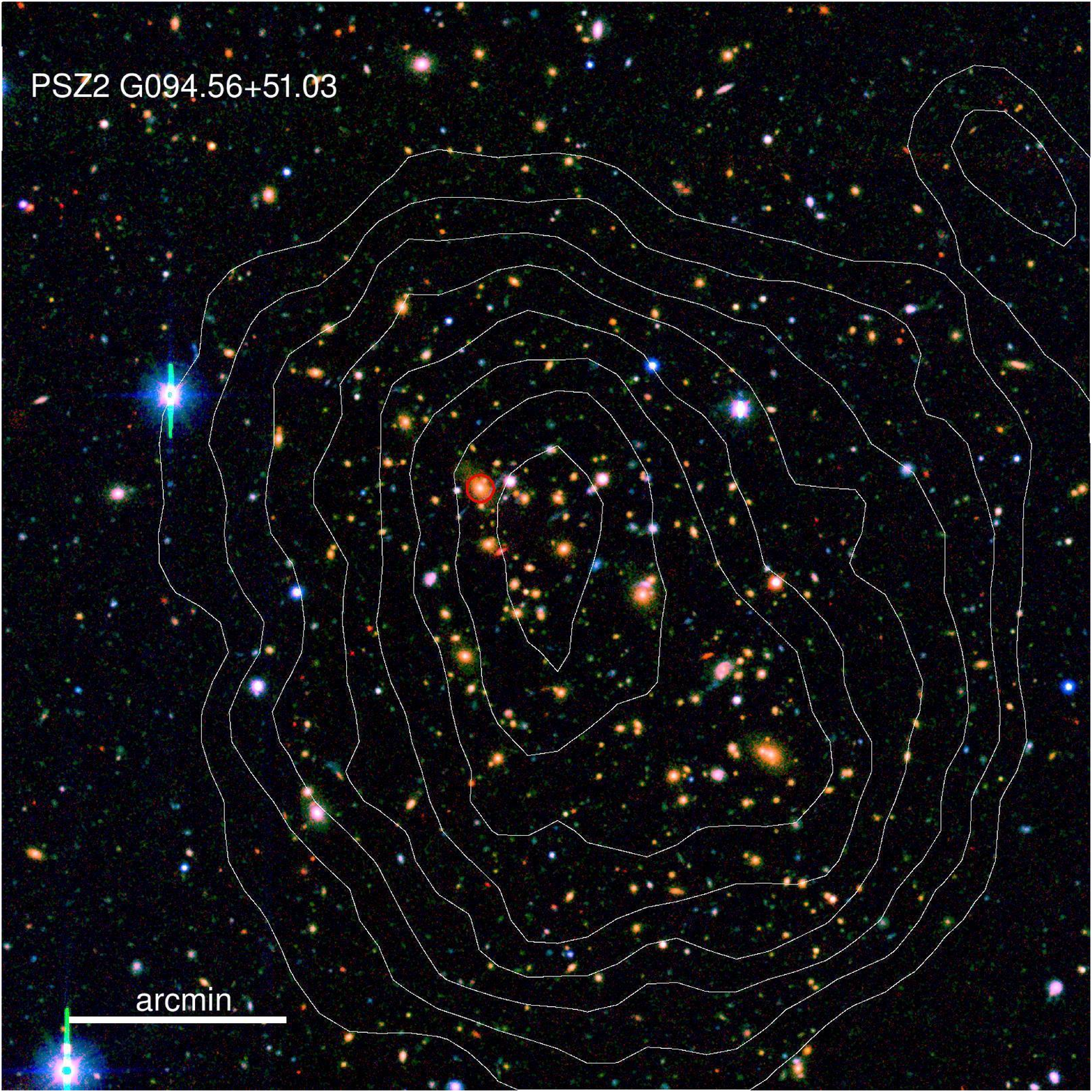}
\end{minipage}
\begin{minipage}{.495\textwidth}
  \centering
  \includegraphics[width=.90\linewidth]{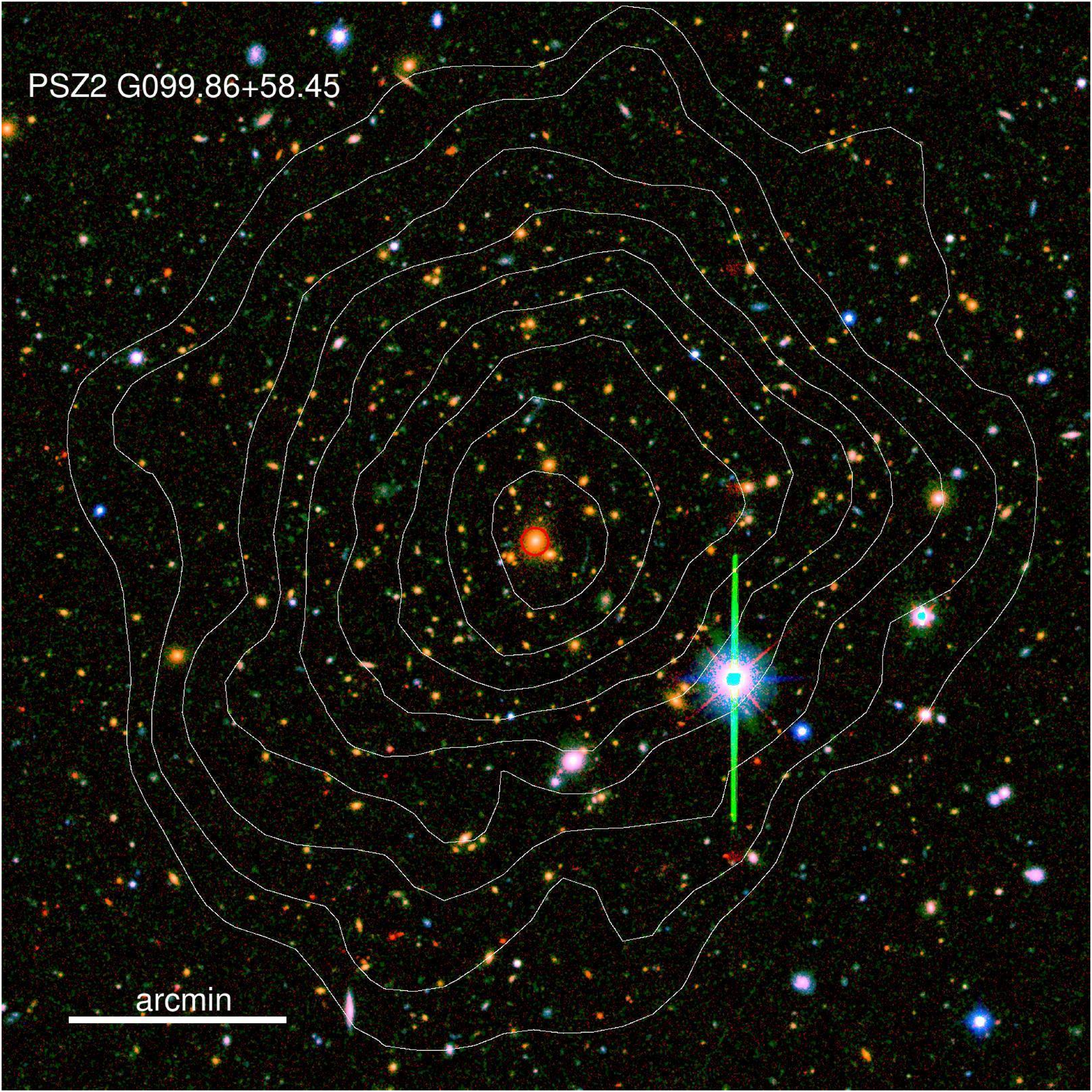}
\end{minipage}
\begin{minipage}{.495\textwidth}
  \centering
  \includegraphics[width=.90\linewidth]{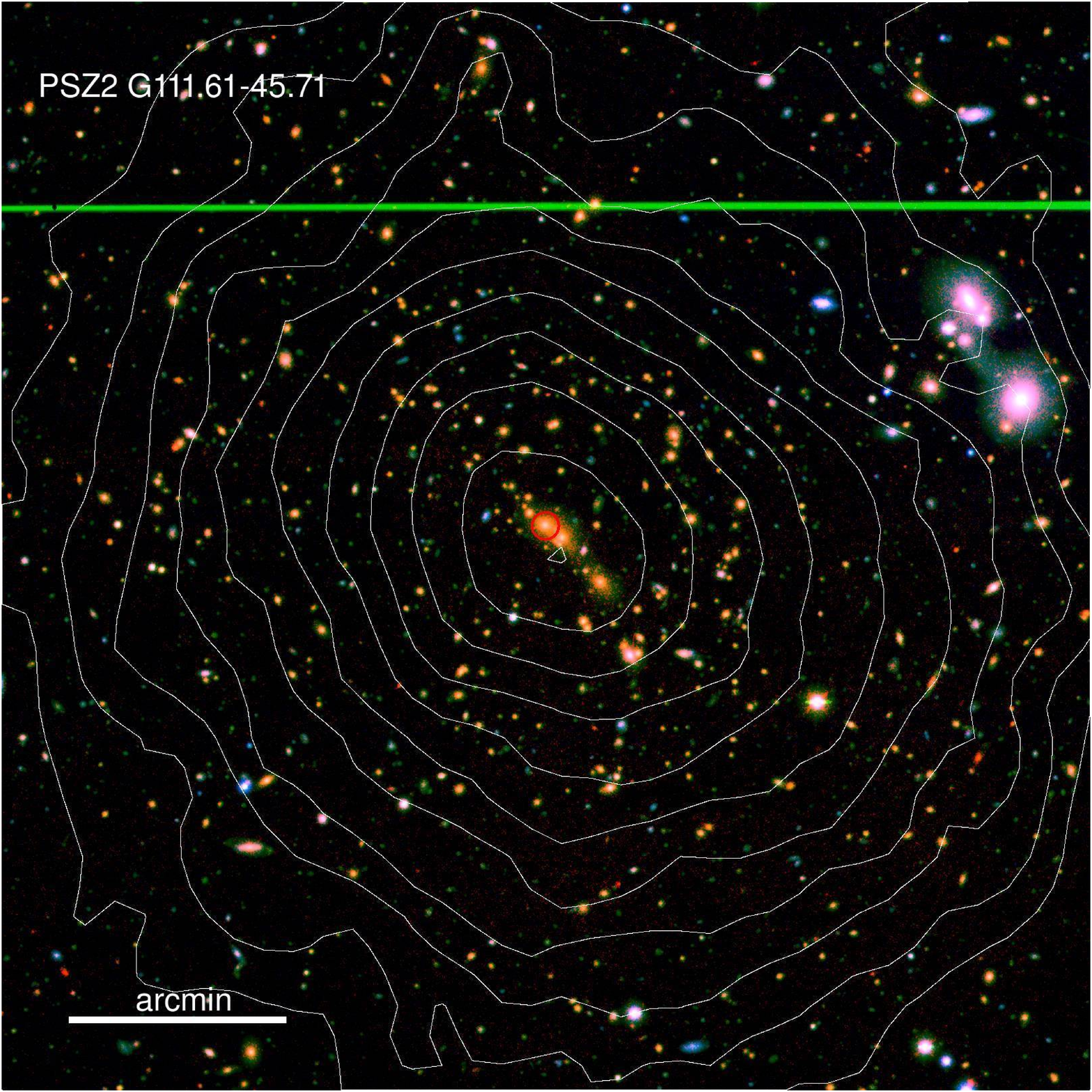}
\end{minipage}
\begin{minipage}{.495\textwidth}
  \centering
  \includegraphics[width=.90\linewidth]{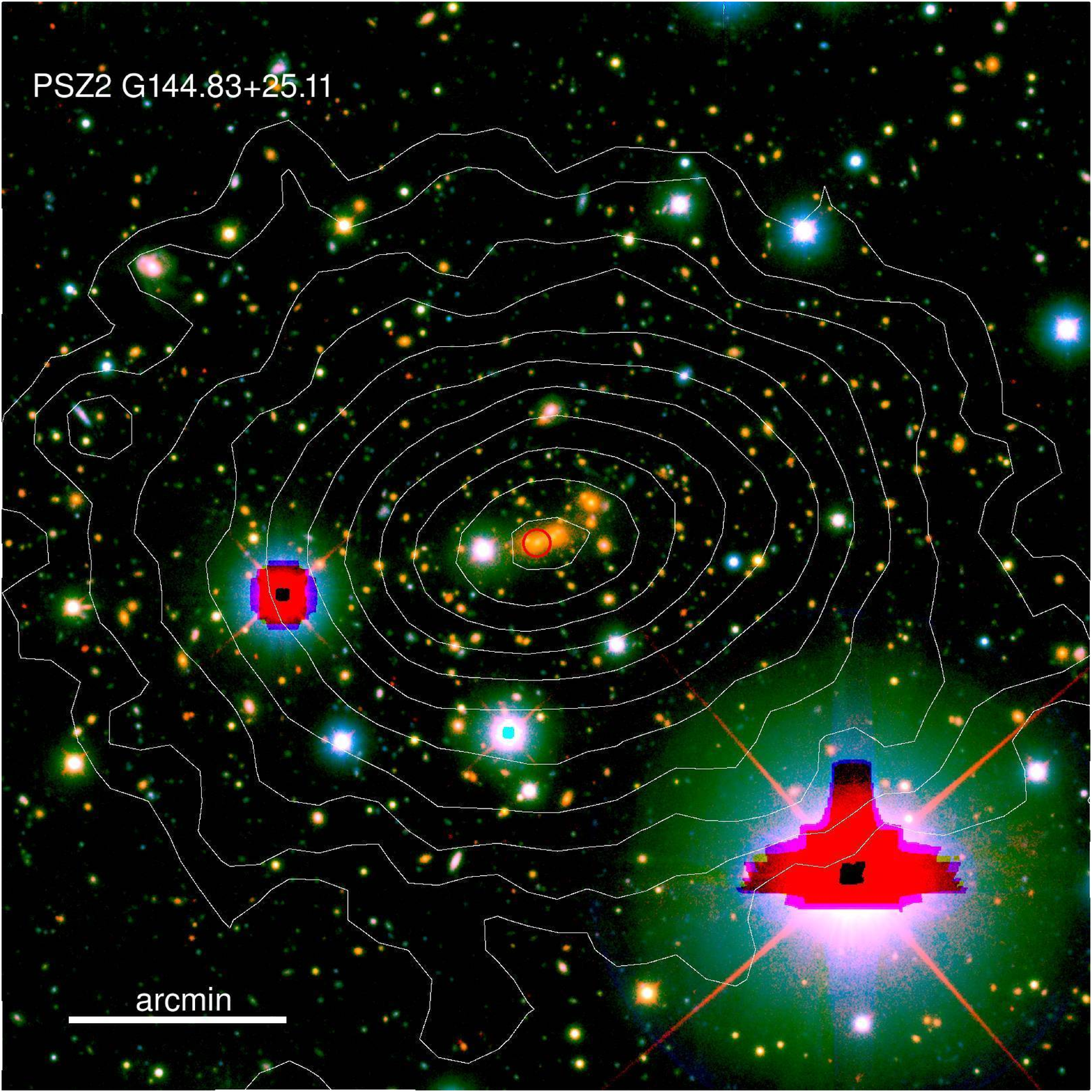}
\end{minipage}
\begin{minipage}{.495\textwidth}
  \centering
  \includegraphics[width=.90\linewidth]{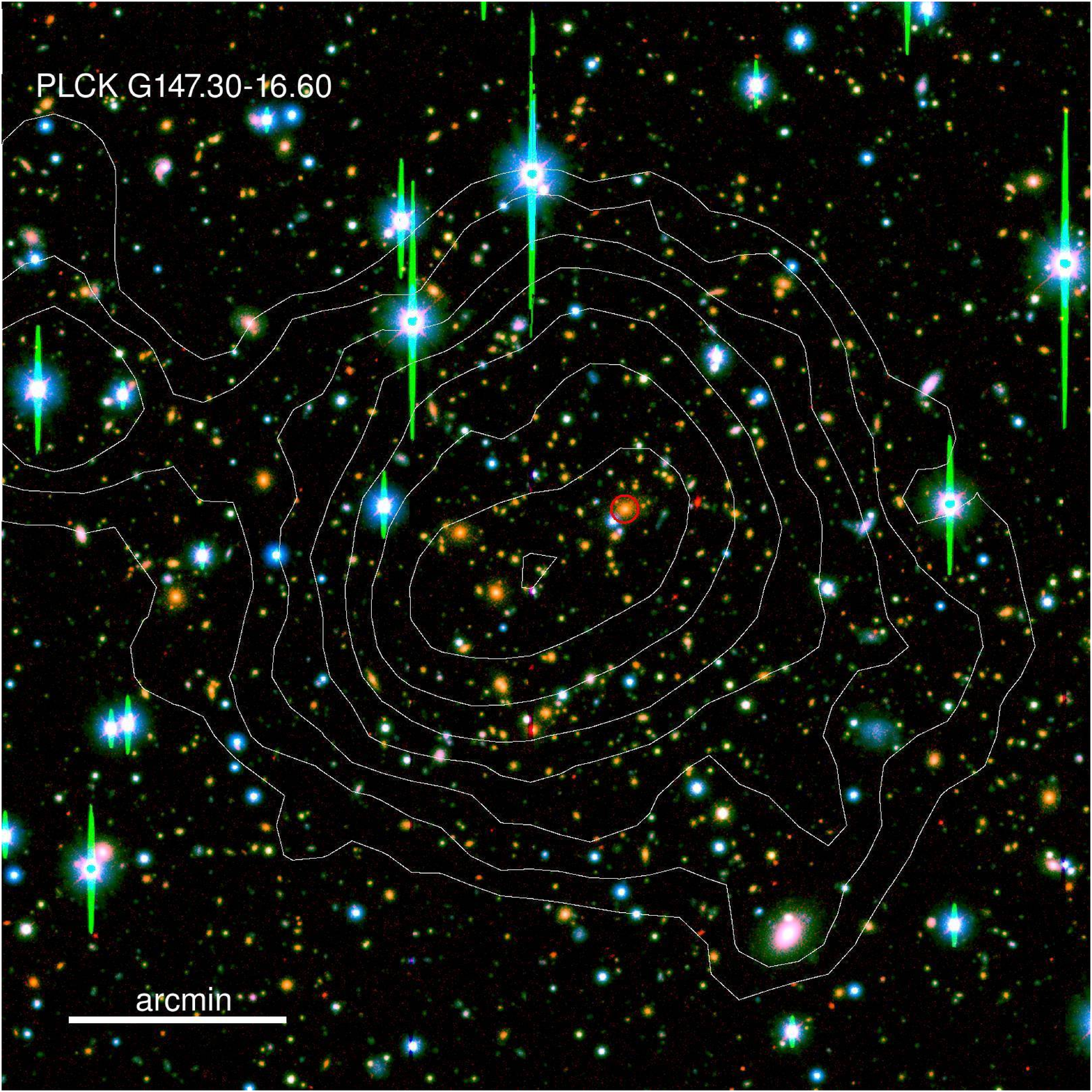}
\end{minipage}
\begin{minipage}{.495\textwidth}
  \centering
  \includegraphics[width=.90\linewidth]{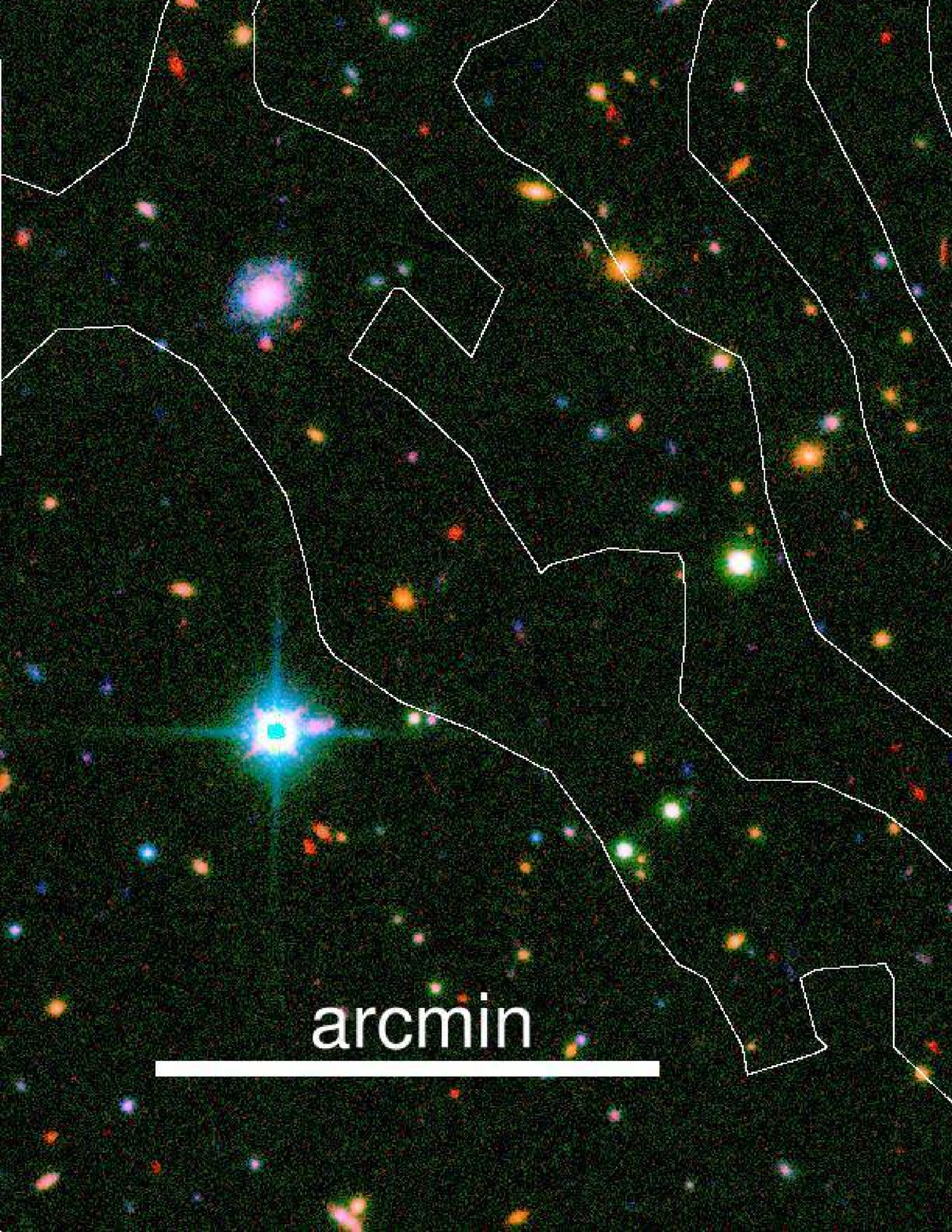}
\end{minipage}
\caption{... continued.}
\label{fig:gallery2}
\end{figure*}

\begin{figure*}
\centering
\begin{minipage}{.495\textwidth}
  \centering
  \includegraphics[width=.90\linewidth]{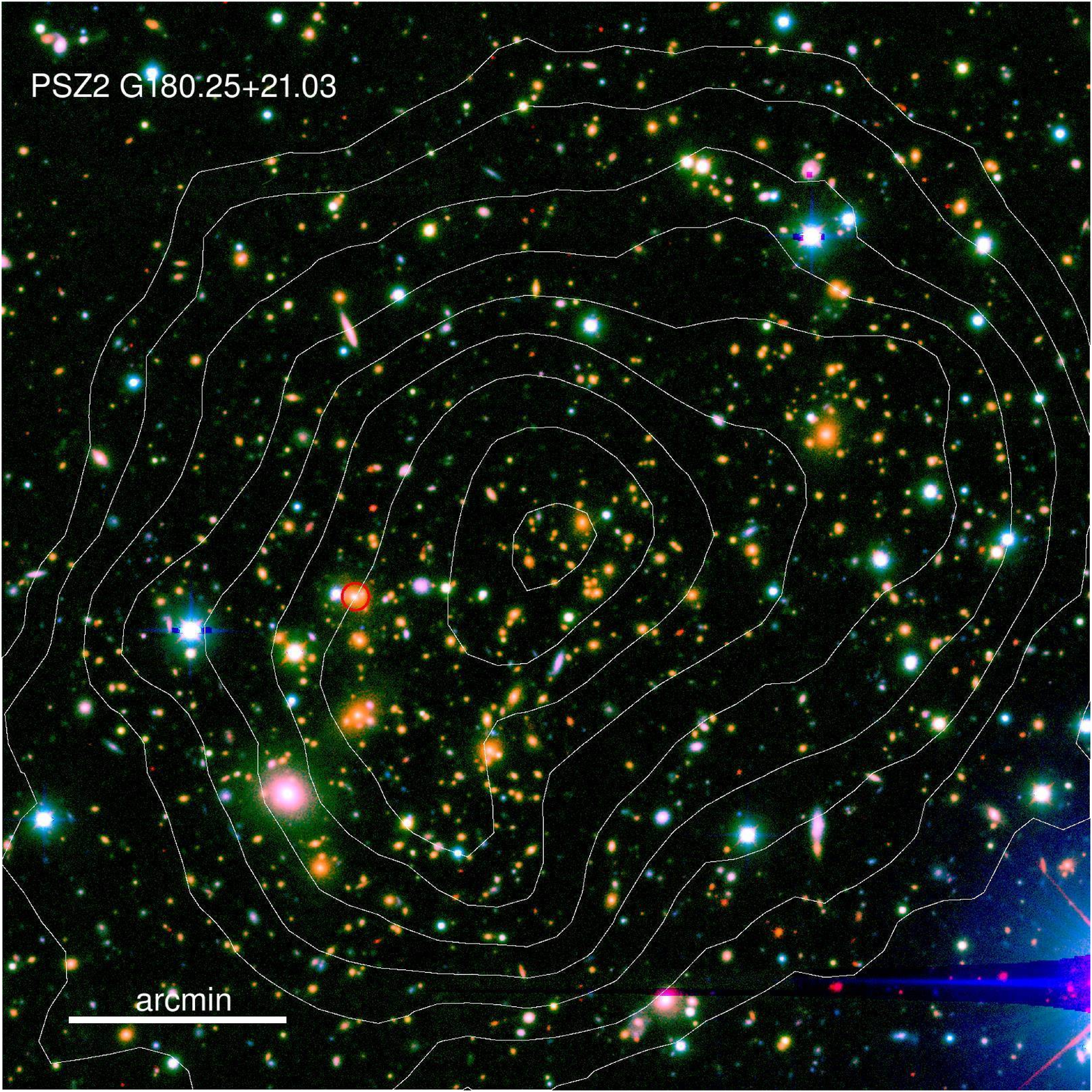}
\end{minipage}
\begin{minipage}{.495\textwidth}
  \centering
  \includegraphics[width=.90\linewidth]{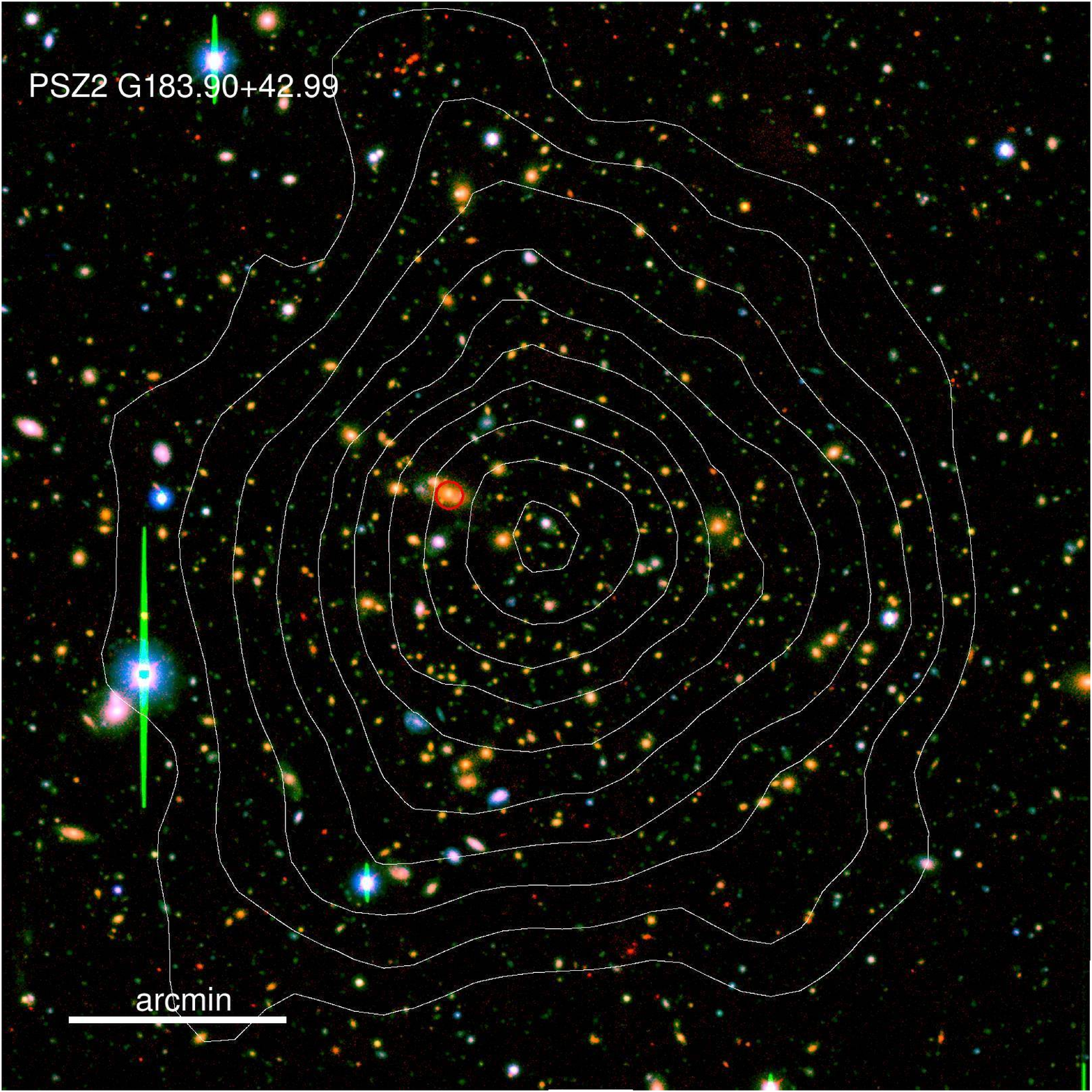}
\end{minipage}
\begin{minipage}{.495\textwidth}
  \centering
  \includegraphics[width=.90\linewidth]{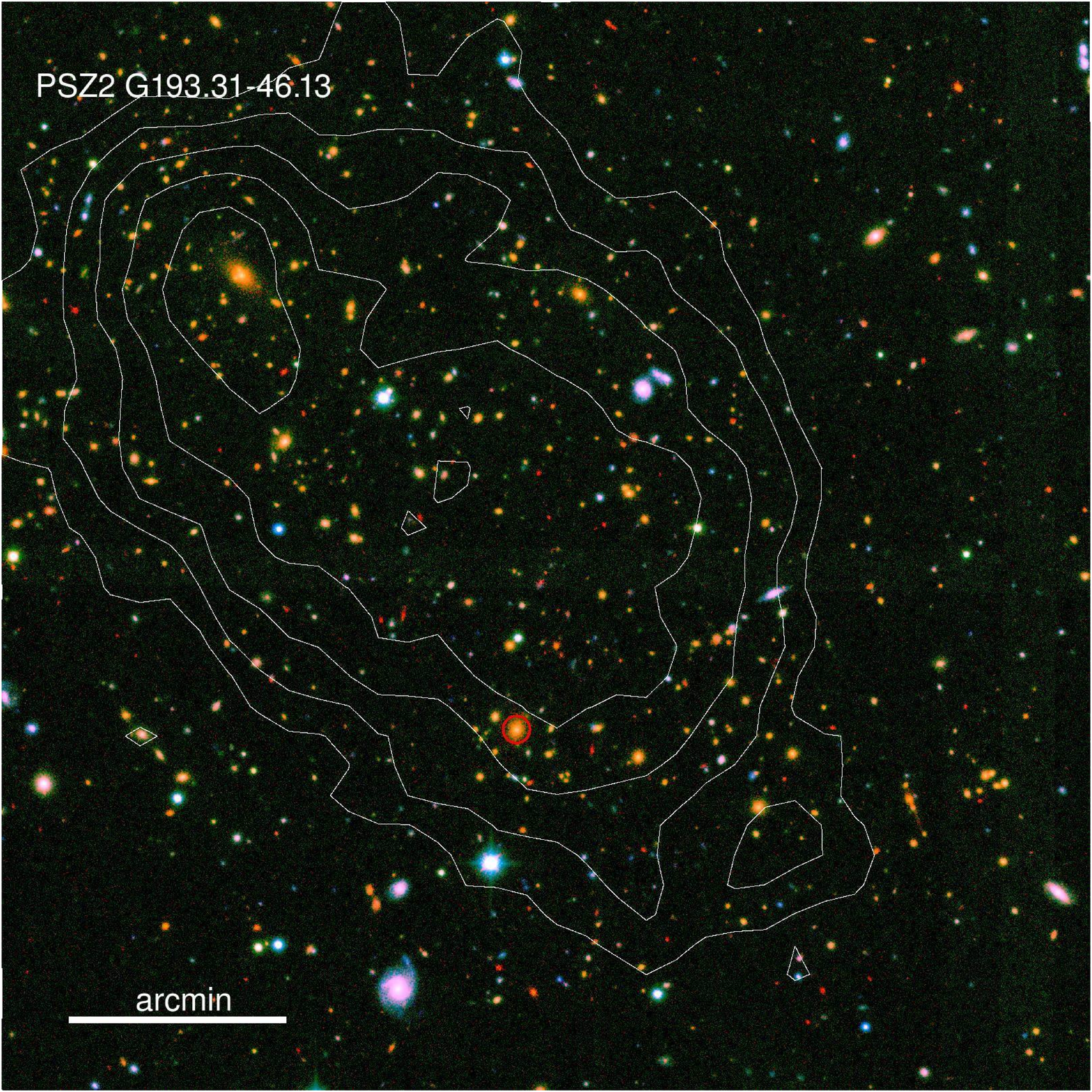}
\end{minipage}
\begin{minipage}{.495\textwidth}
  \centering
  \includegraphics[width=.90\linewidth]{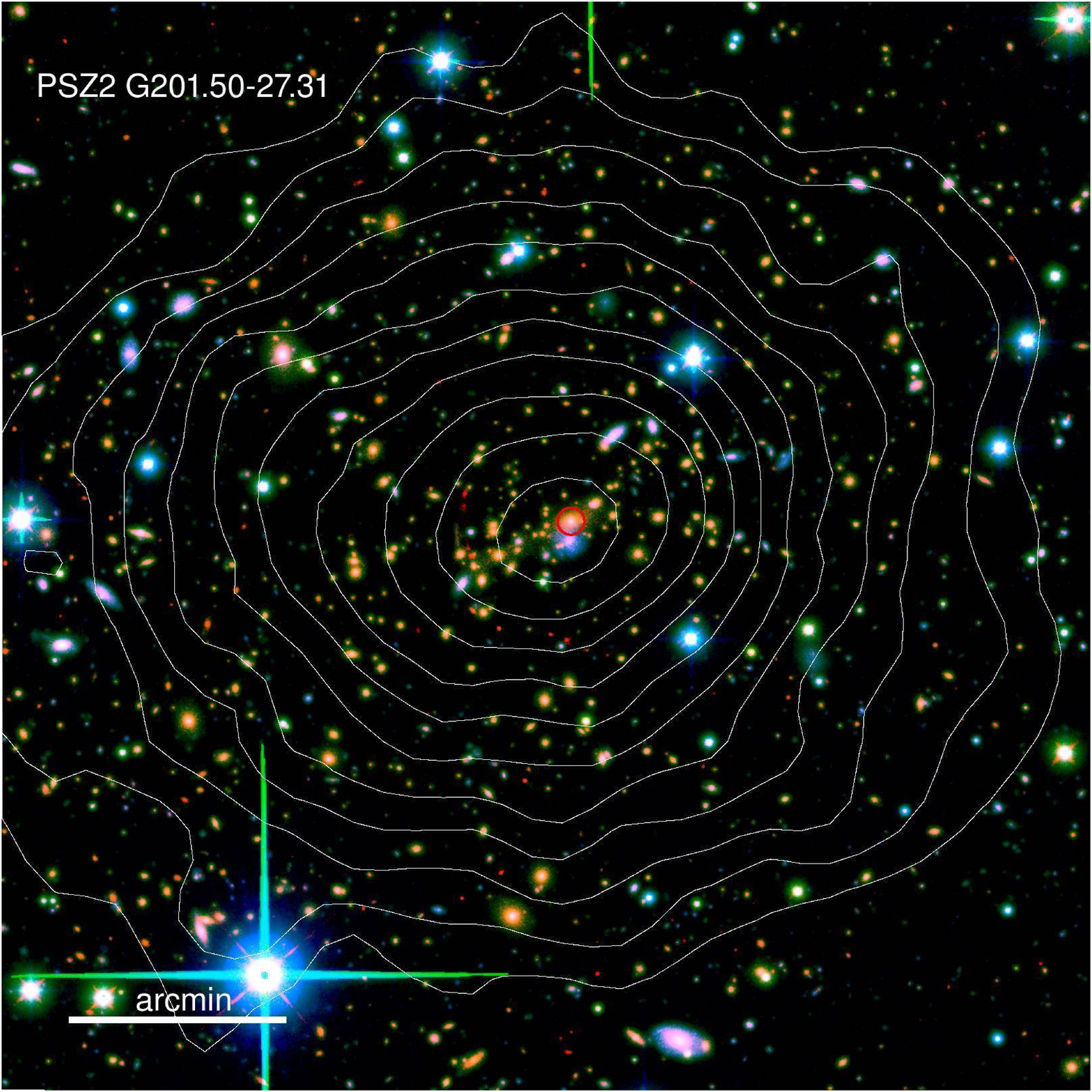}
\end{minipage}
\begin{minipage}{.495\textwidth}
  \centering
  \includegraphics[width=.90\linewidth]{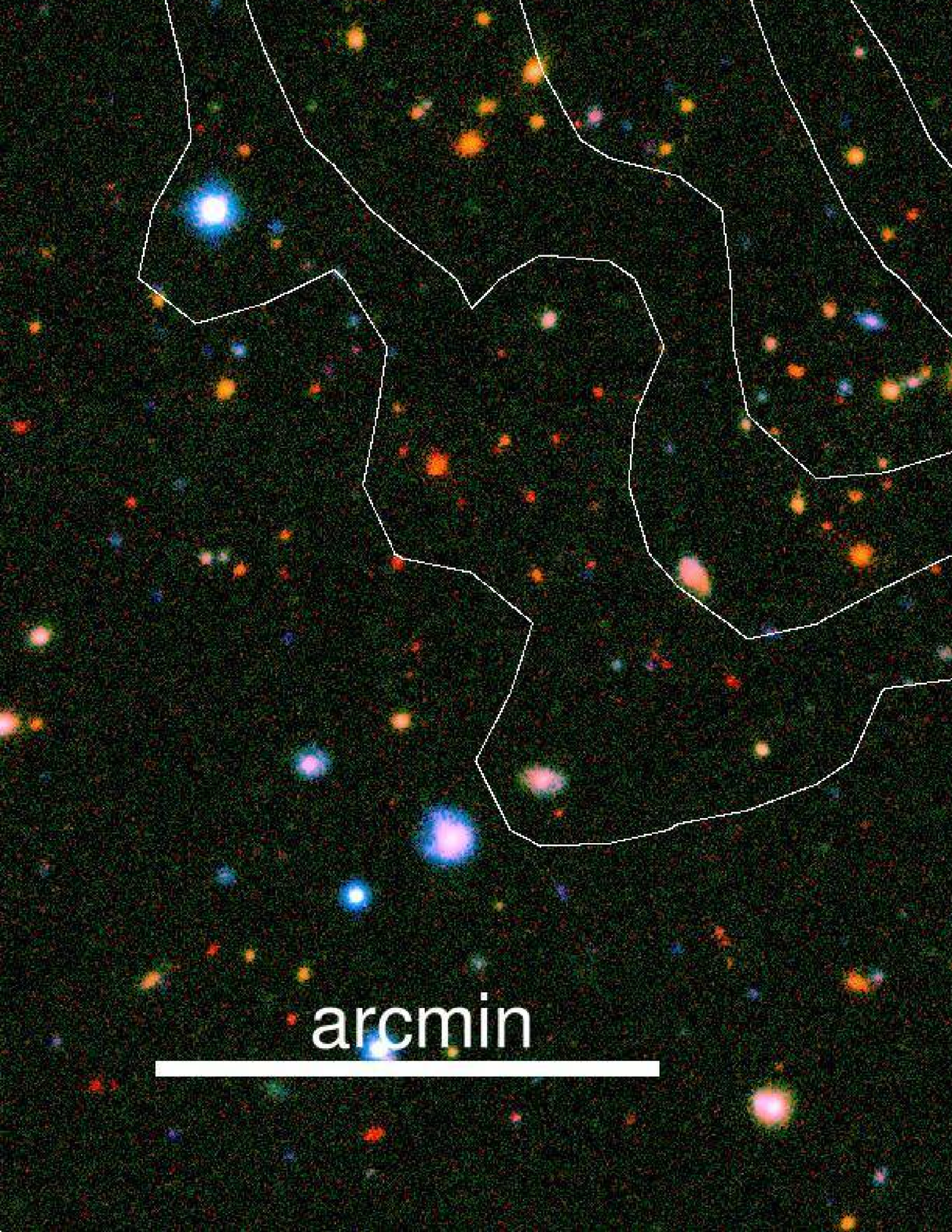}
\end{minipage}
\begin{minipage}{.495\textwidth}
  \centering
  \includegraphics[width=.90\linewidth]{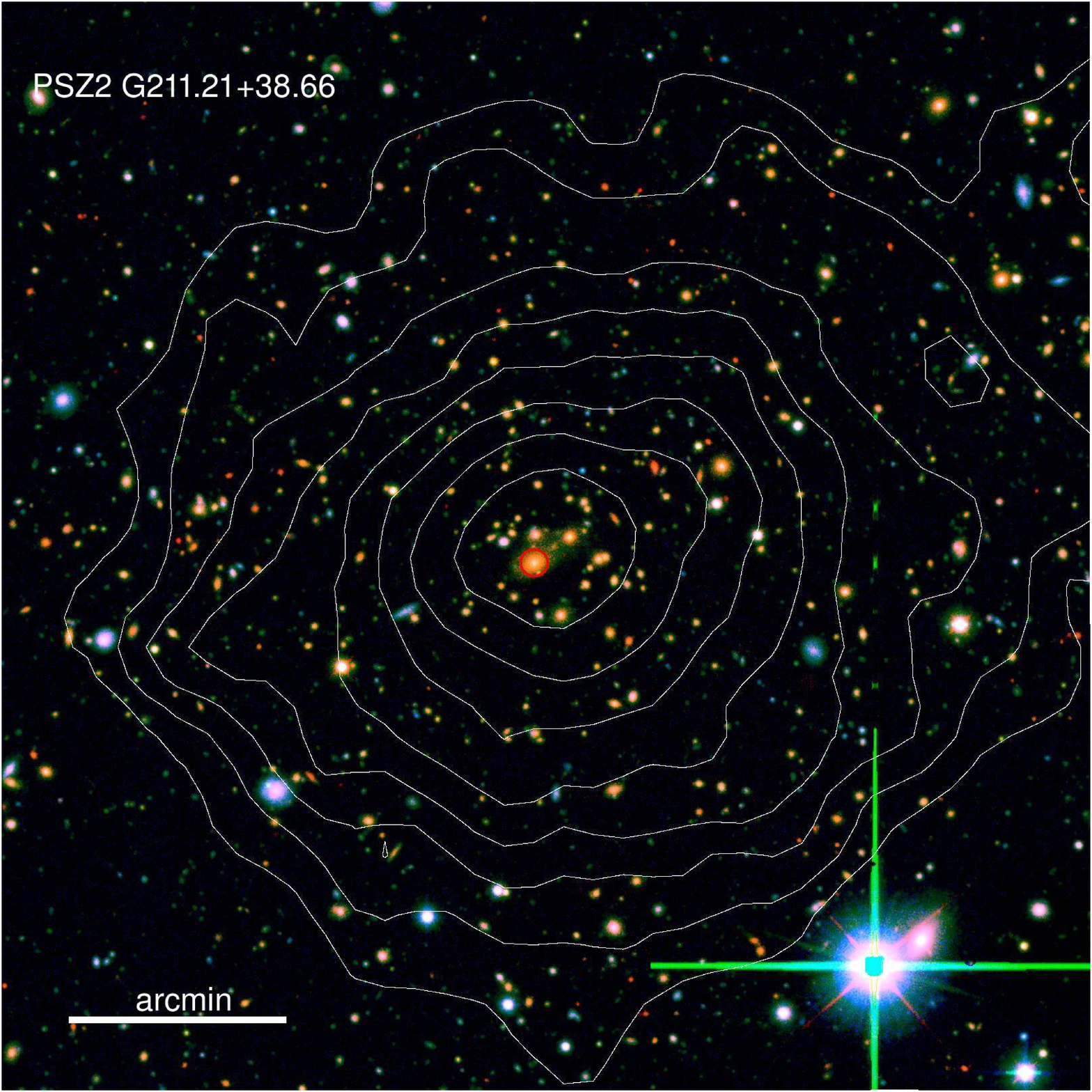}
\end{minipage}
\caption{... continued.}
\label{fig:gallery3}
\end{figure*}

\begin{figure*}
\centering
\begin{minipage}{.495\textwidth}
  \centering
  \includegraphics[width=.90\linewidth]{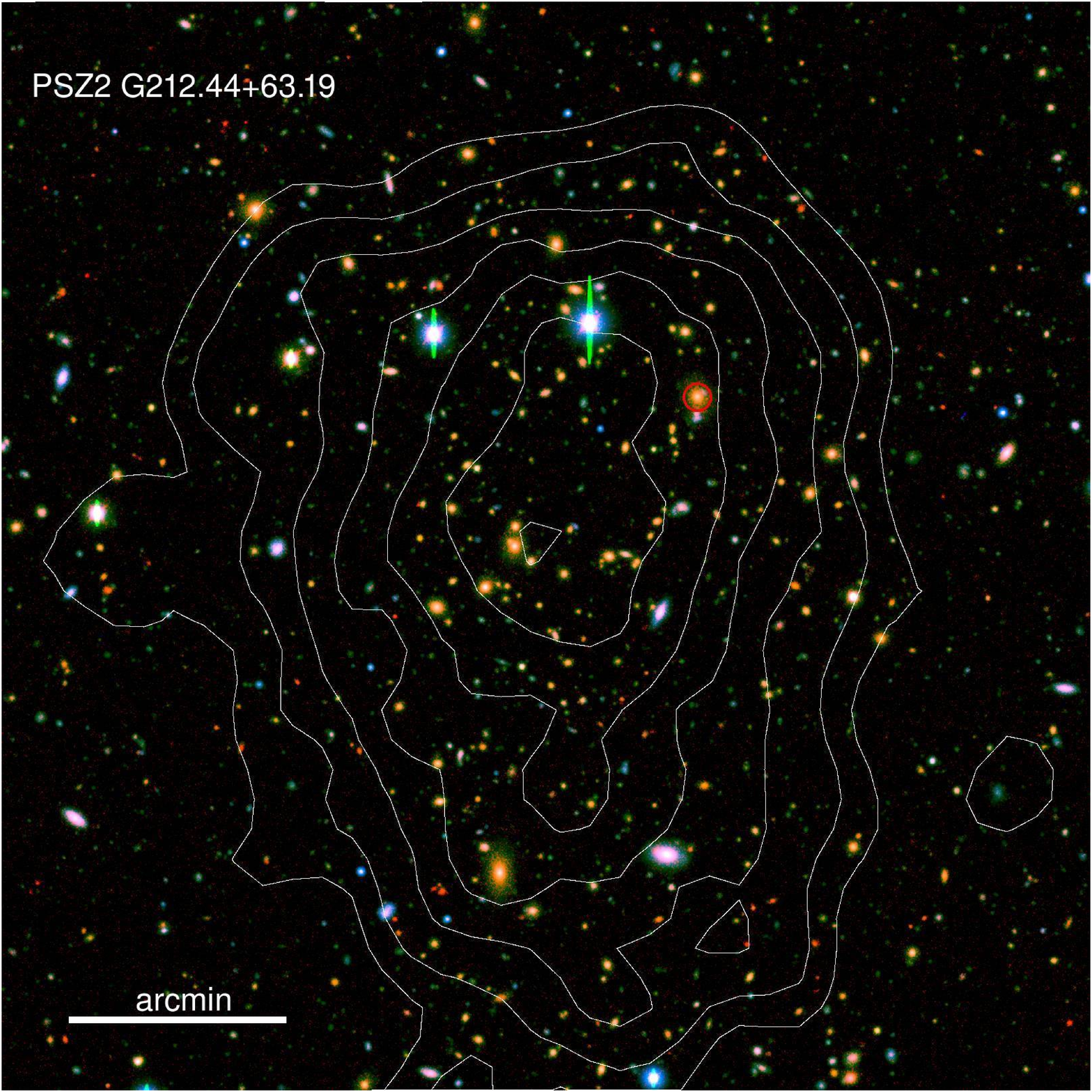}
\end{minipage}
\begin{minipage}{.495\textwidth}
  \centering
  \includegraphics[width=.90\linewidth]{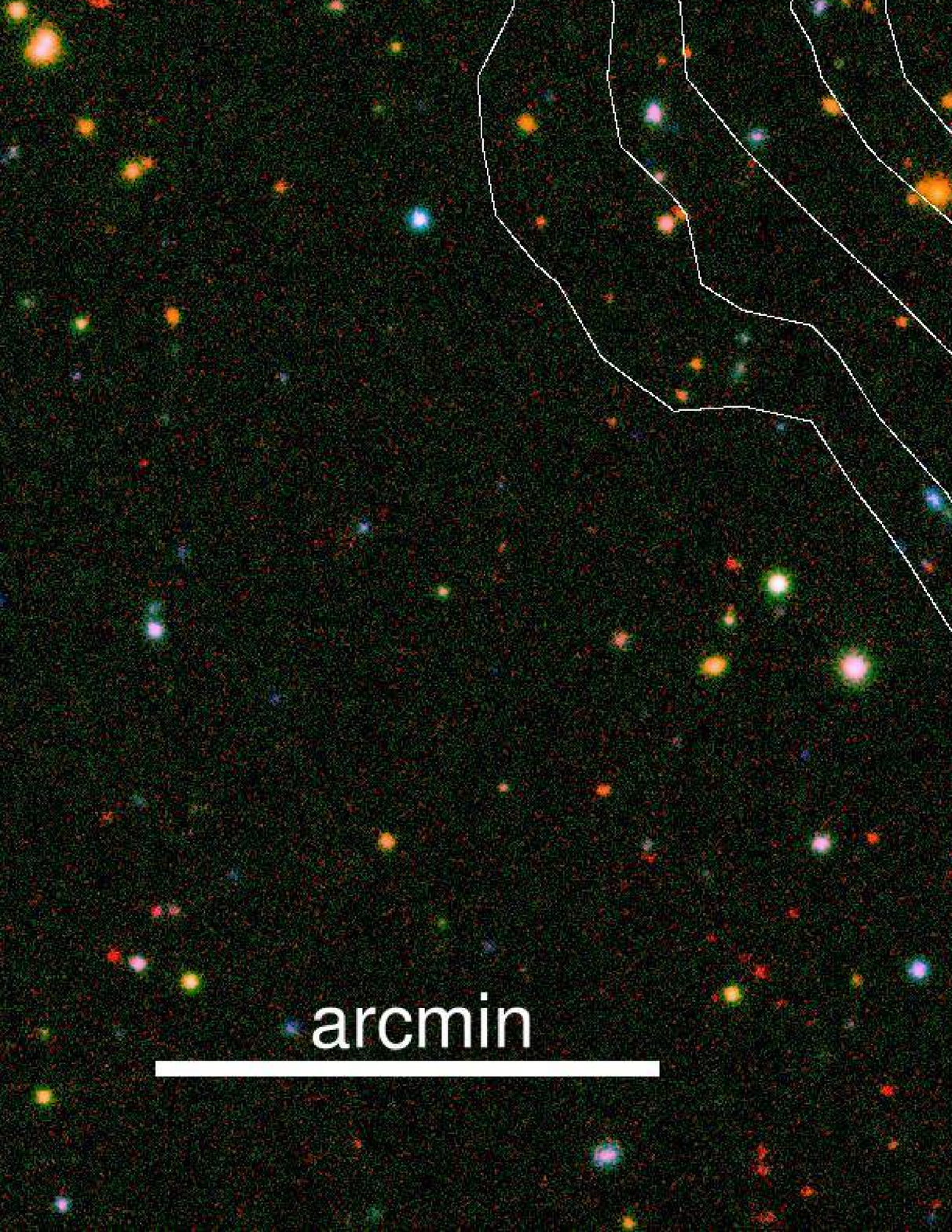}
\end{minipage}
\begin{minipage}{.495\textwidth}
  \centering
  \includegraphics[width=.90\linewidth]{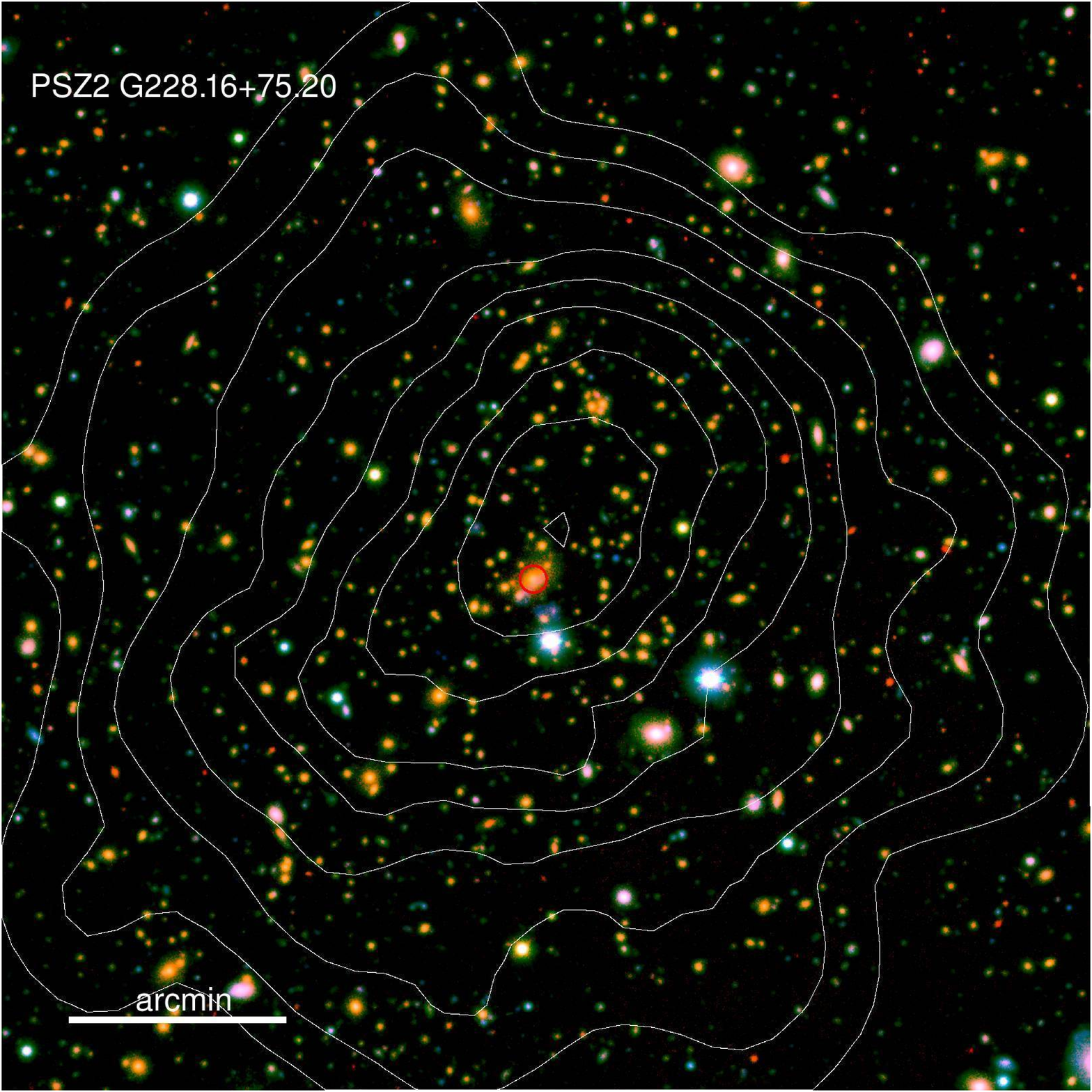}
\end{minipage}
\caption{... continued.}
\label{fig:gallery4}
\end{figure*}

Colour-composite images of all 21 clusters are shown in Figs.~\ref{fig:gallery1}-\ref{fig:gallery4}. They are composed of $g$- or $B$-, $i$- or $I_\mathrm{c}$-, and $\mathrm{K_s}$-band imaging. Regions around bright stars, and their diffraction spikes, are clearly visible here, but these are all masked and not considered in our analysis. 

Overplotted are X-ray surface brightness contours from the deep \textit{XMM-Newton} observations, which we have available for all clusters. They are based on (adaptively-) smoothed surface brightness maps, which are background subtracted, exposure corrected, and from which point sources are excised. Contours are logarithmically spaced with 0.2dex increments. These data form the basis of the X-ray morphological analysis, which is presented in Arnaud et al., in prep.


\end{appendix}
\end{document}